\newcommand{\tom}{\tilde{\omega}}

\documentclass{svjour3} 
\smartqed
\usepackage{times}
\usepackage[dvips]{graphicx}
\usepackage{bm,amsfonts,amssymb,amsmath}
\usepackage{graphicx}\usepackage{subfigure}

\begin{document}

\title{Back-engineering of spiking neural networks parameters.}

\author{H. Rostro, B. Cessac, J.C. Vasquez and T. Vieville}

\institute{H. Rostro, B. Cessac, J.C. Vasquez and Thierry Vieville \at
               INRIA (Neuromathcomp \& Cortex project teams)\\
               2004, Route des Lucioles\\ 
               06902, Sophia-Antipolis, France\\
              Tel.: +33-4-9238-7713\\
              Fax: +33-4-9238-7830\\           
              \email{\{hrostro, bcessac, jcvasquez, tvieville\}@sophia.inria.fr}        
              \and
              B. Cessac \at
              Laboratoire J. A. Dieudonn\'e, U.M.R. C.N.R.S. N° 6621\\
              Universit\'e de Nice Sophia-Antipolis, France
}

\date{Received: date / Accepted: date}
\maketitle

\begin{abstract}
We consider the deterministic evolution of a time-discretized spiking network of neurons with connection weights having delays,
modeled as a discretized neural network of the generalized integrate and fire (gIF) type.
The purpose is to study a class of algorithmic methods allowing to calculate the proper parameters to reproduce exactly a given spike train 
generated by an hidden (unknown) neural network.

This standard problem is known as NP-hard when delays are to be calculated. We propose here a reformulation, now expressed as a Linear-Programming (LP) problem,
thus allowing to provide an efficient resolution.
This allows us to ``back-engineer'' a neural network, i.e. to find out, given a set of initial conditions, which parameters (i.e., connection weights in this case),
allow to simulate the network spike dynamics.

More precisely we make explicit the fact that the back-engineering of a spike train,
is a Linear (L) problem if the membrane potentials are observed and a LP problem if only  spike times are observed, with a gIF model. Numerical robustness is discussed.
We also explain how it is the use of a generalized IF neuron model instead of a leaky IF model that allows us to derive this algorithm.

Furthermore, we point out how the L or LP adjustment mechanism is local to each unit and has the same structure as an ``Hebbian'' rule.
A step further, this paradigm is easily generalizable to the design of input-output spike train transformations.
This means that we have a practical method to ``program'' a spiking network, i.e. find a set of parameters allowing us to exactly reproduce the network output, given an input.

Numerical verifications and illustrations are provided.

\end{abstract}

\section{Introduction}
\label{sec:intro}

Neuronal networks have tremendous computational capacity, but their biological complexity make the exact reproduction of all the mechanisms involved in these networks dynamics essentially impossible, even at the numerical simulation level, as soon as the number of neurons becomes too large. On crucial issue is thus to be able to reproduce the ``output'' of a neuronal network using approximated models easy to implement numerically. The issue addressed here is ``Can we program an integrate and fire network, i.e. tune the parameters, in order to exactly reproduce another network output, on a bounded time horizon, given the input''.\\
This is the issue addressed here.

\subsection*{Calculability power of neural network models}

The main aspect we are interesting here is the \textit{calculability} of neural network models. 
It is known that recurrent neural networks with frequency rates are universal approximators \cite{schafer-zimmermann:06}, 
as multilayer feed-forward networks are \cite{hornik-etal:89}.
This means that neural networks are able to simulate dynamical systems as an example see the very interesting paper of Albers-Sprott using this property to investigate the dynamical stability conjections of  Pales and Smale in the field of dynamical systems theory \cite{albers-sprott:06} or route to chaos in high dimensional systems \cite{albers-sprott-2:06}, not only to approximate measurable functions on a compact domain, 
as originally stated (see, e.g., \cite{schafer-zimmermann:06} for a detailed introduction on these notions).
Spiking neuron networks have been proved to be also universal approximators \cite{maass:01}.

Theoretically, spiking neurons can perform very powerful computations with precise timed spikes.
They are at least as computationally powerful, as the sigmoidal neurons traditionally used in artificial neural networks \cite{maass:97,maass-natschlager:97}.
This results has been shown using a spike-response model (see \cite{maass-bishop:03} for a review) and considering piece-wise linear approximations of the potential profiles.
In this context, analog inputs and outputs are encoded by temporal delays of spikes.
The authors show that any feed-forward or recurrent (multi-layer) analog neuronal network (\`a-la Hopfield, e.g., McCulloch-Pitts) 
can be simulated arbitrarily closely by an insignificantly larger network of spiking neurons. This holds even in the presence of noise \cite{maass:97,maass-natschlager:97}. 
These results highly motivate the use of spiking neural networks, as studied here.

In a computational context, spiking neuron networks are mainly implemented through specific network architectures, 
such as Echo State Networks \cite{jaeger:03} and Liquid Sate Machines \cite{maass-etal:02}, that are called ``reservoir computing'' 
(see \cite{verstraeten-etal:07} for unification of reservoir computing methods at the experimental level). 
In this framework, the reservoir is a network model of neurons (it can be linear or sigmoid neurons, but more usually spiking neurons), with a random topology and a sparse connectivity. The reservoir is a recurrent network, with weights than can be either fixed or driven by an unsupervised learning mechanism.
In the case of spiking neurons (e.g. in the model of \cite{paugam-moisy-etal:08}), the learning mechanism is a form of synaptic plasticity, usually STDP (Spike-Time-Dependent Plasticity), or a temporal Hebbian unsupervised learning rule, biologically inspired.
The output layer of the network (the so-called ``readout neurons'') is driven by a supervised learning rule, generated from any type of classifier or regressor, ranging from a least mean squares rule to sophisticated discriminant or regression algorithms. 
The ease of training and a guaranteed optimality guides the choice of the method.
It appears that simple methods yield good results \cite{verstraeten-etal:07}.
This distinction between a readout layer and an internal reservoir is indeed induced by the fact that only the output of the neuron network activity is constrained, whereas the internal state is not controlled.

\subsection*{Learning the parameters of a neural network model}

In biological context, learning is mainly related to 
synaptic plasticity \cite{gerstner-kistler:02,cooper-intrator-etal:04} and STDP 
(see e.g., \cite{toyoizumi-etal:07} for a recent formalization), as far as spiking neuron networks are concerned. This unsupervised learning mechanism is known to reduce the variability of neuron responses
\cite{bohte-mozer:07} and is related to the maximization of information transmission \cite{toyoizumi-etal:05} and mutual information \cite{chechik:03}.
It has also other interesting computational properties such as tuning neurons to react as soon as possible to the earliest spikes, or segregate the network response
in two classes depending on the input to be discriminated, and more general structuring such as emergence of orientation selectivity
 \cite{guyonneau-vanrullen-etal:04}.

In the present study, the point of view is quite different: we consider supervised learning while,
since ``each spike may matter'' \cite{guyonneau-vanrullen-etal:04,delorme-perrinet-etal:01}, 
we want not only to statistically reproduce the spiking output, but also to reproduce it exactly.

The motivation to explore this track is twofold. On one hand we want to better understand what can be learned at a theoretical level by spiking neuron networks, 
tuning weights and delays. The key point is the non-learnability of spiking neurons \cite{sima-sgall:05}, 
since it is proved that this problem is NP-complete, when considering the estimation of both weights and delays. 
Here we show that we can ``elude'' this caveat and propose an alternate efficient estimation, inspired by biological models.

We also have to notice, that the same restriction apply not only to simulation but, as far as this model is biologically plausible, also holds at the biological level. 
It is thus an issue to wonder if, in biological neuron networks, delays are really estimated during learning processes, or if a weaker form of weight adaptation, as developed now, is considered.

On the other hand, the computational use of spiking neuron networks in the framework of reservoir computing or beyond \cite{schrauwen:07}, at application levels, 
requires efficient tuning methods not only in ``average'', but in the deterministic case. 
This is the reason why we must consider how to {\em exactly} generate a given spike train.

\subsection*{What is the paper about}

In the next section we detail the proposed methods in three steps: first, discussing the neural model considered here, provided underlying assumptions are assumed; 
then, detailing the family of estimation problems corresponding to what is called back-engineering and discussing the related computational properties; finally, making explicit how a general input/output mapping can be ``compiled'' on a spiking neural network thanks to the previous developments.

In the subsequent section, numerical verifications and illustrations are provided, before the final discussion and conclusion.

\section{Problem position: Discretized integrate and fire neuron models.} 
\label{sec:model}

Let us consider a {normalized} and {reduced} ``punctual conductance based generalized integrate and fire'' (gIF) neural unit model 
\cite{destexhe:97} as reviewed in \cite{rudolph-destexhe:06}.
The model is reduced in the sense that both {adaptive} currents and non-linear {ionic} currents are no more explicitly depending on the potential membrane,
but on time and previous spikes only (see \cite{cessac-vieville:08b} for a development). 

Here we follow \cite{cessac:08,cessac-vieville:08,cessac-vieville:08b} after \cite{soula-chow:07} and review how to properly discretize a gIF model. 
The precise derivation is not re-given here, except for one innovative aspect,
whereas the method is only sketched out (see \cite{cessac-vieville:08,cessac-vieville:08b} for details).

\paragraph{Time constrained continuous formulation.}

Let $v$ be the normalized membrane potential, which triggers a spike for $v = 1$ and is reset at $v = 0$. 
The fire regime (spike emission) reads $v(t) = 1 \Rightarrow v(t^+) = 0$.
Let us write $\tom_t = \{ \cdots t_i^n \cdots \}$, the list of spike times $t_i^n < t$.
Here $t_i^n$ is the $n$-th spike-time of the neuron of index $i$. The dynamic of the integrate regime reads:

\begin{displaymath}
\frac{dv}{dt} + \frac{1}{\tau_L} \left[v - E_L\right] +
 \sum_{j} \sum_{n} \rho_j\left(t - t_j^n\right) \left[v - E_j\right]
 = i_m(\tom_t),
 \end{displaymath}
 
Here, $\tau_L$ and $E_l$ are the membrane leak time-constant and reverse potential, 
while $\rho_j()$ and $E_j$ are the spike responses and reverse potentials for excitatory/inhibitory synapses and gap-junctions.
Furthermore, $\rho_j(s)$ is the synaptic or gap-junction response, accounting for the connection delay and time constant;
it is assumed that the response vanishes after a delay $\tau_r$, where $\rho_j(s):$ having the following shape:\\

Finally, $i_m()$ is the reduced membrane current, including simplified adaptive and non-linear ionic current (see \cite{cessac-vieville:08b} for details).

The dynamic of the integrate regime thus writes:\\

\begin{displaymath}
\frac{dv}{dt} + g(t, \tom_t) \, v = i(t, \tom_t),
\end{displaymath}

so that knowing the membrane potential at time $t$, the membrane potential at time $t + \delta$, writes:

\begin{displaymath}\label{IFuSol}\begin{array}{rcl}
 v(t + \delta) &=& \nu(t, t + \delta, \tom_{t}) \, v(t) + \int_{t}^{t + \delta} \nu(s, t + \delta, \tom_{s}) \, i(s, \tom_{s}) \, ds,  \\
\log(\nu(t_0, t_1, \tom_{t_0})) &=& -\int_{t_0}^{t_1} g(s, \tom_{s}) \, d s. \\
\end{array}
\end{displaymath}

The key point is that temporal constraints are to be taken into account \cite{cessac-etal:08a}. Spike-times are bounded by a refractory period $r, r < d_i^{n+1}$,
defined up to some absolute precision $\delta t$, 
while there is always a minimal delay $dt$ for one spike to influence another spike, 
and there might be (depending on the model assumptions) a maximal inter-spike interval $D$ such that either 
 the neuron fires within a time delay $< D$ or remains quiescent forever).
For biological neurons, orders of magnitude are typically, in milliseconds:\\

\centerline{\scriptsize \begin{tabular}{|c|c|c|c|} \hline $r$ & $\delta t$ & $d t$ & $D$ \\ \hline $1$ & $0.1$ & $10^{-[1, 2]}$ & $10^{[3, 4]}$ \\ \hline \end{tabular}}

\paragraph{Network dynamics discrete approximation.}

Combining these assumptions and the previous equations allows one (see \cite{cessac-vieville:08} for technical details)
to write the following discrete recurrence equation for a sampling period $\delta$:
\begin{equation}\label{BMS} V_i[k] = \gamma_i \, V_i[k - 1] \, (1 - Z_i[k - 1]) + \sum_{j = 1}^n \sum_{d = 1}^D W_{ijd} \, Z_j[k - d] + I_{ik},\end{equation}
where $V_i[k] = v_i(k \, \delta)$ and $Z_i[k] = \xi_{[1,+\infty[}(V_i[k])$, where $\xi$ is the indicatrix function, 
$\xi_{A}(x) = 1 \mbox{ if } x \in A \mbox{ and } 0 \mbox{ otherwise}$.

\begin{quotation}

Let us discuss in detail how~(\ref{BMS}) is derived from the previous equations.

The term $(1 - Z_i[k])$ implements the reset mechanism, since this term is equal to $0$ when $V_i[k] \ge 1$.
The interesting technical point is that this equation entirely specifies the integrate and fire mechanism.

The $\gamma_i \equiv \left. \nu(t, t + \delta, \tom_{t})\right|_{t = k \, \delta}$ term takes into account the multiplicative effects of conductances. 
The numerical analysis performed in \cite{cessac-vieville:08} demonstrates that, for numerical values taken from bio-physical models, 
considering here $\delta \simeq 0.1ms$,
this quantity related to the contraction rate, is remarkably constant, with small variations within the range:
\begin{displaymath}
\gamma_i \in [0.965, 0.995] \simeq 0.98,
\end{displaymath}
considering random independent and identically distributed Gaussian weights.
It has been numerically verified that taking this quantity constant over time and neurons does not significantly influence the dynamics. 
This the reason why we write $\gamma_i$ as a constant here.
This corresponds to a ``current based'' (instead of ``conductance based'') approximation of the connections dynamics.

The additive current 
\begin{displaymath}I_{ik}
\equiv \left. \int_{t}^{t + \delta} \nu(s, t + \delta, \tom_{s}) \, \left(i_m(\tom_s) + \frac{E_L}{\tau_L}\right) \, ds \right|_{t = k \, \delta}
\simeq \left. \delta \, \gamma_i \, \left(i_m(\tom_t) + \frac{E_L}{\tau_L}\right) \right|_{t = k \, \delta}
\end{displaymath}
accounts for membrane currents, including leak. 
The right-hand size approximation assume $\gamma_i$ is constant. 
Furthemore, we have to assume that the additive currents are independent from the spikes.
This means that we neglect the membrane current non-linearity and adaptation.

On the contrary, the term related to the connection weights $W_{ijd}$ is not straightforward to write and now requires to use the previous numerical approximation. 
Let us write:
\begin{displaymath}
\begin{array}{rcl} W_{ij}[k - k_j^n] 
 &\equiv& \left. E_j \, \int_{t}^{t + \delta} \nu(s, t + \delta, \tom_{t}) \, \rho_j\left(t - t_j^n\right) ds\right|_{t = k \, \delta, t_j^n =  k_j^n \, \delta} \\
 &\simeq& \left. E_j \, \delta \, \gamma_i \, \rho_j \left(t - t_j^n\right) \right|_{t = k \, \delta, t_j^n =  k_j^n \, \delta},
\end{array}
\end{displaymath}
assuming $\nu(s, t + \delta, \tom_{t}) \simeq \gamma_i$ as discussed previously.
This allows us to consider the spike response effect at time $t_j^n =  k_j^n \, \delta$ as a function only of $k - k_j^n$. 
The response $W_{ij}[d]$ vanishes after a delay $D, \tau_r = D \, \delta$, as stated previously.
We assume here that $\delta < \delta t$ i.e. that the spike-time precision allows to define the spike time as $k_j^n, t_j^n =  k_j^n \, \delta$ 
(see \cite{cessac-vieville:08,cessac-vieville:08b} for an extensive discussion).
We further assume that only zero or one spike is fired by the neuron of index $j$, during a period $\delta$, which is obvious as soon as $\delta < r$.

This allows to write $W_{ijd} = W_{ij}[d]$ so that:

\begin{displaymath}
\begin{array}{rcl} \sum_{j=1}^n W_{ij}[k - k_j^n] 
 &=& \sum_{d=1}^D \sum_{j=1}^n W_{ij}[d] \xi_{\{k_j^n\}}(k - d) \\
 &=& \sum_{d=1}^D W_{ij}[d] \xi_{\{k_j^1\cdots, k_j^n, \cdots \}}(k - d) \\
 &=&  \sum_{d=1}^D W_{ijd}  \, Z_j[k - d]
\end{array}
\end{displaymath}
~\\since $Z_j[l] = \xi_{\{k_j^1\cdots, k_j^n, \cdots \}}(l)$ is precisely equal to $1$ on spike time and $0$ otherwise, which completes the derivation of~(\ref{BMS}).

\end{quotation}

\paragraph{Counting the model's degrees of freedom.}

\begin{figure}[!htbp]
\begin{center}
\includegraphics[width = 1\textwidth, height=4cm]{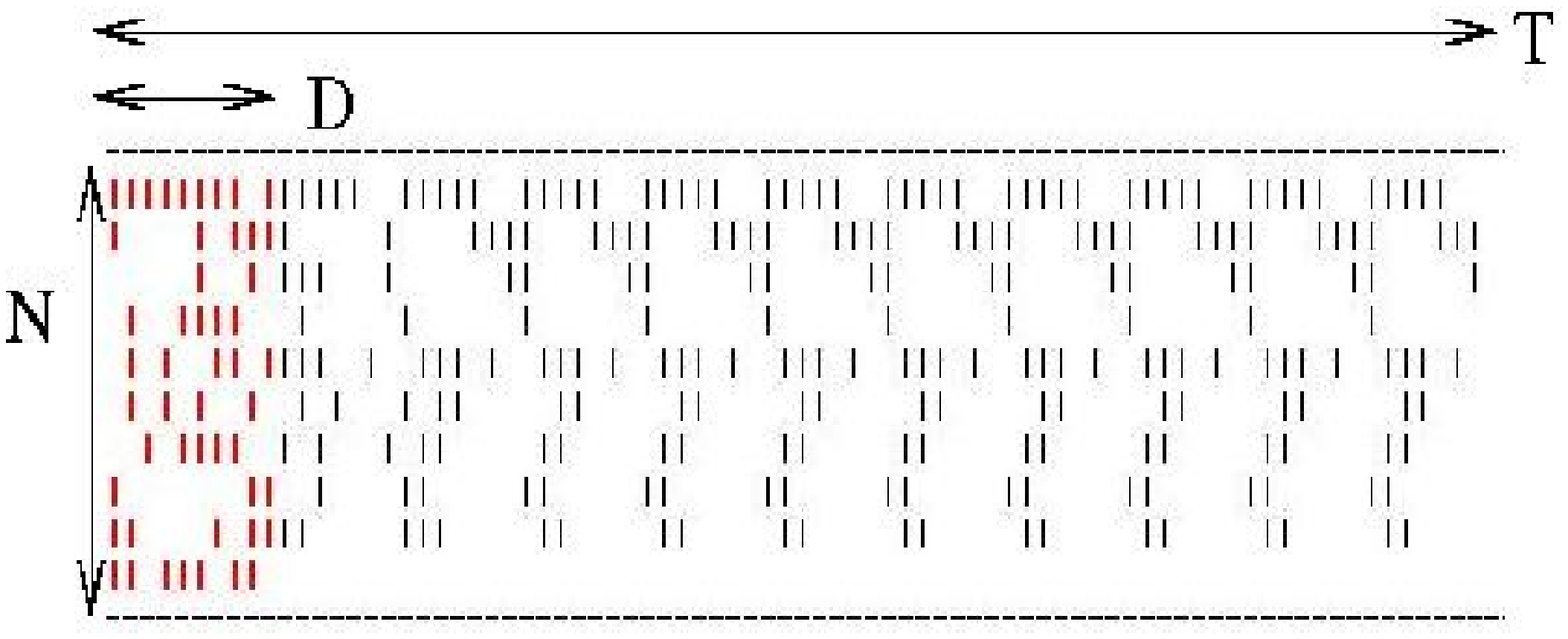}
\caption{\label{fig:raster} Schematic representation of a raster of $N$ neurons observed during a time interval $T$ after an initial conditions interval $D$ (in red).
This example corresponds to a periodic raster, but non-periodic raster are also considered. See text for further details.}
\end{center}
\end{figure}

Let us consider a network of $N$ units, whose dynamics is defined by~(\ref{BMS}), generating a raster of the form schematized in Fig.~\ref{fig:raster}. 

In order to determine the dynamics of the neural network, we require the knowledge of the initial condition.
Here, due to the particular structure of equation~(\ref{BMS}) with a delay $D$, the initial condition is the piece of trajectory $V_i[k], k \in \{0, D\{$.
The notation $k \in \{0, D\{$ stands for $0 \leq k < D$.
In fact, it is sufficient to consider $V_i[0]$ and $Z_j[k], k \in \{0, D\{$, entirely defining $V_i[k], k \in \{0, D\{$ from~(\ref{BMS}).

If the neuron $i$ has fired at least once, the dependence in the initial condition is removed thanks to the reset mechanism. This means that its state does not depend
on $V_i[0]$ any more, as soon as spikes are known. We thus can further assume $V_i[0] = 0$, for the sake of simplicity.

The initial state is thus defined by $N$ numerical values and $D \times N$ binary values.

The dynamics is parametrized by the weights $W_{ijd}$ thus $N \times N \times D$ values. Here it is assumed that the $\gamma_i$ are known and constant, 
while $I_{ik}$ are also known, as discussed in the sequel.

When the potential and/or spikes are observed during a period of $T$ samples, $N \times T$ numerical/binary values are measured.

\section{Methods: Weights and delayed weights estimation} 
\label{sec:estimation}

With the assumption that $V_i[0] = 0$ discussed previously, (\ref{BMS}) writes:
\begin{equation} \label{BMS0}
V_{i}[k] = \sum_{j = 1}^{N} \sum_{d = 1}^{D} W_{ijd} \sum_{\tau = 0}^{\tau_{ik}} \gamma^{\tau} \, Z_{j}[k - \tau - d] + I_{ik\tau}
\end{equation}
writing $I_{ik\tau} = \sum_{\tau = 0}^{\tau_{ik}} \gamma^{\tau} \, I_{i(k-\tau)} $ with:
\\ \centerline{$\tau_{ik} = k - \mbox{arg min}_{l > 0} \{Z_{i}[l - 1] = 1\}$,} \\ 
the derivation of this last form being easily obtained by induction from~(\ref{BMS}).
Here $\tau_{ik}$ is the delay from the last spiking time, i.e., the last membrane potential reset.
If no spike, we simply set $\tau_{ik} = k$.

This equation shows that there is a direct explicit linear relation between spikes and membrane potential. 
See \cite{cessac:08} for more detailed about the one-to-one correspondence between the spike times and the membrane potential trajectory that seems 
defined by~(\ref{BMS}) and~(\ref{BMS0}).
Here, we use~(\ref{BMS0}) in a different way.

Let us now discuss how to retrieve the model parameters from the observation of the network activity.  
We propose different solutions depending on the paradigm assumptions.

\subsection*{Retrieving weights and delayed weights from the observation of spikes and membrane potential}

Let us assume that we can observe both the spiking activity $Z_i[k]$ and the membrane potential $V_i[k]$. Here, (\ref{BMS0}) writes in matrix form:
\begin{equation}\label{L} \mathbf{C}_i \, \mathbf{w}_i = \mathbf{d}_i \end{equation}
with: 
\[ \begin{array}{rclcl}
\mathbf{C}_i &=&
 \left( \begin{array}{ccc}
  \ldots & \ldots & \ldots \\
  \ldots &  \sum_{\tau = 0}^{\tau_{ik}} \gamma^{\tau} Z_{j}[k - \tau - d]  & \ldots \\
  \ldots & \ldots & \ldots
  \end{array} \right) &\in& R^{T - D \times N \, D}, \\
\mathbf{d}_i &=& (\ldots \quad V_{i}[k] - I_{ik\tau}\quad \ldots)^\dagger &\in& R^{T - D}, \\
\mathbf{w}_i &=& (\ldots \quad W_{ijd} \quad \ldots)^\dagger &\in& R^{N \, D}. \\
\end{array} \]
writing ${\bf u}^\dagger$ the transpose of ${\bf u}$.

Here, $\mathbf{C}_{i}$ is defined by the neuron spike inputs, $\mathbf{d}_i$ is defined by the neuron membrane potential outputs and membrane currents, 
and the network parameter by the weights vector $\mathbf{w}_i$.

More precisely, $\mathbf{C}_i$ is a rectangular matrix with: \begin{itemize}
 \item $N \, D$ columns, corresponding to product with the $N \, D$ unknowns $W_{ijd}$, for $j \in \{1, N\}$ and $d \in \{1, D\}$,
 \item $T - D$ rows, corresponding to the $T - D$ measures $V_{i}[k]$, for $k \in \{D, T\}$, and,
 \item $T - D \times N \, D$ coefficients corresponding to the raster, i.e. the spikes $Z_{j}[k]$.
\end{itemize}
The weights are thus directly {\em defined by a set of linear equalities for each neuron}. Let us call this a Linear (L) problem. 

Furthermore, the equation defined in~(\ref{L}), concerns only the weights of one neuron of index $i$.
It is thus a weight estimation local to a neuron, and not global to the network.
Furthermore, the weight estimation is given by the observation of the input $Z_i[k]$ and output $V_i[k]$. 
These two characteristics correspond to usual Hebbian-like learning rules architecture.
See \cite{gerstner-kistler:02} for a discussion.

Given a general raster (i.e., assuming $\mathbf{C}_i$ is of full rank $\min(T-D, N\,D)$): \begin{itemize}

 \item This linear system of equations has always solutions, in the general case, if:
   \begin{equation} \label{Lsol} N > \frac{T - D}{D} = O\left(\frac{T}{D}\right) \Leftrightarrow D > \frac{T}{N+1} = O\left(\frac{T}{N}\right)\Leftrightarrow D \, (N+1) > T.\end{equation} 
 This requires enough non-redundant neurons $N$ or weight profile delays $D$, with respect to the observation time $T$. 
 In this case, given any membrane potential and spikes values, \textit{there are always weights able to map the spikes input onto the desired potential output}. 

 \item On the other hand, if $N \, D \leq T - D$, then the system has no solution in the general case. 
  This is due to the fact that we have a system with more equations than unknowns, thus with no solution in the general case.
  However, there is obviously a solution if the potentials and spikes have been generated by a neural network model of the form of~(\ref{BMS}).
\end{itemize}

If $\mathbf{C}_i$ is not of full rank, this may correspond to several cases, e.g.: \begin{itemize}
 \item Redundant spike pattern: some neurons do not provide linearly independent spike trains.
 \item Redundant or trivial spike train: for instance with a lot of bursts (with many $Z_{j}[k] = 1$) or a very sparse train (with many $Z_{j}[k] = 0$).
  Or periodic spike trains.
\end{itemize}

Regarding the observation duration $T$, it has been demonstrated in \cite{cessac:08,cessac-vieville:08} that the dynamic of an integrate and fire neural network is 
generically\footnote{Considering a basic leaky integrate and fire neuron network the result is true except for a negligible set of parameters. 
Considering an integrate and fire neuron model with conductance synapses the result is true, providing synaptic responses have a finite memory.} periodic.
This however depends on parameters such as external current or synaptic weights, while periods can be larger than any accessible computational time.

In any case, several choices of weights $\mathbf{w}_i$ (in the general case a $D \, (N+1) - T$ dimensional affine space) 
may lead to the same membrane potential and spikes. The problem of retrieving weights from the observation of spikes and membrane potential 
may thus have many solutions.

The particular case where $D = 1$ i.e. where there is no delayed weights but a simple weight scalar value to define a connection strengths is included in this framework.

\subsection*{Retrieving weights and delayed weights from the observation of spikes}

Let us now assume that we can observe the spiking activity $Z_i[k]$ only (and not the membrane potentials) which corresponds to the usual assumption, 
when observing a spiking neural network. 

In this case, the value of $V_i[k]$ is not known, whereas only its position with respect to the firing threshold is provided:

\begin{displaymath}
Z_{i}[k] = 0 \Leftrightarrow  V_i[k] < 1 \mbox{ and } Z_{i}[k] = 1 \Leftrightarrow  V_i[k] \geq 1,
\end{displaymath}
which is equivalent to write the condition:

\begin{displaymath}
e_{ik} = (2 \, Z_{i}[k] - 1) \, (V_i[k] - 1) > 0.
\end{displaymath}
the case $V_i[k] = 1$ being excluded here. Note that the case $V_i[k] = 1$ corresponds to a singularity in the dynamics leading to effects such as sensibility to perturbations. In this section, we exclude this case setting $e_{ik} > 0$
(at the numerical level, this yields $e_{ik} > \epsilon$, for some $\epsilon > 0$).

If the previous condition $e_{ik} > 0$ is verified for all time index $k$ and all neuron index $i$,
then the spiking activity of the network exactly corresponds to the desired firing pattern.
However, in this case, the membrane potential value may differ from one setting to another. 

Expanding~(\ref{BMS0}), with the previous condition allows us to write, in matrix form:
\begin{equation}\label{LP} \mathbf{e}_i = \mathbf{A}_i \, \mathbf{w}_i + \mathbf{b}_i > 0 \end{equation}
writing: 
\[ \begin{array}{rclcl}
\mathbf{A}_i &=&
 \left( \begin{array}{ccc}
  \ldots & \ldots & \ldots \\
  \ldots &  (2 \, Z_{i}[k] - 1) \, \sum_{\tau = 0}^{\tau_{jk}} \gamma^{\tau} Z_{j}[k - \tau - d]  & \ldots \\
  \ldots & \ldots & \ldots
  \end{array} \right) &\in& R^{T - D \times N \, D}, \\
\mathbf{b}_i &=& (\ldots \quad (2 \, Z_{i}[k] - 1) \, (I_{ik\tau} - 1) \quad \ldots)^\dagger &\in& R^{T - D}, \\
\mathbf{w}_i &=& (\ldots \quad W_{ijd} \quad \ldots)^\dagger &\in& R^{N \, D}, \\
\mathbf{e}_i &=& (\ldots \quad (2 \, Z_{i}[k] - 1) \, (V_i[k] - 1) \quad \ldots)^\dagger &\in& R^{T - D}, \\
\end{array} \]
thus $\mathbf{A}_i = \mathbf{D}_i \, \mathbf{C}_i$ where $\mathbf{D}_i$ is the non-singular $R^{T - D \times T - D}$ diagonal matrix with 
$\mathbf{D}_i^{kk} = 2 \, Z_{i}[k] - 1 \in \{-1, 1\}$.

The weights are now thus directly {\em defined by a set of linear inequalities for each neuron}. This is thus a Linear Programming (LP) problem. 
See \cite{darst:90} for an introduction and \cite{bixby:92} for the detailed method used here to implement the LP problem.

Furthermore, the same discussion about the dimension of the set of solutions applies to this new paradigm except that we now have to consider a {\em simplex} of solution, 
instead of a simple {\em affine sub-space}.

A step further, $0 \leq e_{ik}$ is the ``membrane potential distance to the threshold''. 
Constraining the $e_{ik}$ is equivalent to constraining the membrane potential value $V_i[k]$.

It has been shown in \cite{cessac:08} how: 
\begin{equation} \label{dBMS} |\mathbf{e}|_\infty = \min_{i} \inf_{k \geq 0} e_{ik} \end{equation}
can be interpreted as a ``edge of chaos'' distance, the smallest $|\mathbf{e}|$ the higher the dynamics complexity, and the orbits periods.

On the other hand, the higher $e_{ik}$, the more robust the estimation. 
If $e_{ik}$ is high, sub-threshold and sup-threshold values are clearly distinct.
This means that numerical errors are not going to generate spurious spikes or cancel expected spikes.

Furthermore, the higher $|\mathbf{e}|_\infty$ the smaller the orbits period \cite{cessac:08}.
As a consequence, the generated network is expected to have rather minimal orbit periods.

In the sequel, in order to be able to use an efficient numerical implementation, we are going to consider a weaker but more robust norm, than $|\mathbf{e}|_\infty$:
\begin{equation} \label{daBMS} |\mathbf{e}_i|_1 = \sum_{k} e_{ik} \end{equation}
We are thus going to maximize, for each neuron, the sum, thus, up to a scale factor, the average value of $e_{ik}$.

Let us now derive a bound for $e_{ik}$. 
Since $0 \leq V_i[k] < 1$ for sub-threshold values and reset as soon as $V_i[k] > 1$, it is easily bounded by:

\begin{displaymath}
V_i^{min} = \sum_{j d, W_{ijd} < 0} W_{ijd} \leq V_i[k] \leq V_i^{max} = \sum_{j d, W_{ijd} > 0} W_{ijd}
\end{displaymath}
and we must have at least $V_i^{max} > 1$ in order for a spike to be fired while $V_i^{min} \leq 0$ by construction. 
These bounds are attained in the high-activity mode when either all excitatory or all inhibitory neurons fire. 
From this derivation, $e^{max} > 0$ and we easily obtain:

\begin{displaymath}
0 < e_{ik} \leq e^{max} = \max_i(1 - V_i^{min}, V_i^{max} - 1)
\end{displaymath}
thus an explicit bound for $e_{ik}$.

Collecting all elements of the previous discussion, the present estimation problem writes:
\begin{equation}\label{LP1} 
 \max_{\mathbf{e}_i, \mathbf{w}_i} \sum_k e_{k}, \mbox{ with, } 0 < e_{ik} \leq e^{max}, \mbox{ and, } \mathbf{e}_i = \mathbf{A}_i \, \mathbf{w}_i + \mathbf{b}_i 
\end{equation}
which is a standard bounded linear-programming problem.

The key point is that a LP problem can be solved in polynomial time, thus is not a NP-complete problem, subject to the curse of combinatorial complexity.
In practice, this LP problem can be solved using the Simplex method (which is, in principle, NP-complete in the worst case, but) in practice, as fast as, when not faster,
than polynomial methods.

\subsection*{Retrieving signed and delayed weights from the observation of spikes}

In order to illustrate how the present method is easy to adapt to miscellaneous paradigms, 
let us now consider the fact that the weight emitted by each neuron have a fixed sign, either positive for excitatory neurons, or negative for inhibitory neurons.
This additional constraint, known as the ``Dale principle'' \cite{strata-harvey:99}, is usually introduced to take into account the fact, that synaptic weights signs are fixed
by the excitatory or inhibitory property of the presynaptic neuron. 

Although we do not focus on the biology here, it is interesting to notice that this additional constraint is obvious to introduce in the present framework, writing:
\\ \centerline{$W_{ijd} = S_{ijd} \, W^\bullet_{ijd}, \mbox{ with } S_{ijd} \in \{-1, 1\}, \mbox{ and } W^\bullet_{ijd} \geq 0$} \\
thus separating the weight sign $S_{ijd}$ which is a-priory given and the weight value $W^\bullet_{ijd}$ which now always positive.

Then, writing:
\\ \centerline{$\mathbf{A}^\bullet{ijkd} = \mathbf{A}_{ijkd} \, S_{ijd}$} \\
the previous estimation problem becomes:
\begin{equation}\label{LP2} 
 \max_{\mathbf{e}_i, \mathbf{w}^\bullet_i} \sum_k e_{k},
    \mbox{ with, } 0 < e_{ik} \leq e^{max}, 0 \leq W^\bullet_{ijd} \leq 1, \mbox{ and, } \mathbf{e}_i = \mathbf{A}^\bullet_i \, \mathbf{w}^\bullet_i + \mathbf{b}_i 
\end{equation}
which is still a similar standard linear-programming problem.

\subsection*{Retrieving delayed weights and external currents from the observation of spikes}

In the previous derivations, we have considered the membrane currents $I_{ik}$ are inputs, i.e. are known in the estimation. 
Let us briefly discuss the case where they are to be estimated too.

For adjustable non-stationary current $I_{ik}$, the estimation problem becomes trivial. 
An obvious solution is $W_{ijd} = 0, I_{ik} = 1 + a \, (Z_{i}[k] - 1/2)$ for any $a > 0$,
since each current value can directly drive the occurrence or inhibition of a spike, without any need of the network dynamics.

Too many degrees of freedom make the problem uninteresting: adjusting the non-stationary currents leads to a trivial problem.

To a smaller extends, considering adjustable stationary currents $I_{i}$ also ``eases'' the estimation problem, providing more adjustment variables. 
It is obvious to estimate not only weights, but also the external currents, since the reader can easily notice that yet another
linear-programming problem can be derived. 

This is the reason why we do not further address the problem here, and prefer to explore in details a more constrained estimation problem.

\subsection*{Considering non-constant leak when retrieving parametrized delayed weights}

For the sake of simplicity and because this corresponds to numerical observations, we have assumed here that the neural leak $\gamma$ is constant. 
The proposed method still works if the leak varies with the neuron and with time i.e. is of the form $\gamma_{it}$, 
since this is simply yet another input to the problem. The only difference is that, in~(\ref{BMS0}) and the related equations, 
the term $\gamma^\tau$ is to be replaced by products of $\gamma_{it}$.

However, if $\gamma$ is a function of the neural dynamics, thus of $W_{ijd}$, the previous method must be embedded in a non linear estimation loop.
Since we know from \cite{cessac-vieville:08} that this dependency is numerically negligible in this context, we can propose the following loop:

\begin{enumerate}
\item Fix at step $t=0$, $\gamma_{it}^0$, to initial values.
\item $k$- Estimate the weights $W_{ijd}$, given leaks $\gamma_{it}^k$ at $k = 0, 1, ..$.
\item $k$- Re-simulate the dynamics, given the weights and to obtain corrected values $\tilde{\gamma}_{it}^k$.
\item $k$- Smoothly modify $\gamma_{it}^{k+1} = (1 - \upsilon) \, \gamma_{it}^{k+1} + \upsilon \, \tilde{\gamma}_{it}^k$
\item $k+1$ Repeat step 2,$k$- for $k+1$ the convergence of this non-linear relaxation method being guarantied for sufficiently small $\upsilon$. 
See \cite{vieville-lingrand-etal:01} for an extended discussion about this class of methods.
\end{enumerate}

This shows that considering models with leaks depending on the dynamics itself 
is no more a LP-problem, but an iterative solving of LP-problems.

\subsection*{Retrieving parametrized delayed weights from the observation of spikes}

In order to further show the interest of the proposed method, let us now consider that the {\em profile} of the weights is fixed, i.e. that 
\\ \centerline{$W_{ijd} = W^\circ_{ij} \, \alpha_\tau(d)$ with, e.g., $\alpha(d) = \frac{d}{\tau} \, e^{-\frac{d}{\tau}}$} \\
thus the weights is now only parametrized by a magnitude $W^\circ_{ij}$, while the temporal profile is known.

Here $\alpha_\tau(d)$ is a predefined synaptic profile, while $\tau$ is fixed by biology (e.g., $\tau = 2 \, ms$ for excitatory connections and $\tau = 10 ms$ for inhibitory ones).
Let us note that the adjustment of $\tau$ would have been a much more complex problem, as discussed previously in the non-parametric case.

This new estimation is defined by:
\begin{equation}\label{LP3} \mathbf{e}_i = \mathbf{A}^\circ_i \, \mathbf{w}^\circ_i + \mathbf{b}_i > 0 \end{equation}
writing: 
\[ \begin{array}{rclcl}
\mathbf{A}^\circ_i &=&
 \left( \begin{array}{ccc}
  \ldots & \ldots & \ldots \\
  \ldots &  (2 \, Z_{i}[k] - 1) \, \sum_d \sum_{\tau = 0}^{\tau_{jk}} \gamma^{\tau} Z_{j}[k - \tau - d]  \, \alpha(d) & \ldots \\
  \ldots & \ldots & \ldots
  \end{array} \right) &\in& R^{T - D \times N} \\
\mathbf{w}^\circ_i &=& (\ldots \quad W_{ij} \quad \ldots)^\dagger &\in& R^{N} \\
\end{array} \]
thus a variant of the previously discussed mechanisms.

This illustrates the nice versatility of the method. Several other variants or combinations could be discussed 
(e.g. parametrized delayed weights from the observation of spikes and potential, ..), 
but they finally leads to the same estimations.

\subsection*{About retrieving delays from the observation of spikes}

Let us now discuss the key idea of the paper.

In the previous derivations, we have considered delayed weights, i.e. a quantitative weight value $W_{ijd}$ at each delay $d \in \{1, D\}$.

Another point of view is to consider a network with adjustable synaptic delays. Such estimation problem may, e.g., correspond to the ``simpler'' model:

\begin{displaymath}
V_i[k] = \gamma_i \, V_i[k - 1] \, (1 - Z_i[k - 1]) + \sum_{j = 1}^n W_{ij} \, Z_j[k - d_{ij}] + I_{ik},
\end{displaymath}
where now the weights $W_{ij}$ and delays $d_{ij}$ are to estimated. 

As pointed out previously, the non-learnability of spiking neurons is known \cite{sima-sgall:05}, i.e. the previous estimation is proved to be NP-complete. 
We have carefully checked in \cite{sima-sgall:05} that the result still apply to the present setup.
This means that in order to ``learn'' the proper parameters we have to ``try all possible combinations of delays''. 
This is intuitively due to the fact that each delay has no ``smooth'' effect on the dynamics but may change the whole dynamics in a unpredictable way.

We see here that the estimation problem of delays $d_{ij}$ seems not compatible with usable algorithms, as reviewed in the introduction.

We propose to elude this NP-complete problem by considering {\em another} estimation problem.
Here we do not estimate {\em one} delay (for each synapse) but consider connection weights at several delay and 
then estimate a weighted pondering of their relative contribution. 
This means that we consider a {\em weak} delay estimation problem.

Obviously, the case where there is a weight $W_{ij}$ with a corresponding delay $d_{ij} \in \{0, D\}$ is a particular case of considering several delayed weights $W_{ijd}$
(corresponding to have all equal weights to zero except at $d_{ij}$, i.e., $W_{ijd} = \mbox{ if } d = d_{ij} \mbox{ then } W_{ij} \mbox{ else } 0$).
In other words, with our weaker model, we are still able to estimate a neural network with adjustable synaptic delays. 

We thus do not restrain the neural network model by changing the position of the problem, but enlarge it.
In fact, the present estimation provides a smooth approximation of the previous NP-complete problem.

We can easily conjecture that the same restriction also apply of the case where the observation of spikes and membrane potential is considered.

We also have to notice, that the same restriction apply not only to simulation but, as far as this model is biologically plausible, also true at the biological level. 
It is thus an issue to wonder if, in biological neural network, delays are really estimated during learning processes, or if a weaker form of weight adaptation, 
as discussed in this paper, is considered.

\section{Methods: Exact spike train simulation}
\label{sec:reservoir}

\subsection{Introducing hidden units to match any raster}

\subsubsection*{Position of the problem}

Up to now, we have assumed that a raster $\bar{Z}_i[k], i \in \{1, N\}, k \in \{1, T\}$ is to be generated by a network whose dynamics is defined by~(\ref{BMS}),
with initial conditions $\bar{Z}_j[k], j \in \{1, N\}, k \in \{1, D\}$ and $V_j[0] = 0$. In the case where a solution exists, we have discussed how to compute it. 

We have seen a solution always exists, {\em in the general case}, if the observation period is small enough, i.e., $T < O(N \, D)$. 
Let us now consider the case where $T \gg O(N \, D)$.

In this case, there is, in general, no solution. This is especially the case when the raster has not been generated by a network given by~(\ref{BMS}), e.g.,
in the case when the raster is random.

What can we do then ? For instance, in the case when the raster is entirely random and is not generated by a network of type~(\ref{BMS}) ?

\begin{figure}[!htbp]
\begin{center}
\includegraphics[width = 1\textwidth, height=6cm]{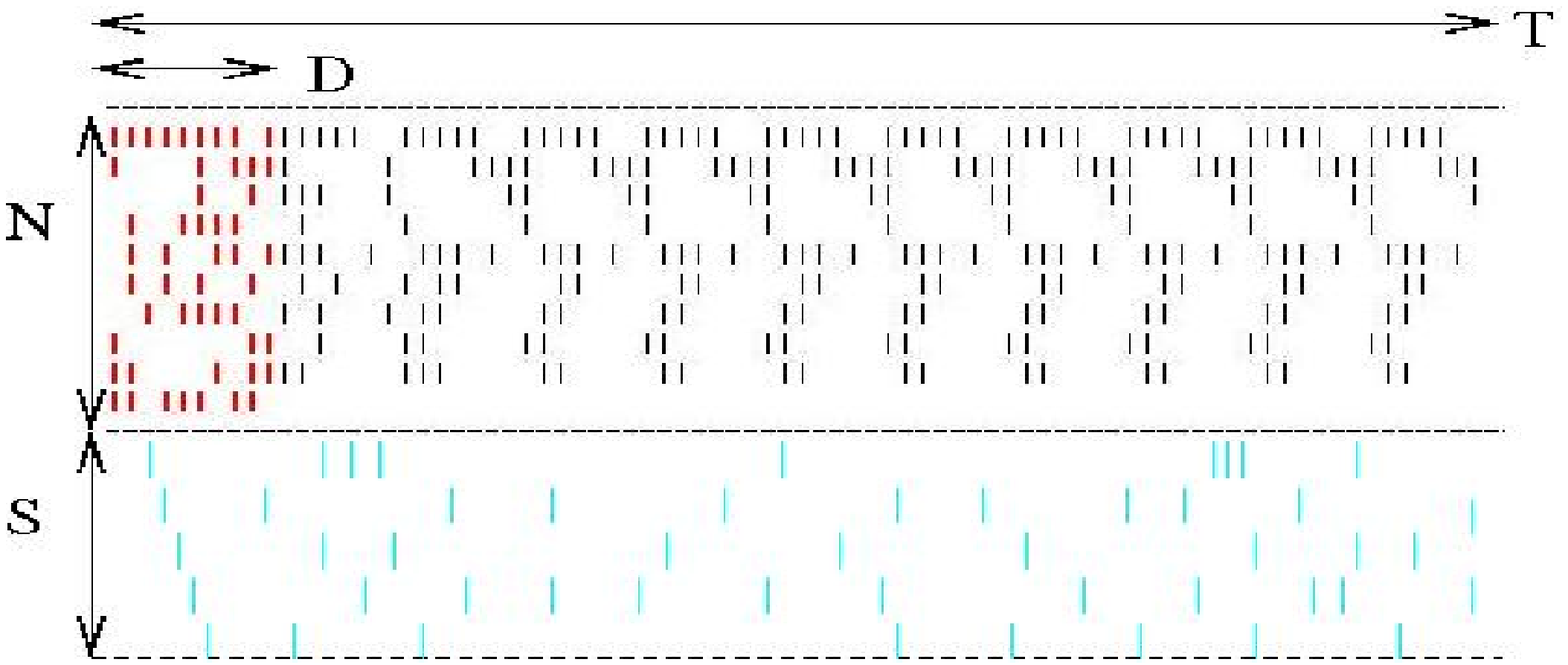}
\caption{\label{fig:haster} Schematic representation of a raster of $N$ output neuron observed during a time interval $T$ after an initial conditions interval $D$,
with an add-on of $S$ hidden neurons, in order increase the number of degree of freedom of the estimation problem. See text for further details.}
\end{center}
\end{figure}

The key idea, borrowed from the reservoir computing paradigm reviewed in the introduction, is to add a reservoir of ``hidden neurons'', 
i.e., to consider not $N$ but $N + S$ neurons.
The set of $N$ ``output'' neurons is going to reproduce the expected raster $\bar{Z}_i[k]$ and the set of $S$ ``hidden'' neurons to increase the number of degree of freedom
in order to obtain $T < O((N + S) \, D)$, thus being able to apply the previous algorithms to estimate the optimal delayed weights. 
Clearly, in the worst case, it seems that we have to add about $S = O(T / D)$ hidden neurons. This is illustrated in Fig.~\ref{fig:haster}.

In order to make this idea clear, let us consider a trivial example.

\subsubsection*{Sparse trivial reservoir}

\begin{figure}[!htbp]
\begin{center}
\includegraphics[width=1 \textwidth, height=4cm]{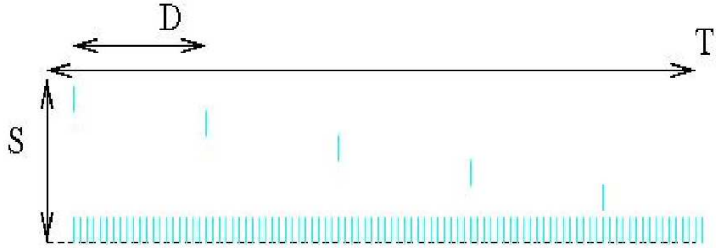}
\caption{\label{fig:sparser} Schematic representation of a sparse trivial set of hidden neurons, allowing to generate any raster of length $T$. See text for further details.}
\end{center}
\end{figure}

Let us consider, as illustrated in Fig.~\ref{fig:sparser}, $S = T/D + 1$ hidden neurons of index $i' \in \{0, S\}$ each neuron firing once at $t_{i'} = i' \, D$, 
except the last once always firing (in order to maintain a spiking activity), thus:

\begin{displaymath}
Z_{i'}[k] = \delta(i' \, D - k), 0 \leq i' < S, Z_{S}[k] = 1
\end{displaymath}
where $\delta(k) = \xi_{\{0, 0\}}(k)$ is the Kronecker symbol, as shown in Fig.~\ref{fig:sparser}.

Let us choose:
\begin{displaymath}\begin{array}{rcl}
W_{SS1}   &>& 1 \\
W_{i'S1}  &=& \frac{1 - \gamma'}{1 - \gamma^{'t_{i'}-1/2}} \\
W_{i'i'1} &<& -\frac{\gamma^{2 \, t_{i'}} - \gamma^{'T}}{\gamma^{'T} \, (1 - \gamma^{'t_{i'}})} < 0\\
W_{i'j'd} &=& 0 \mbox{ otherwise } \\
\end{array}
\end{displaymath}
with initial conditions $Z_{i'}[k] = 0, i' \in \{0, S\{ \mbox{ and } Z_S[k] = 1, k \in \{1, D\}$, while $I_{i'k} = 0$.

A straight-forward derivation over equation~(\ref{BMS}) allows to verify that this choice allows to generate the specified $Z_{i'}[k]$.
In words, as the reader can easily verify, it appears that:
\begin{itemize}
\item the neuron of index $S$ is always firing since (though $W_{SS1}$) a positive internal loop maintains its activity;
\item the neurons of index $i' \in \{0, S\{$, whose equation writes: 
\begin{displaymath}
V_{i'}[k] = \gamma' \, V_{i'}[k - 1] \, (1 - Z_{i'}[k - 1]) + W_{i'S1} + W_{i'i'1} \, Z_{i'}[k - 1]
\end{displaymath}
is firing at $t_{i'}$ integrating the constant input $W_{i'S1}$;
\item the neurons of index $i' \in \{0, S\{$, after firing is inhibited (though $W_{i'i'1}$)  by a negative internal loop, 
thus reset at value negative low enough not to fire anymore before $T$. We thus generate $Z_{i'}[k]$ as expected.
\end{itemize}

Alternatively, the use of the firing neuron of index $S$ can be avoided by introducing a constant current $I_{i'k} = W_{i'S1}$. 

However, without the firing neuron of index $S$ or some input current, the sparse trivial raster {\em can not} be generated, although $T < O(N \, D)$. 
This comes from the fact that the activity is too sparse to be self-maintained.
This illustrates that when stating that ``a solution exists, {\em in the general case}, if the observation period is small enough, i.e., $T < O(N \, D)$'', 
a set of singular cases, such as this one, was to be excluded.

The hidden neurons reservoir raster being generated, it is straight-forward to generate the output neuron raster, considering:

\begin{itemize}
\item no recurrent connection between the $N$ output neurons, i.e., $W_{ijd} = 0, i \in \{1, N\}, j \in \{1, N\}, d \in \{1, D\}$, 
\item no backward connection from the $N$ output neurons to the $S$ hidden neurons i.e., $W_{i'jd} = 0, i' \in \{0, N\{, j \in \{1, N\}, d \in \{1, D\}$, 
\item but forward excitatory connections between hidden and output neurons: 
\begin{displaymath}
W_{ij'd} = (1 + \epsilon) \, \bar{Z}_i[j' \, D + d] \quad \textrm{for some small} \quad \epsilon > 0
\end{displaymath}
\end{itemize}
yielding, from~(\ref{BMS}) :
\\ \centerline{$\begin{array}{rcl} V_i[k] 
 &=& \sum_{j' = 1}^n \sum_{d = 1}^D W_{ij'd} \, Z_{j'}[k - d] \\
 &=& \sum_{j' = 1}^n \sum_{d = 1}^D (1 + \epsilon) \, \bar{Z}_i[j' \, D - d] \, \delta(j' \, D - (k - d)) \\
 &=& (1 + \epsilon) \, \bar{Z}_i[k]) \\
\end{array}$} \\
setting $\gamma = 0$ for the output neuron and $I_{i'k} = 0$, so that $Z_i[k] = \bar{Z}_i[k]$, i.e., 
the generated spikes $Z_i[k]$ correspond to the desired $\bar{Z}_i[k]$, as expected.

\subsubsection*{The linear structure of a network raster}

The previous construction allows us to state: 
 {\em given any raster of $N$ neurons and observation time $T$, there is always a network of size $N + T/D + 1$ with weights delayed up to $D$, 
which exactly simulates this raster}. What do we learn from this fact ?

This helps to better understand how the reservoir computing paradigm works: {\em Although it is not always possible to simulate any raster plot using a ``simple''
integrate and fire model such as the one defined in~(\ref{BMS}), adding hidden neurons allows to embed the problem in a higher-dimensional space where a solution 
can be found.}

This results is induced by the fact that we have made explicit, in the previous section, that learning the network weights is essentially a linear (L or LP) problem.  
With this interpretation, a neuron spiking sequence is a vector in this linear space, while a network raster is a vector set.
Designing a ``reservoir'' simply means choosing a set of neurons which spiking activity {\em spans the space of expected raster}.
We are going to see in the next section that this point of view still holds in our framework when considering network inputs.

This linear-algebra interpretation further explains our ``trivial sparse'' choice: 
We have simply chosen a somehow canonical orthonormal basis of the raster linear space. 
One consequence of this view is that, {\em given a network raster, any other network raster which is a linear combination of this raster} can be generated by the same network,
by a simple change of weights.
This is due to the fact that a set of neurons defining a given raster corresponds to the set of vectors spanning the linear space of all possible raster 
generated by this network. 
Generating another raster corresponds to a simple change of generating vectors in the spanning set. 
This also allows us to define, for a given raster linear space, a minimal set of generating neurons, i.e. a vectorial basis. 
The ``redundant'' neurons are those which spiking sequence is obtained by feed-forward connections from other neurons. 

We must however take care of the fact the numerical values of the vector are binary values, not real numbers. 
This is a linear space over a finite field, whereas its scalar product is over the real numbers.

\subsubsection*{On optimal hidden neuron layer design}

In the previous paragraph, we have fixed the hidden neuron spiking activity, choosing a sparse ad-hoc activity. 
It is clearly not the only one solution, very likely not the best one.

We thus may consider the following problem: 
 given $N$ output neurons and $S$ hidden neurons what are the ``best'' weights ${\bf W}$ and hidden neuron activity $Z_{j'}[k]$ allowing to reproduce the output raster.

By ``best'', we mean optimal in a precise sense defined in the previous section. In that case, instead of having to solve a LP-problem, as specified in~(\ref{LP1}), 
we have to consider a much more complicated problem now:
\begin{itemize}
\item not a linear but bi-linear problem (since we have to consider the products of weights to estimate with spike activity to estimate, as readable in~(\ref{BMS});
\item not a standard linear programming problem with real values to estimate, but a mixed integer programming problem with both integer values to estimate.
\end{itemize}

This has a dramatic consequence, since such problem is known as being NP-hard, thus not solvable in practice, as discussed previously for the estimating of delays.

This means that we can not consider this very general question, but must propose heuristics in order to choose or constraint the hidden neuron activity,
and then estimate the weights, given the output and hidden neuron's spikes, in order to still consider a LP-problem.

Let us consider one of such heuristic.

\subsubsection*{A maximal entropy heuristic}

Since we now understand that hidden neuron activity must be chosen in order to span as much as possible the expected raster space, 
and since we have no a-priory information about the kind of raster we want to generate (we target here a ``general'' algorithm), 
the natural idea is to randomly choose the neuron activity with a maximal randomness.

Although it is used here at a very simple level, this idea is profound and is related to random sampling and sparse approximation of complex signal in noise
(see \cite{tropp:04a,tropp:04b} for a didactic introduction), leading to greedy algorithms and convex relaxation \cite{tropp:06,tropp-etal:06}. 
Since inspired by these elaborated ideas, 
the proposed method is simple enough to be described without any reference to such formalisms.

In this context, maximizing the chance to consider a hidden neuron with a spiking activity independent from the others, 
and which adds new independent potential information, simply corresponds to choose the activity ``as random as possible''.
This corresponds to a so called Bernouilli process, i.e., simply to randomly choose each spike state independently with equi-probability.

Since we want to simulate the expected raster with a minimal number of hidden neuron, we may consider the following algorithmic scheme: 
\begin{enumerate}
 \item Starts with no hidden but only output neurons.
 \item Attempts to solve~(\ref{LP1}) on hidden (if any) and output neurons, 
	in order to obtain weights which allows the reproduction of the expected raster on the output neurons.
 \item If the estimation fails, add a new hidden neuron and randomly draw its spiking activity
 \item Repeat step 2 and 3 until an exact reproduction of the expected raster is obtained
\end{enumerate}

Clearly, adding more and more random points to the family of generating elements must generate a spanning family after a finite time, 
since randomly choosing point in an affine space, there is no chance to always stay in a given affine sub-space.
This means that we generate a spanning family of neuron after a finite time, with a probability of one.
So that the algorithm converges.

What is to be numerically experimented is the fact we likely obtain a somehow minimal set of hidden neurons or not. This is going to be experimented in section~\ref{sec:results}.

\subsection{Application: Input/Output transfer identification} 
\label{sec:application}

 Let us now describe the main practical application of the previous algorithmic development, which is to ``program'' a spiking network in order to generate a given spike train 
or realize a given input/output spike train function. 

 In the present context, this means finding the ``right'' or the ``best'' spiking network parameters in order to map an input's set onto an output's set.

 What is pointed out here, is the fact that the previous formalism does not only apply to the simulation of a unique, input less, fully connected network, 
but is applicable to a much wider set of problems. 

 In order to make this explicit, let us consider the following specification of spiking neural networks with units defined by the recurrent equation~(\ref{BMS}). 
\begin{description}

 \item[connectivity] We now do not assume a fully connected network but consider a connection graph ${\cal K}$, 
i.e., some connections weights are zero. This writes:

\begin{displaymath}\forall i, j \notin {\cal K} \subseteq \{0, N\{^2, \forall d, W_{ijd} = 0\end{displaymath}
or in matrix form:
\begin{equation} \label{topo} \mathbf{K} \, \mathbf{w} = 0 \end{equation}
for a diagonal matrix $\mathbf{K}$ with $\mathbf{K}_{ijd\,ijd} = 0 \mbox{ if } i, j \in {\cal K} \mbox{ and } 1 \mbox{ otherwise}$.

 \item[input current] We consider that any neurons can be driven by an input current $I_{ik}$, thus defining an ``analog'' input.

 \item[input spikes] We have also to consider that the network can also be driven by external incoming spikes. In order to implement this feature in the present framework, 
we use the following trick.

For each spike train input, we add an ``input spiking neuron'', thus with its state corresponding to the incoming spike train.
This is obviously implementable in the previous network by considering $V_i[k] = I_{ik}$, 
with $I_{ik}$ being a binary value $I_{ik} \in \{0, 1 + \epsilon\}$ for some small $\epsilon > 0$.
This corresponds to neuron defined by~(\ref{BMS}) but with $\gamma_i = 0$ and $\forall j, d, W_{ijd} = 0$, thus a trivial connectivity as introduced previously.

This tricks allows to reduce the spike input specification to a current input specification.

 \item[output neurons] We consider that a subset of the neurons are output neurons, thus with a state which is {\em readout} and thus must be constrained,
as defined in~(\ref{LP}). Other neurons are hidden neurons as discussed in the previous section.

As discussed previously, the best heuristics at this level of knowledge is to randomly generate the hidden neuron required activity.

 \item[weighted estimation] We further consider that depending on neuron and time the estimation requirement is not homogeneous, whereas there are
times and neurons for which the importance of potential to threshold distance estimation differs from others. This generalized estimation is 
obvious to introduce in our formalism, defining:
\begin{equation} \label{daBMSw} |\mathbf{e}_i|_{1, \Lambda} = \sum_{k} \Lambda_{ik} \, e_{ik}, \Lambda_{ik} \geq 0 \end{equation}
for some metric $\Lambda$.

\end{description}

We further consider a ``supervised learning paradigm'' is the following sense. We now consider a family of $L$ input current or spikes vectors:
\begin{displaymath}
{\bf I}^l = (\ldots \quad I_{ik}^l \quad \ldots)^\dagger \in R^{N \times T - D},
\end{displaymath}
to be mapped on family of output spike trains:

\begin{displaymath}
{\bf Z}^l = (\ldots \quad Z_{i}[k]^l \quad \ldots)^\dagger \in R^{N \times T - D},
\end{displaymath}
given initial states:

\begin{displaymath}{\bf Z}_0^l = (\ldots \quad Z_{i}[k]^l \quad \ldots)^\dagger \in R^{N \times D}, k \in \{0, D\{,
\end{displaymath}
for $l \in \{0, L\{$ and would like to find the right weights $\mathbf{w}$ allowing to perform this input/output mapping.
The estimation problem is in fact strictly equivalent to~(\ref{LP}) now concatenating the input information (except the initial states), thus writing:
\[ \begin{array}{rclcl}
\mathbf{A}_i &=&
 \left( \begin{array}{ccc}
  \ldots & \ldots & \ldots \\
  \ldots &  (2 \, Z_{i}[k]^l - 1) \, \sum_{\tau = \tau_{jk}}^{0} \gamma^{\tau} Z_{j}[k - \tau - d]^l  & \ldots \\
  \ldots & \ldots & \ldots
  \end{array} \right) &\in& R^{L \, (T - D) \times N \, D}, \\
\mathbf{b}_i &=& (\ldots \quad (2 \, Z_{i}[k]^l - 1) \, (I_{ik\tau}^l - 1) \quad \ldots)^\dagger &\in& R^{L \, (T - D)}. \\
\end{array} \]

This formalism, thus now allows us to find an exact input/output mapping, adding hidden neurons in order to have enough degree of freedom to obtain a solution.

\subsection{Application: approximate Input/Output transfer identification} \label{approximate}

Let us finally discuss how to apply the previous formalism to approximate transfer identification. We consider, in our context, deterministic alignment metrics 
defined on spike times \cite{victor-purpura:96,victor:05}. 

\paragraph{Using alignment metric.}

In this context, the distance between two finite spike trains ${\cal F}$, ${\cal F}'$ is defined in terms of the minimum cost of transforming one spike train into another.
Two kinds of operations are defined: \begin{itemize}
\item spike insertion or spike deletion, the cost of each operation being set to $1$
\item spike shift, the cost to shift from $t_i^n \in {\cal F}$ to $t_i^{'m} \in {\cal F}'$ being set to $|t_i^n - t_i^{'m}| / \tau$ for a time constant $\tau$.
\end{itemize}

Although computing such a distance seems subject to a combinatorial complexity, it appears that quadratic algorithms are available, 
with the same complexity in the discrete and continuous case.

For small $\tau$, the distance approaches the number of non-coincident spikes, since instead of shifting spikes it is cheaper to insert/delete non-coincident spikes, the distance being always bounded by the number of spikes in both trains.

For high $\tau$, the distance basically equals the difference in spike number (rate distance), while for two spike trains with the same number of spikes, 
there is always a time-constant $\tau$ large enough for the distance to be equal to $\sum_n |t_i^n - t_i^{'n}| / \tau$.

It appears that the LP algorithms initial phase \cite{darst:90,bixby:92} which attempts to find an initial solution before optimizing it, 
generates a divergence between the obtained and expected raster, this divergence being zero, if a solution exists. Furthermore, this divergence can be related to
the present alignment metric, for high $\tau$, and on-going work out of the scope of this subject develops this complementary aspect.

When considering spike trains with more than one unit, an approach consists to sum the distances for each alignment unit-to-unit. Another point of view is to consider that a spike can ``jump'', with some cost, from one unit in ${\cal F}$ to another unit in ${\cal F}'$. The related algorithmic complexity is no more quadratic but to the power of the number of units \cite{aronov:03}.

This family of metrics include aligments not only on spike times, but also on inter-spike intervals, or metrics which are sensitive to patterns of spikes, etc... They have been fruitfully applied to a variety of neural systems, in order to characterize neuronal variability and coding \cite{victor:05}. For instance, in a set of neurons, that act as coincidence detectors, with integration time (or temporal resolution) $\tau$, spike trains will have similar postsynaptic effects if they are similar w.r.t. this metric.
Furthermore, this metric generalizes to metric with causality (at a given time, the cost of previous spikes alignment decreases with the obsolescence of the spike)
and non-linear shift's costs \cite{cessac-etal:08a}.

\paragraph{Application to approximate identification.}

Our proposal is to re-visit the maximal entropy heuristic algorithm defined previously and consider having a correct identification,
if the distance between the expected and obtained raster is not zero (equality) but {\em below a threshold}.

This allows to not only address:
 \begin{itemize}
 \item the {\em exact} estimation problem, i.e. find an exact input/output mapping if and only if there is one, but also
 \item the {\em approximate} estimation problem, i.e. to find an approximate input/output mapping that minimizes a given distance.
\end{itemize}

This however, is a trivial strategy because the alignment distance is not used to find the optimal solution, but only to check wether this solution is acceptable.
The reason of such a choice is easy to explain: alignment metrics, as it, are highly non-differenciable with respect to the network parameters.
Therefore variational methods, considering. e.g., the criterion gradient in order to drive the parameters value towards a local optimum do not apply here.

Several alternatives exist. One considers the spike time defined by the equation $V_{\bf W}(t_i^n) = \theta$, where $V$ is the membrane potential,
$\theta$ is the firing threshold, and ${\bf W}$ are the network parameters to adjust \cite{schrauwen:07}. From the implicit function theorem it is obvious to relate
locally $d {\bf W}$ and $d t_i^n$, thus derive a parameter gradient with respect to the spike times. However, such method is numerically ill-defined, since the threshold
value $\theta$ is not significant, but only a modeling trick.

Another alternative is to consider convolution metrics \cite{cessac-rostro-etal:08b}, in order to relate the spike train to a differentiable signal, 
thus in a context where variational methods can be applied.
One instantiation of this idea considers an abstract specification of the input/output methods, using piece-wise linear SRM models \cite{gerstner-kistler:02}.
This algebraic simplification allows to implement variational methods \cite{vieville-rochel:06,kornprobst-vieville-etal:06} in order to specify the 
network input/output relations. Such methods, however, are restrained to a given neuronal model and to a given coding (temporal coding in this case) of
a continuous information using spike times. Such methods must thus be generalized in order to be applied in the present context.

When the present method fails, it still provides an approximate solution with a ``maximal'' number of correct spikes, by the virtue of the~(\ref{LP1}) minimization 
mechanism. Each ``false'' state corresponds to $e_{ik} < 0$ (i.e. either a spurious spike or a missing spike), and it is an easy exercise to relate this to a 
particular alignment metric. We conjecture that this is a interesting track in order to generalize the present work, from exact estimation, to exact or approximate
estimation.

\section{Results} 
\label{sec:results}

\subsection{Retrieving weights from the observation of spikes and membrane potential}

In a first experiment, we consider the linear problem defined in~(\ref{L}) and use the singular value decomposition (SVD) mechanism \cite{gantmatcher:77} in order to 
obtain a solution in the least-square sense. Here the well-established {\tt GSL\footnote{\tt http://www.gnu.org/software/gsl}} library SVD implementation is used.

This allows us to find: \begin{itemize}
 \item if one or more solution, the weights of minimal magnitude $|{\bf w}_i|^2 = \sum_{jd} W_{ijd}^2$;
 \item if no exact solution, the solution which minimizes $\sum_{k} (V_i[k] - \tilde{V}_i[k])^2$ where $\tilde{V}_i[k]$ is the membrane potential predicted by the estimation.
\end{itemize}

\subsubsection*{The master and servant paradigm.}

We have seen that, if $D \, (N+1) > T$, i.e., if the observation time is small enough for any raster there is a solution. 
Otherwise, there is a solution if the raster is generated by a model of the form of~(\ref{BMS}). 
We follow this track here and consider a master/servant paradigm, as follows:
\begin{enumerate}
\item In a first step we randomly choose weights and generate a ``master'' raster.
\item The corresponding output raster is submitted to our estimation method (the ``servant''), while the master weights are hidden.
The weights are taken from a normal distribution ${\cal N}(0, \frac{\sigma^2}{N})$ with $70\%$ excitatory connections and $30\%$ for inhibitory one. 
The standard deviation $\sigma \in [1, 10]$ has been chosen in order to obtain an interesting dynamics, as discussed in \cite{cessac:08}.
\end{enumerate}
The algorithm defined in~(\ref{Lsol}) or in~(\ref{LP1}) thus receives a set of spikes for which we are sure that a solution exists. 
Therefore it can be used and leads to a solution with a raster which must exactly correspond to the master input raster.

Note that this does not mean that the servant is going to generate the raster with the same set of weights, since several solutions likely exist in the general case.
Moreover, except for the paradigm~(\ref{Lsol}),the master and servant potential $V_i[k]$ are expected to be different, 
since we attempt to find potentials which distance to the threshold is maximal, in order to increase the numerical robustness of the method.

\begin{figure}[!htbp]
\begin{center}
\includegraphics [width=1 \textwidth, height=4cm]{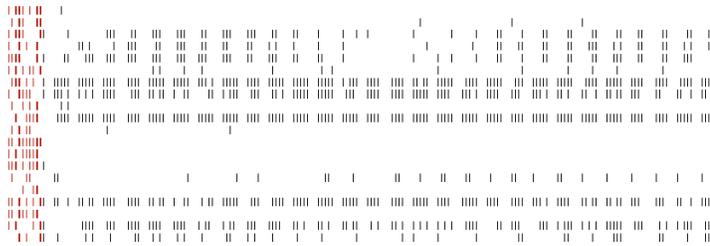}
\caption{\label{fig:Ratser01L} A ``chaotic'' dynamics with 20 neurons fully connected within network and simulation time $T = 200$, using both excitatory ($70\%$) and inhibitory ($30\%$) connections, with $\sigma= 5$ (weight standard-deviation). 
After estimation, we have checked that master and servant generate \bf{exactly} the same raster plot, thus only show the servant raster, here and in the sequel.}
\end{center}
\end{figure}

\begin{figure}[!htbp]
\begin{center}
\includegraphics[width=1 \textwidth, height=4cm]{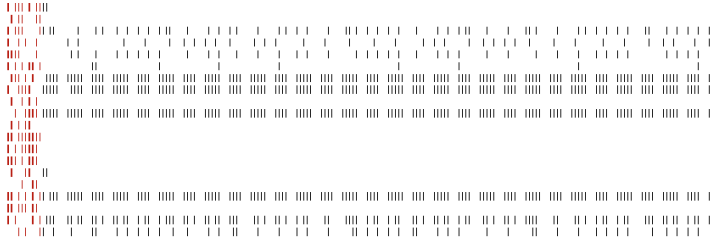}
\caption{\label{fig:Ratser02L} A ``periodic'' dynamics with 20 neurons fully connected within network and simulation time $T = 200$, using both excitatory ($70\%$) and inhibitory ($30\%$) connections, with $\sigma= 5$. In this figure a periodic dynamics (9 periods of period 17) is observed, with 20 neurons fully connected within network and simulation time $T = 200$. Again the master and servant rasters are the same.}
\end{center}
\end{figure}

This is going to be the validation test of our method. As an illustration we show two results in Fig.~\ref{fig:Ratser01L} and Fig.~\ref{fig:Ratser02L} for two different dynamics. 
The former is ``chaotic'' in the sense that the period is higher than the observation time. 

In the non trivial case in Fig.~\ref{fig:Ratser01L}, it is expected that only one weight's set can generate such a non-trivial raster,
since, as discussed before, we are in the ``full rank'' case, thus with a unique solution.
We observe the same weights for both master and servant in this case, as expected.
This would not the case for simpler periodic raster, e.g. in Fig.~\ref{fig:Ratser02L},
where the weight's estimation by the servant differs from the master's weights, since several solutions are possible.

Retrieving weights from the observation of spikes {\em and} membrane potential has been introduced here in order to explain and validate the general method in a 
easy to explain case.
Let us now turn to the more interesting cases where only the observation of spikes are available.

\subsection{Retrieving weights from the observation of spikes}

In this setup we still use the master / servant paradigm, but now consider the LP problem defined previously. The numerical solutions are derived thanks to
the well-established improved simplex method as implemented in {\tt GLPK\footnote{\tt http://www.gnu.org/software/glpk}}.

\begin{figure}[!htbp]
\begin{center}
\includegraphics[width=1 \textwidth, height=8cm]{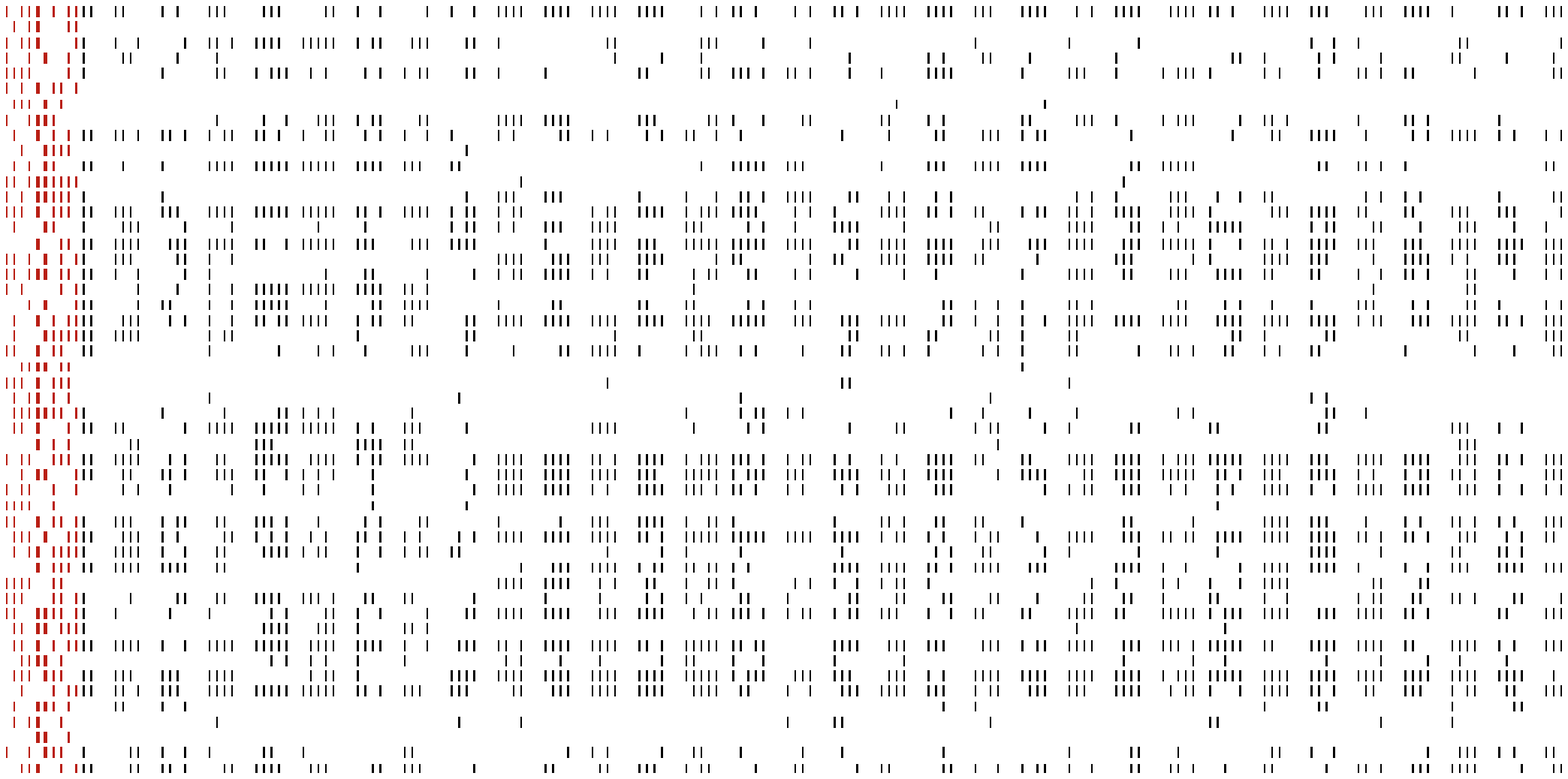}
\caption{\label{fig:Raster01LPC} Example of rather complex ``chaotic'' dynamics retrieved by a the LP estimation defined in~(\ref{LP1}) using the master / servant paradigm with 50 neurons fully connected, initial conditions $D = 10$ and a time observation $T = 200$, used here to validate the method.}
\end{center}
\end{figure}

\begin{figure}[!htbp]
\begin{center}
\includegraphics[width=1 \textwidth, height=6cm]{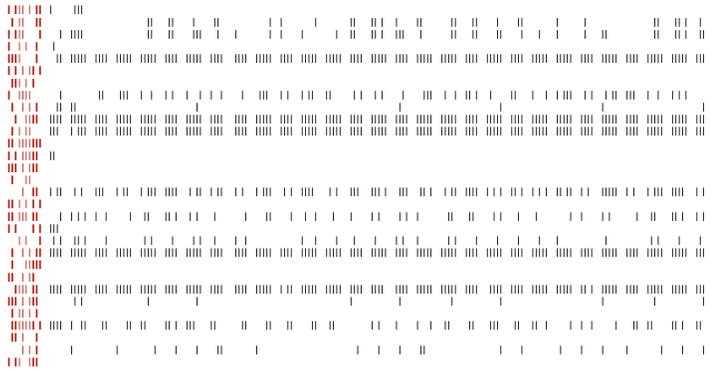}
\caption{\label{fig:Raster01LPP} Example of periodic dynamics retrieved by a the LP estimation defined in~(\ref{LP1}) using the master / servant paradigm, 
here a periodic raster of period 29 is observed during 4.76 periods. ($N = 30$, $T = 150$ and $D = 10$)
As expected from by the theory, the estimated dynamics remains periodic after the estimation time, thus corresponding to a parsimonious estimation.}
\end{center}
\end{figure}

 As an illustration we show two results in Fig.~\ref{fig:Raster01LPC} and Fig.~\ref{fig:Raster01LPP} for two different dynamics. 
Interesting is the fact that, numerically, the estimated weights correspond to a parsimonious dynamics in the sense that 
the servant raster period tends to be minimal:

\begin{itemize}
\item if the master raster appears periodic, the servant raster is, with same period;
\item if the master raster appears aperiodic (i.e., ``chaotic'') during the observation interval, the servant raster is periodic with a period close (but not always identical) 
to the observation time $T$. This is coherent with the theoretical analysis of such networks \cite{cessac:08,cessac-vieville:08},
ans is futher investigated in the sequel.
\end{itemize}

\subsection{Retrieving delayed weights from the observation of spikes}

\begin{figure}[!htbp]
\begin{center}
\includegraphics[width=1\textwidth,height=4cm]{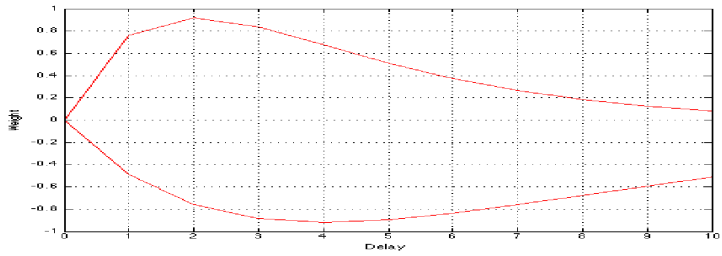}
\caption{\label{fig:Profil} Weights distribution (positive and negative) used to generate delayed weights, with $D = 10$.}
\end{center}
\end{figure}

In this next setup we still consider the same master/servant paradigm, for $N = 50$ units, with a leak $\gamma = 0.95$ and an external current $I = 0.3$,
but in this case the master delayed weight profile has the standard form shown in Fig.~\ref{fig:Profil}.

\begin{figure}[!htbp]
\centering
\subfigure{\includegraphics[width = 0.4\linewidth, totalheight = 0.15\textheight]{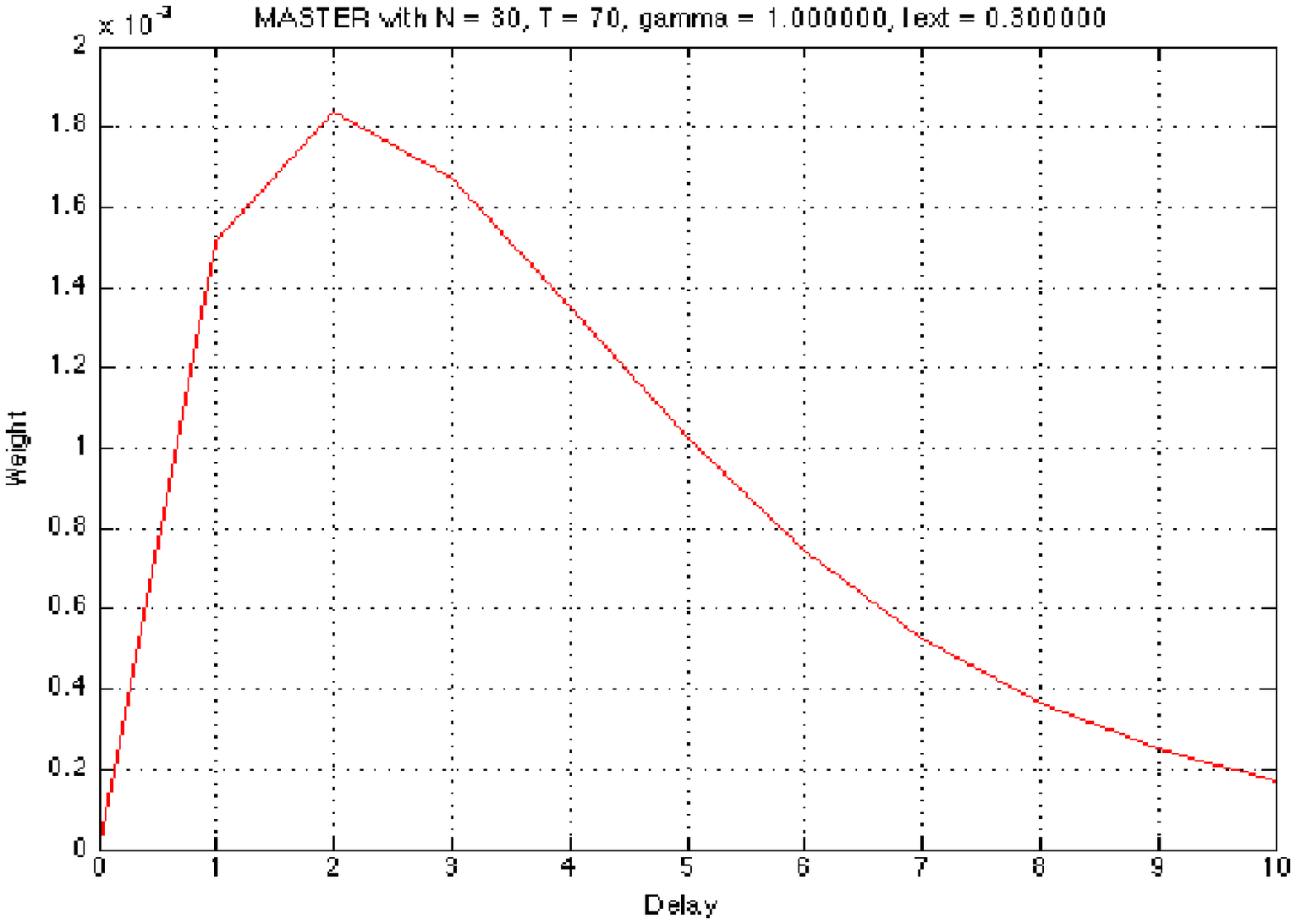}}
\hspace{0.1in}
\subfigure{\includegraphics[width = 0.4\linewidth, totalheight = 0.15\textheight]{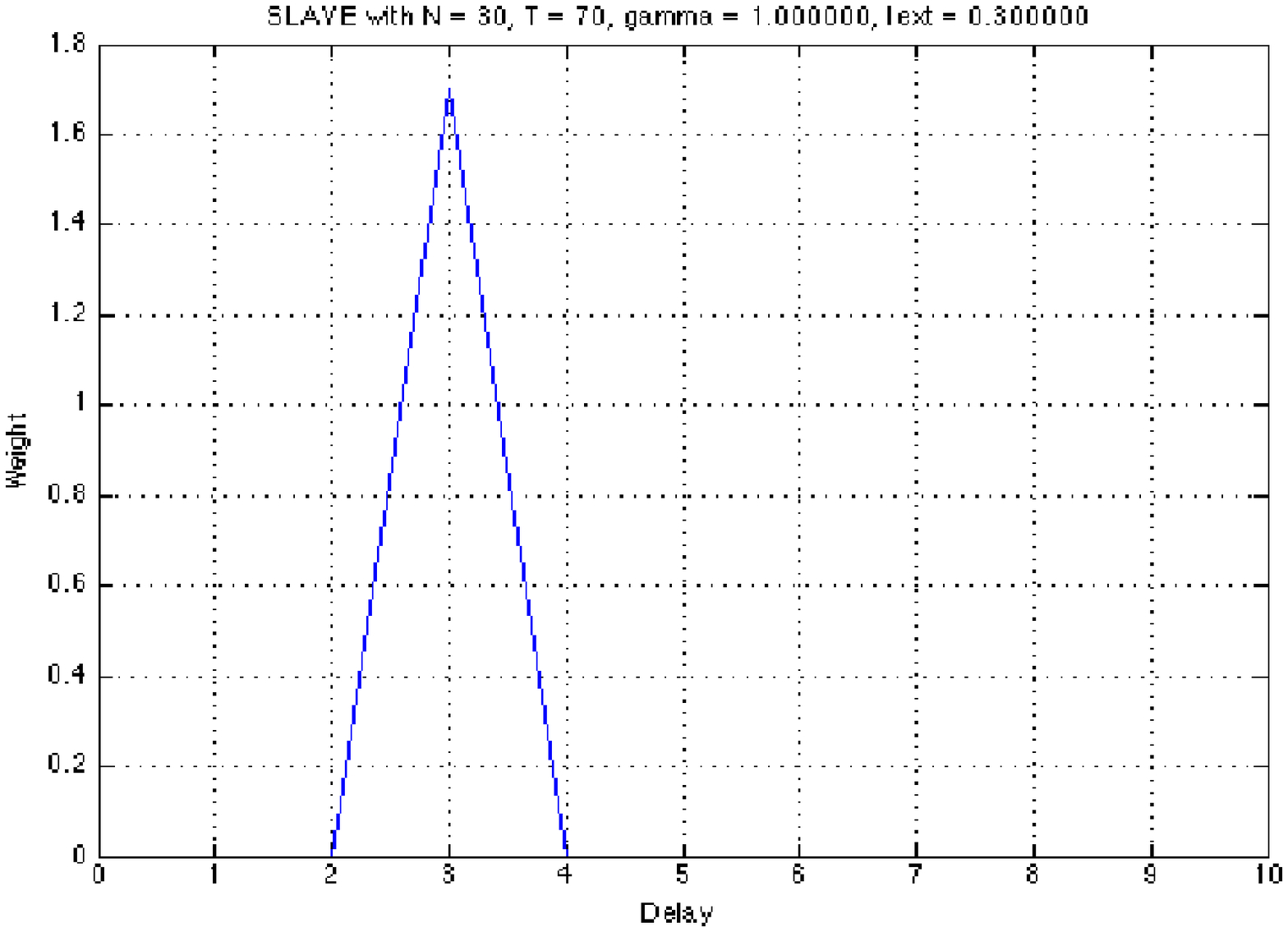}} \\
\centering
\subfigure{\includegraphics[width=1\textwidth, totalheight = 0.15\textheight]{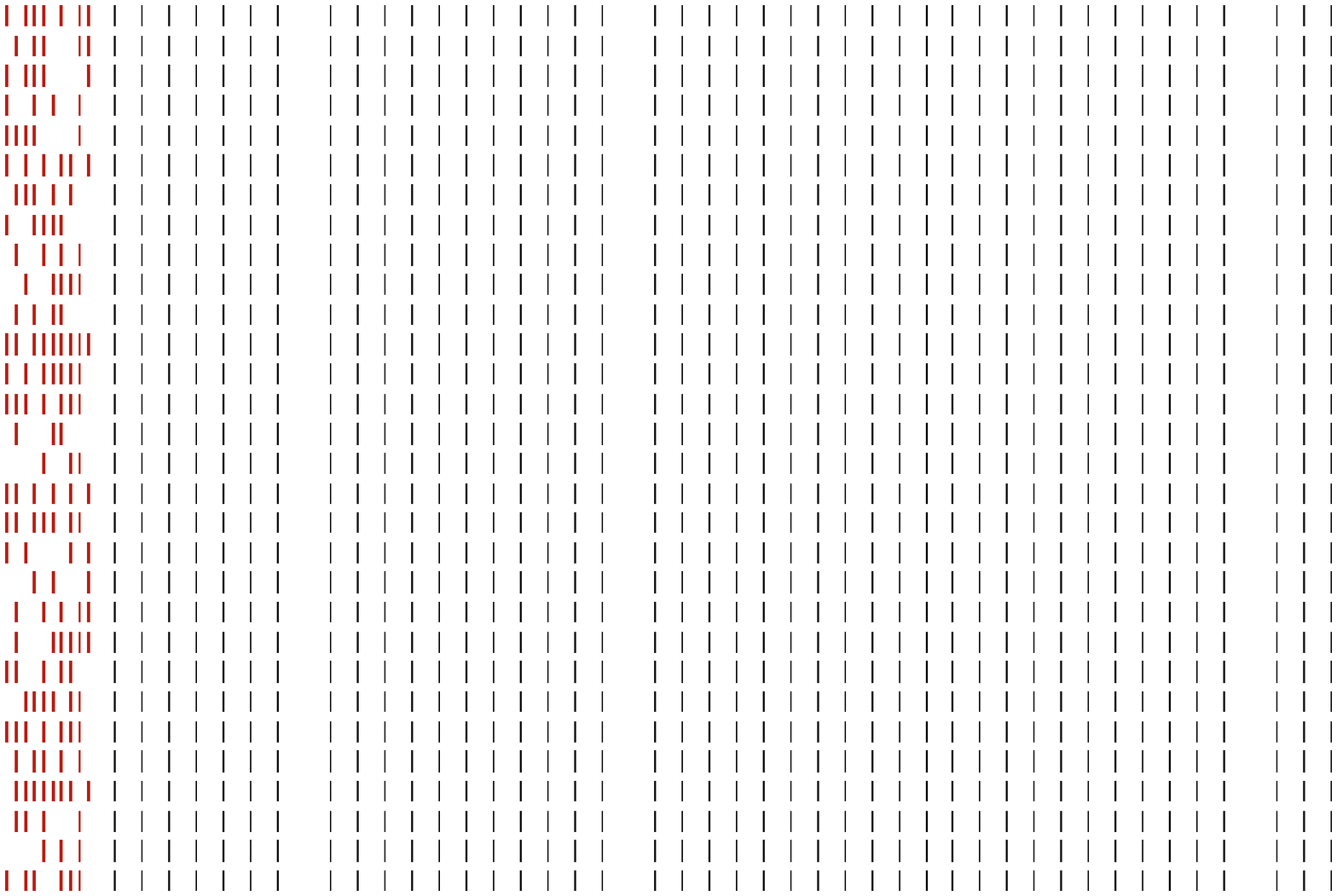}}
\centering
\caption{\label{fig:Raster01D} An example of trivial dynamics obtained with excitatory weights profiles shown in the top-left view (master weight's profile), 
with $N = 30$, $\gamma = 0.98$, $D = 10$ $T = 70$.
The estimated weights profile (servant weight's profile) is shown in the top-right view. All weights have the same shape and value.
To generate such trivial periodic raster, shown in the bottom view, only weights with a delay equal to the period have not zero values.
This corresponds to a minimal edge of the estimation simplex, this parsimonious estimation being a consequence of the chosen algorithm.}
\end{figure}

\begin{figure}[!htbp]
\centering
\subfigure{\includegraphics[width = 0.4\linewidth]{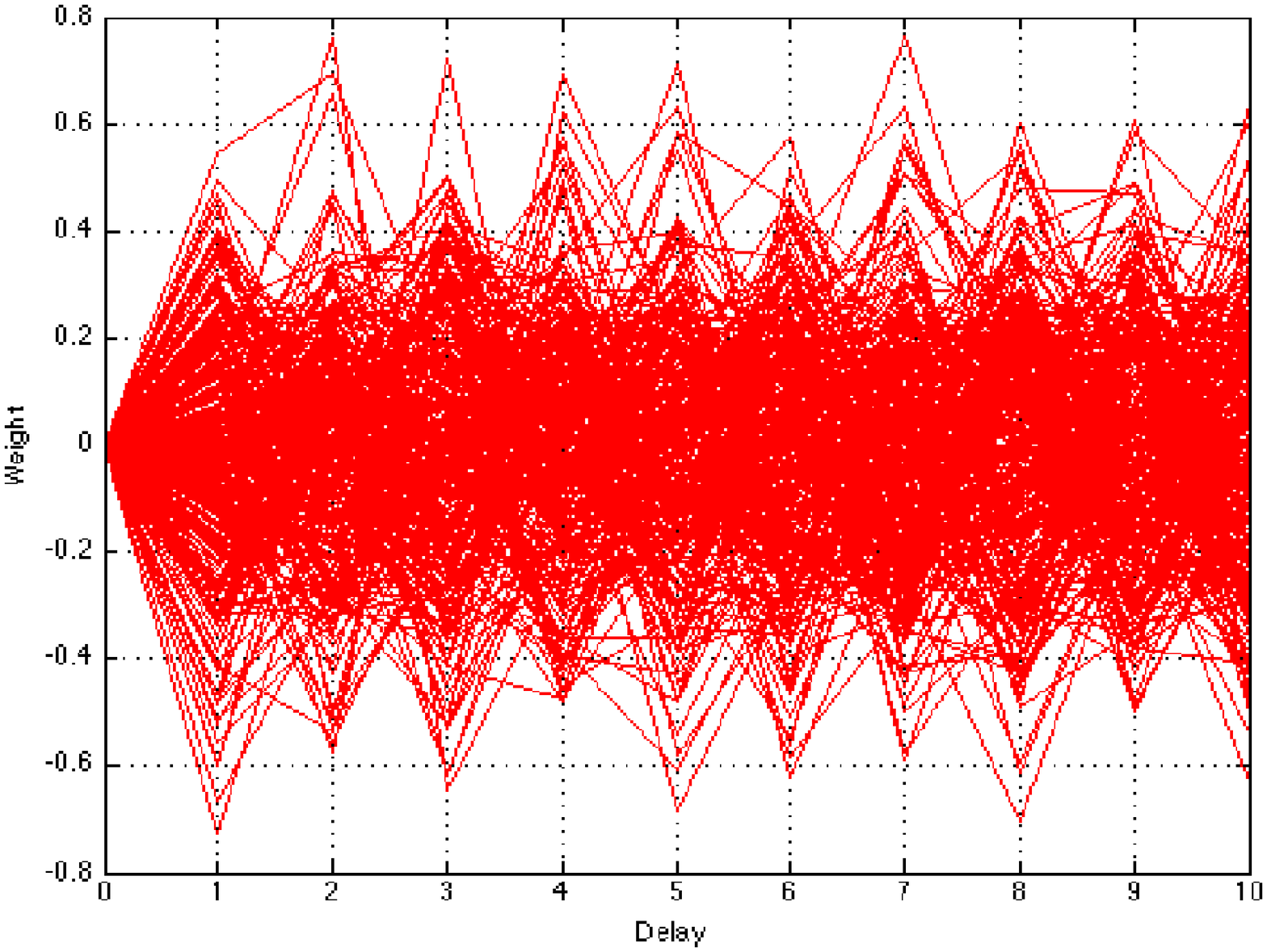}}
\hspace{0.1in}
\subfigure{\includegraphics[width = 0.4\linewidth]{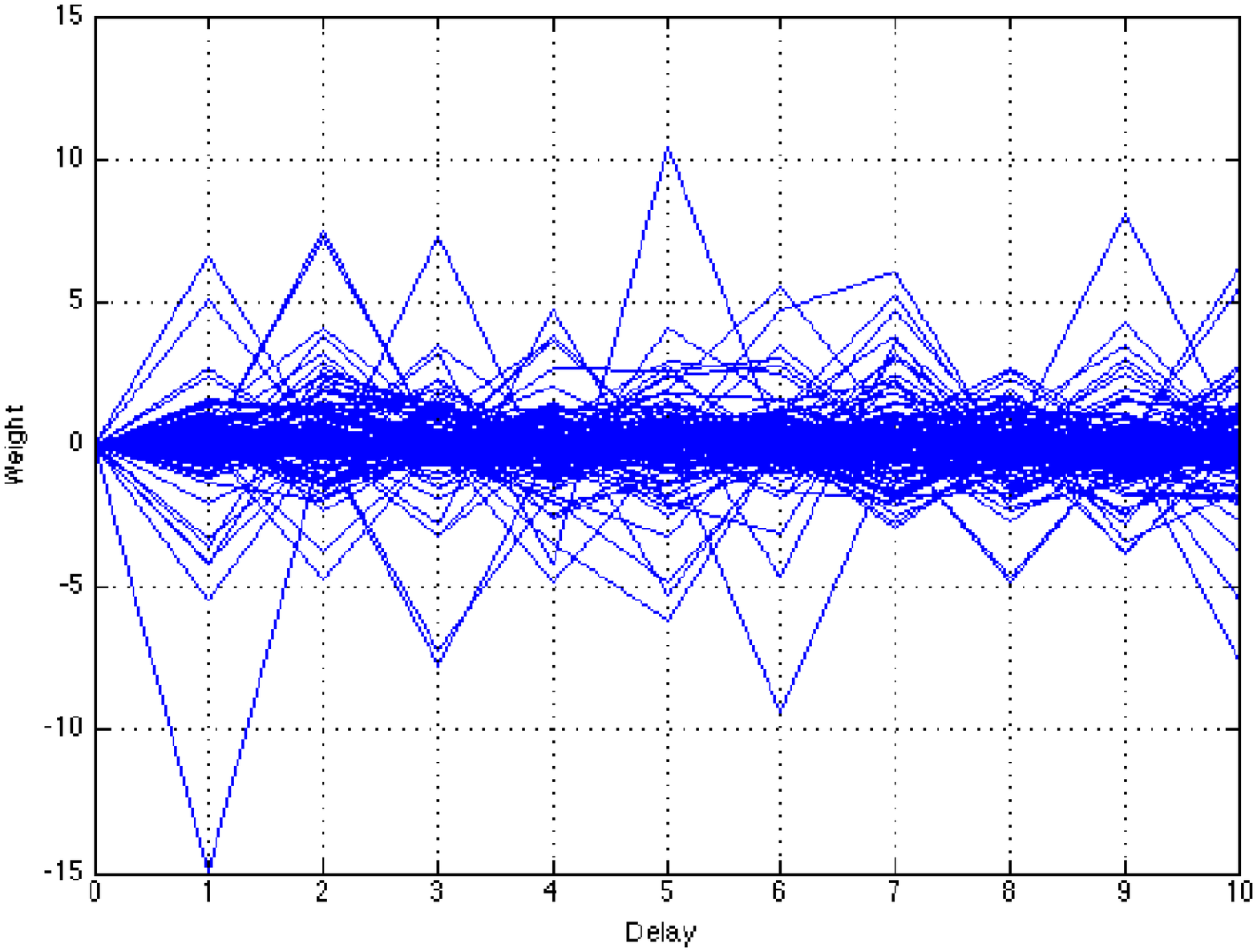}} \\
\centering
\subfigure{\includegraphics[width = 1\linewidth, height = 1.5in]{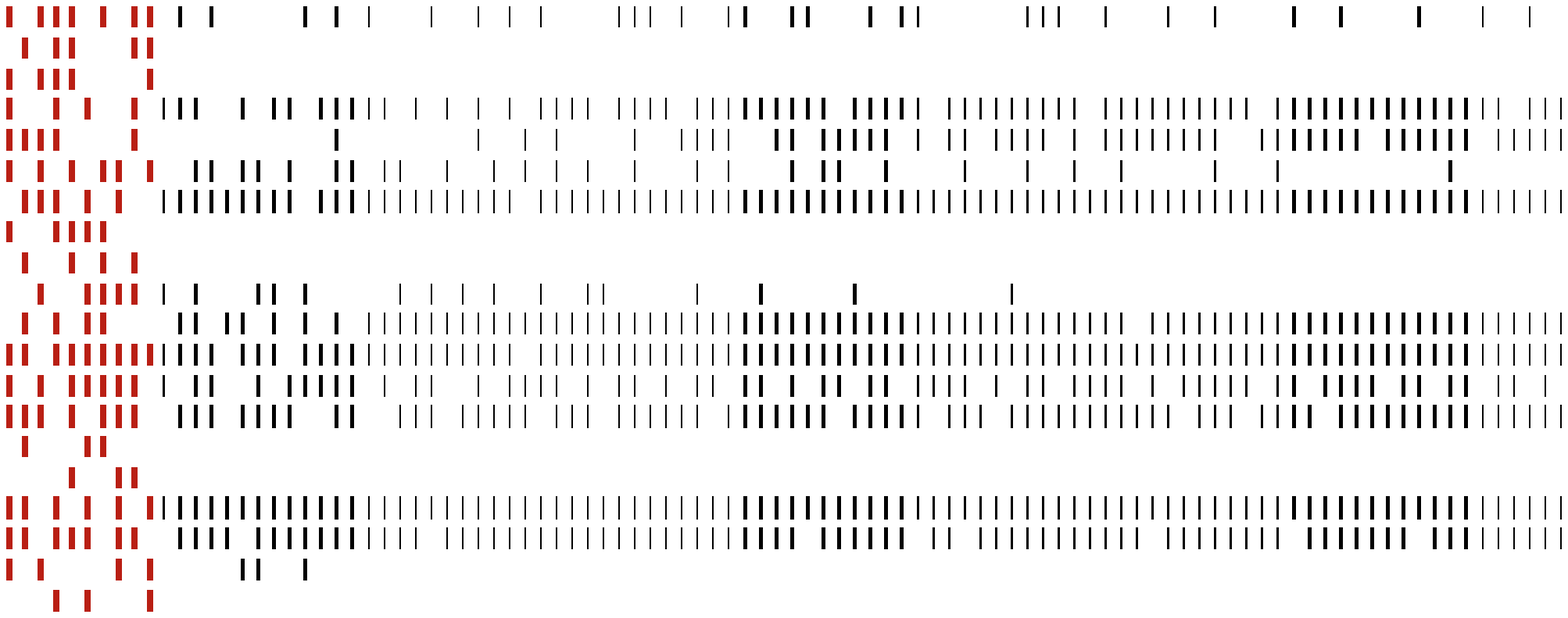}}
\centering
\caption{\label{fig:Raster02D} An example of non-trivial dynamics,with $N = 30$, $\gamma = 0.98$, $D = 10$ $T = 100$. 
Profiles corresponding to the master's excitatory profiles are superimposed in the top-left figure, 
those corresponding to the master's inhibitory profiles are superimposed in the top-left figure.
The estimated raster is shown in the bottom view.
This clearly shows that, in the absence of additional constraint, the optimal solution corresponds to wide distribution of weight's profiles.}
\end{figure}

\begin{figure}[!htbp]
\centering
\subfigure{\includegraphics[width = 0.4\linewidth, totalheight = 0.1\textheight]{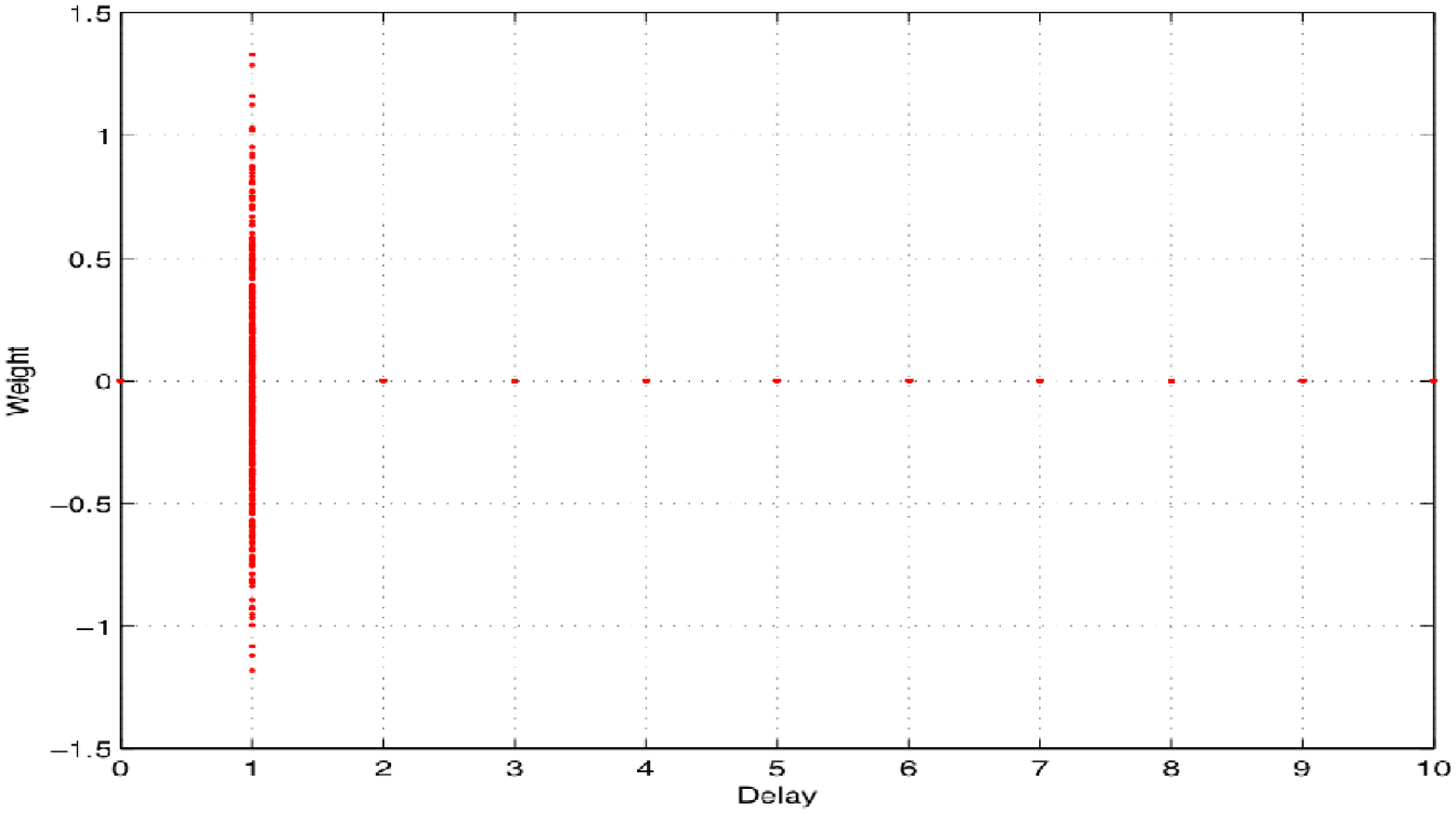}}
\hspace{0.1in}
\subfigure{\includegraphics[width = 0.4\linewidth, totalheight = 0.1\textheight]{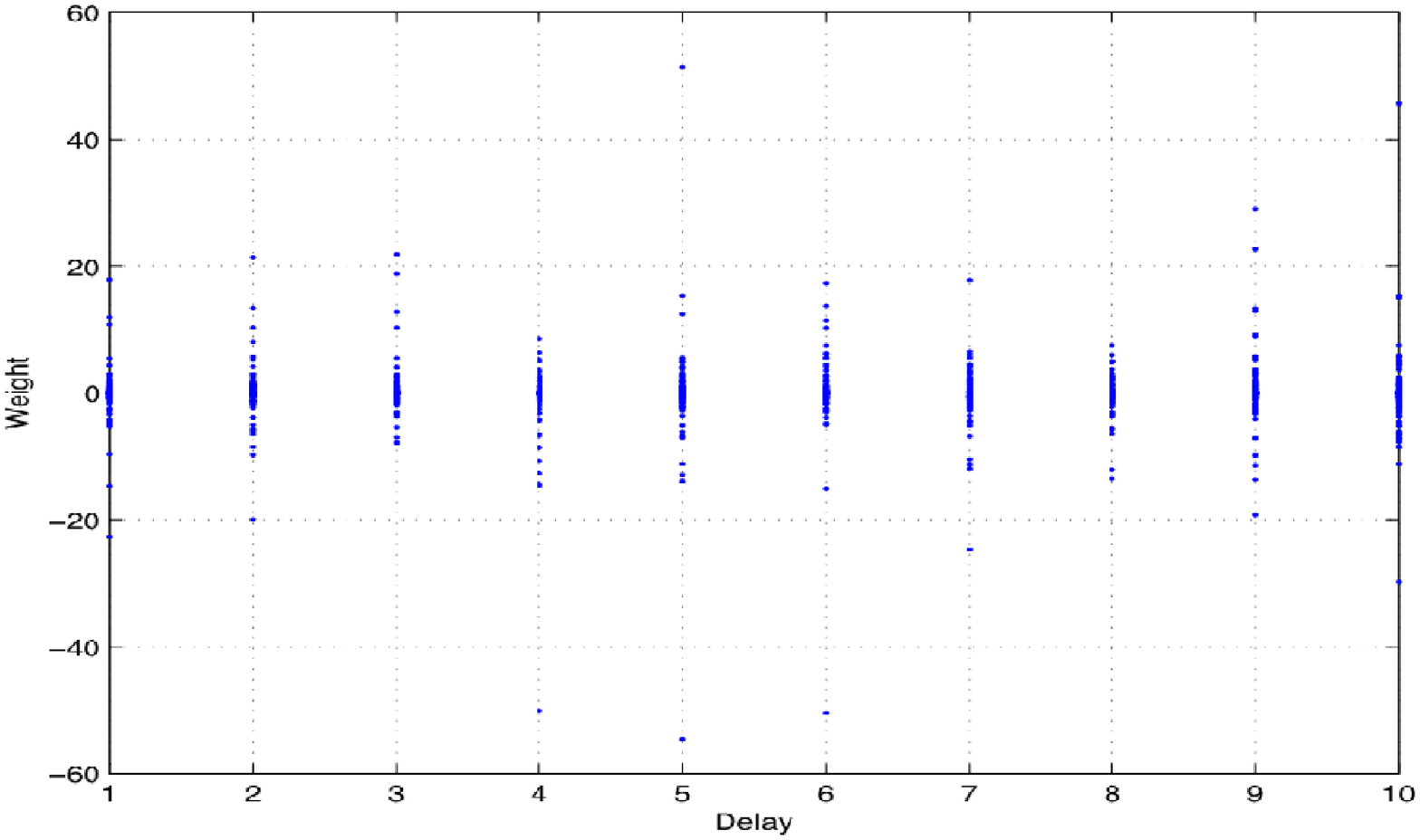}} \\
\centering
\subfigure{\includegraphics[width = 0.4\linewidth, totalheight = 0.1\textheight]{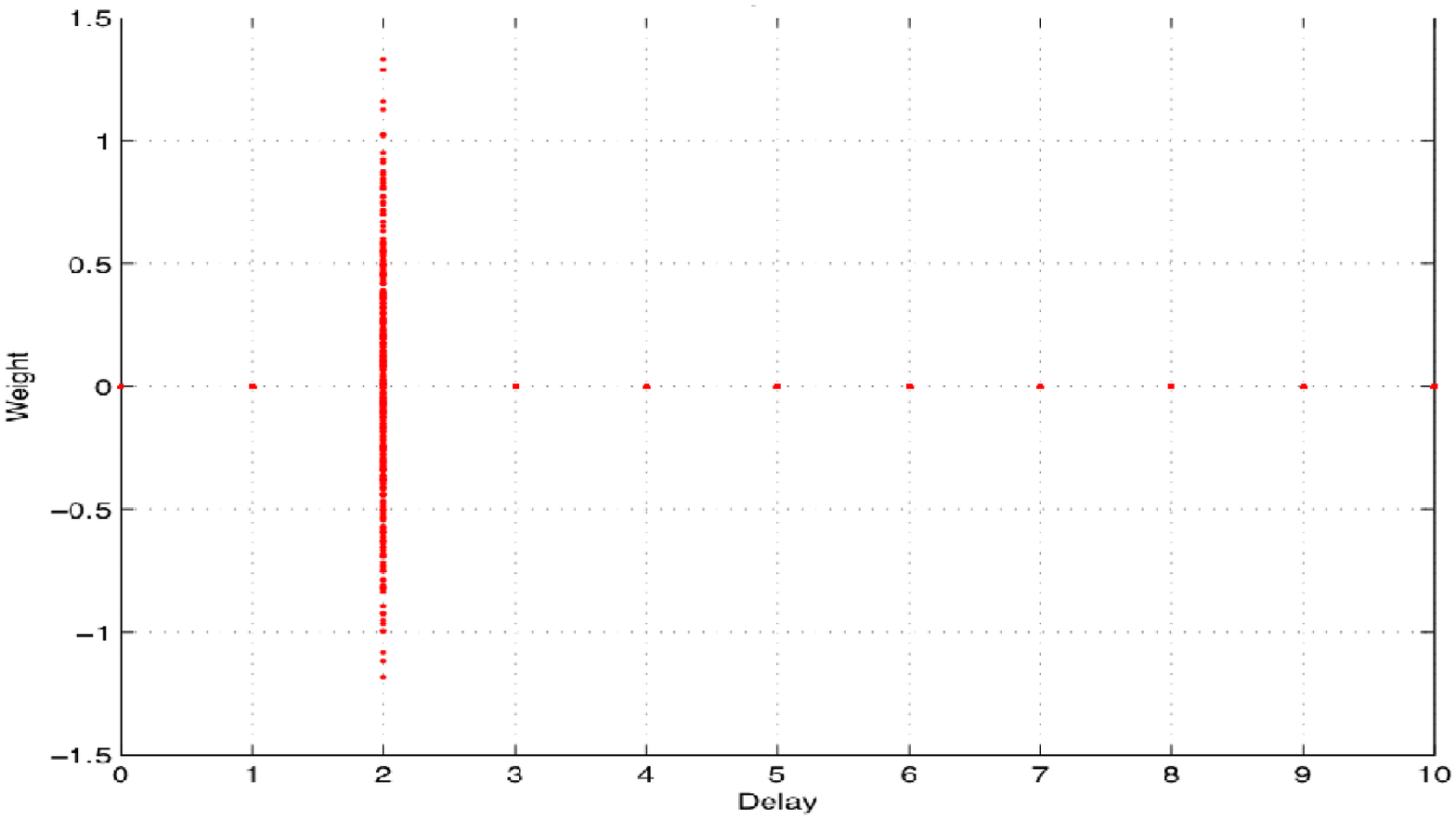}}
\hspace{0.1in}
\subfigure{\includegraphics[width = 0.4\linewidth, totalheight = 0.1\textheight]{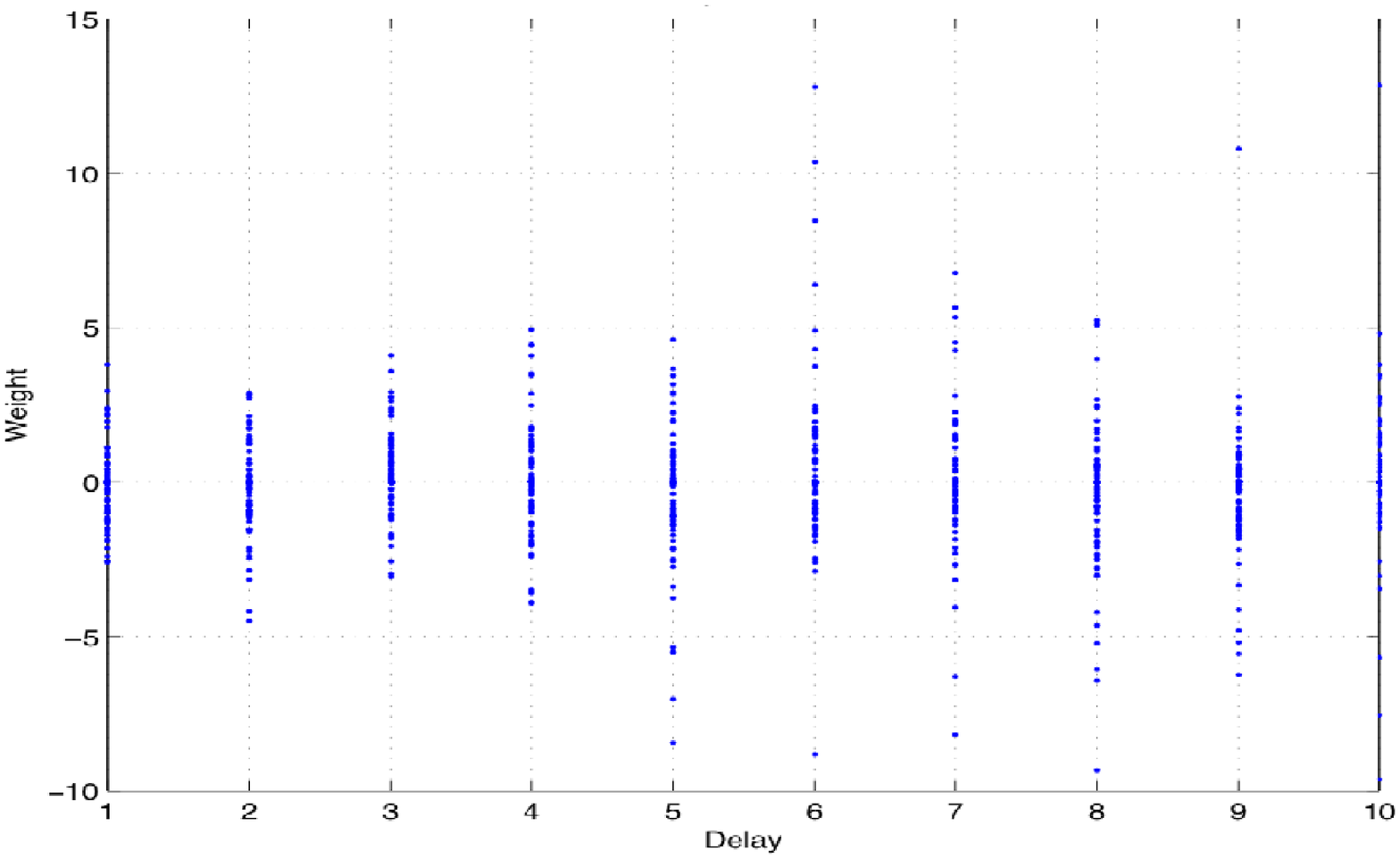}} \\
\centering
\subfigure{\includegraphics[width = 0.4\linewidth, totalheight = 0.1\textheight]{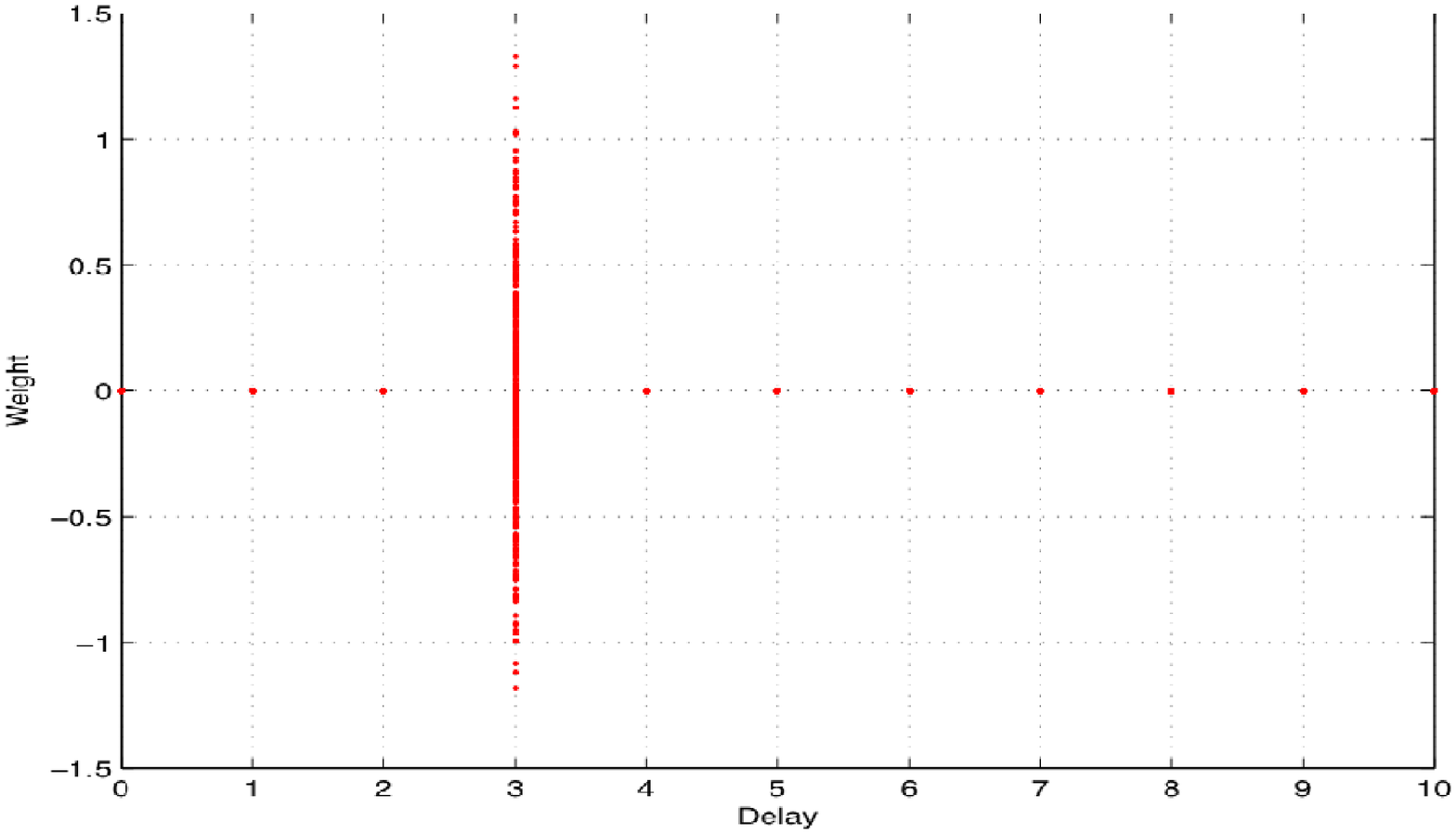}}
\hspace{0.1in}
\subfigure{\includegraphics[width = 0.4\linewidth, totalheight = 0.1\textheight]{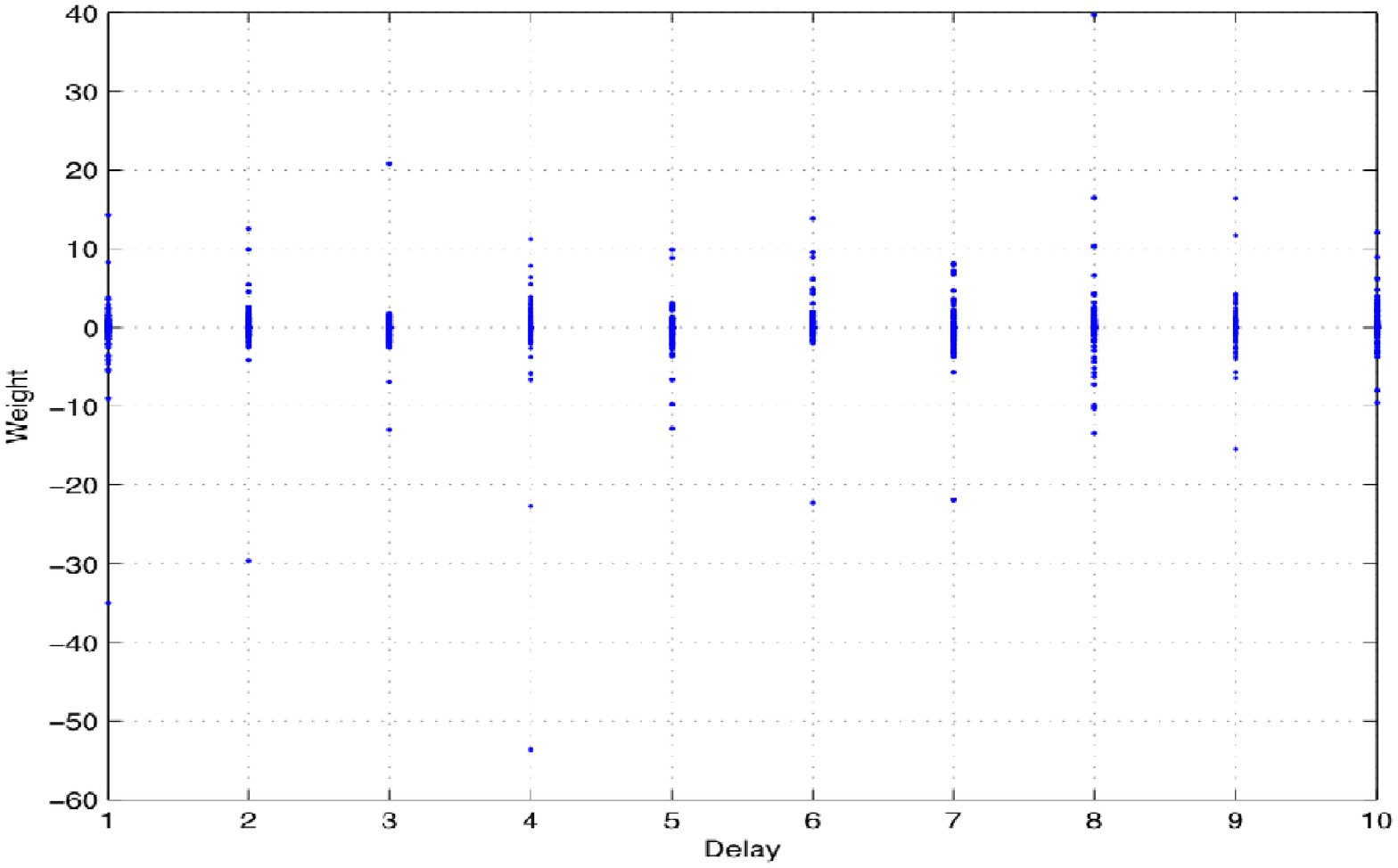}} \\
\centering
\subfigure{\includegraphics[width = 0.4\linewidth, totalheight = 0.1\textheight]{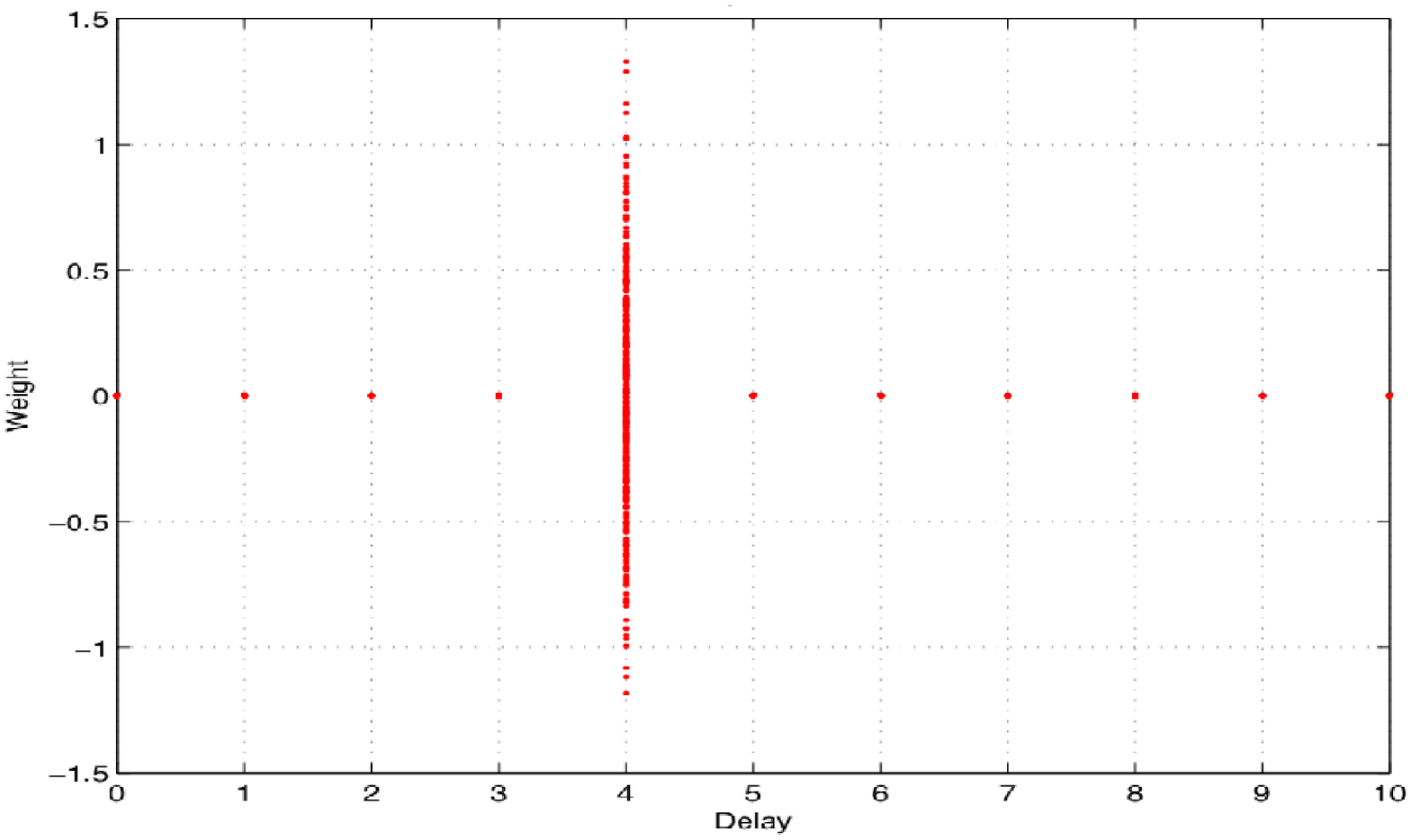}}
\hspace{0.1in}
\subfigure{\includegraphics[width = 0.4\linewidth, totalheight = 0.1\textheight]{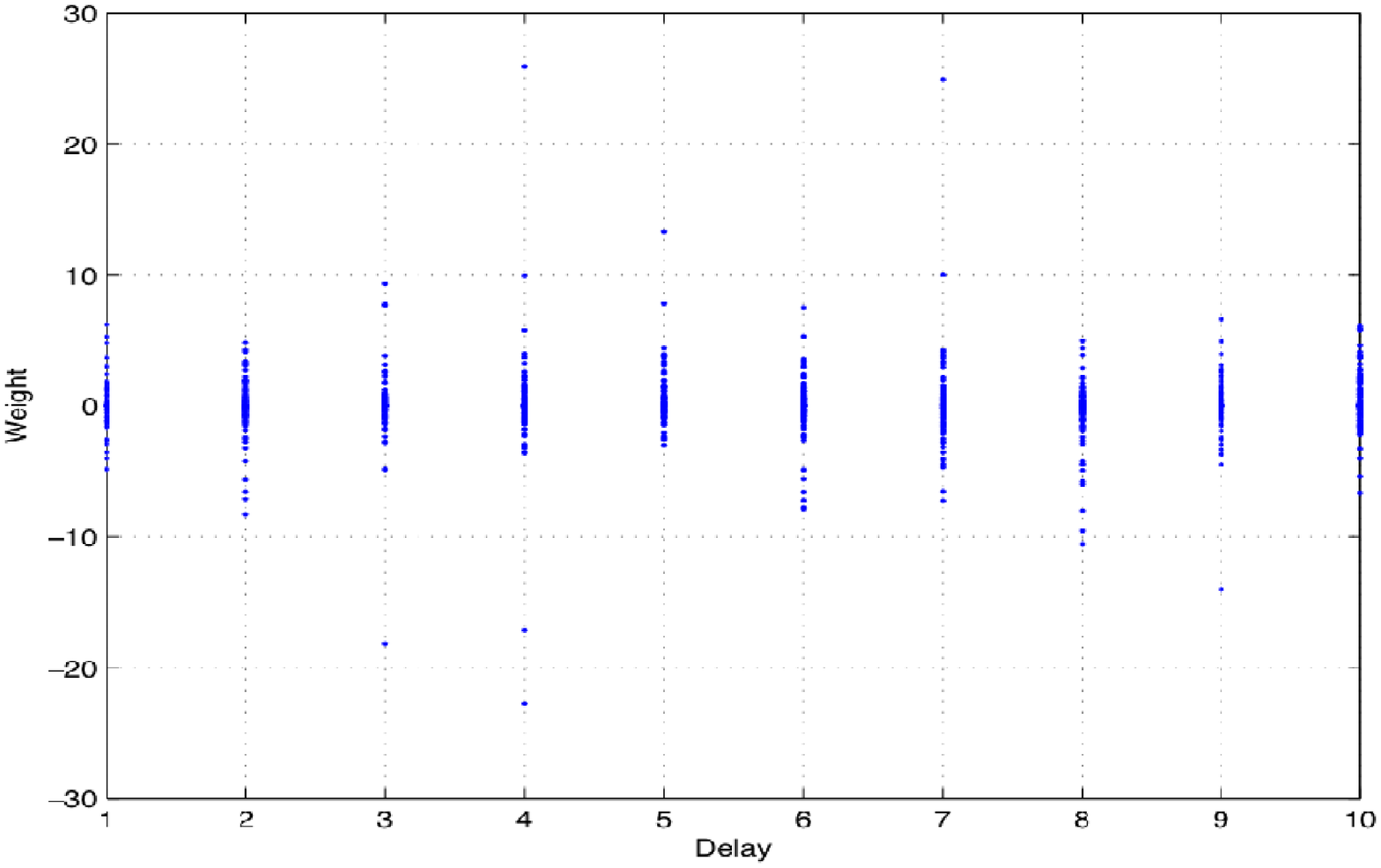}} \\
\caption{\label{fig:Raster03D} In this figure we show that whatever be the weights and delays in the master, 
with $N = 20$, $\gamma = 0.98$, $D = 10$ $T = 100$,
the estimator use all the weights and delays for calculate the raster, in order to obtain an optimal solution.}
\end{figure}

In the case of trivial dynamics, it is observed that the estimated servant weight distribution is sparse as illustrated in Fig.~\ref{fig:Raster01D}. 
However, as soon as the dynamics is not trivial, the proposed algorithm uses all delayed weight parameters in order to find an optimal solution,
without any correspondence between the master initial weight distribution and the servant estimated weight distribution. 
This is illustrated in Fig.~\ref{fig:Raster03D}, where instead of the standard profiles shown in Fig.~\ref{fig:Profil}, 
a ``Dirac'' profile has been used in the master, while the estimated weights are distributed at all possible delays. 
In order to complete this illustration an non trivial dynamics is shown in Fig.~\ref{fig:Raster02D}.

\subsubsection*{On the complexity of the generated network}

Here we maximize~(\ref{daBMS}) which is in relation with~(\ref{dBMS}).
The latter was shown to be a relevant numerical measure of the network complexity \cite{cessac:08,cessac-vieville:08}.
We thus obtain, among networks which generate exactly the required master, the ``less complex'' network, in the sense of~(\ref{daBMS}).
A very simple way to figure out how complex is the servant network is to observe its generated raster after $T$, i.e., after the period of time where it matches exactly the 
required master's raster. They are indeed the same before $T$.

After $T$, in the case of delayed weights, we still observe that if the original raster is periodic, the generated raster is still periodic with the same period.

\begin{figure}[!htbp]
\centering
\subfigure[]{\includegraphics[width = 1\linewidth, height=4cm]{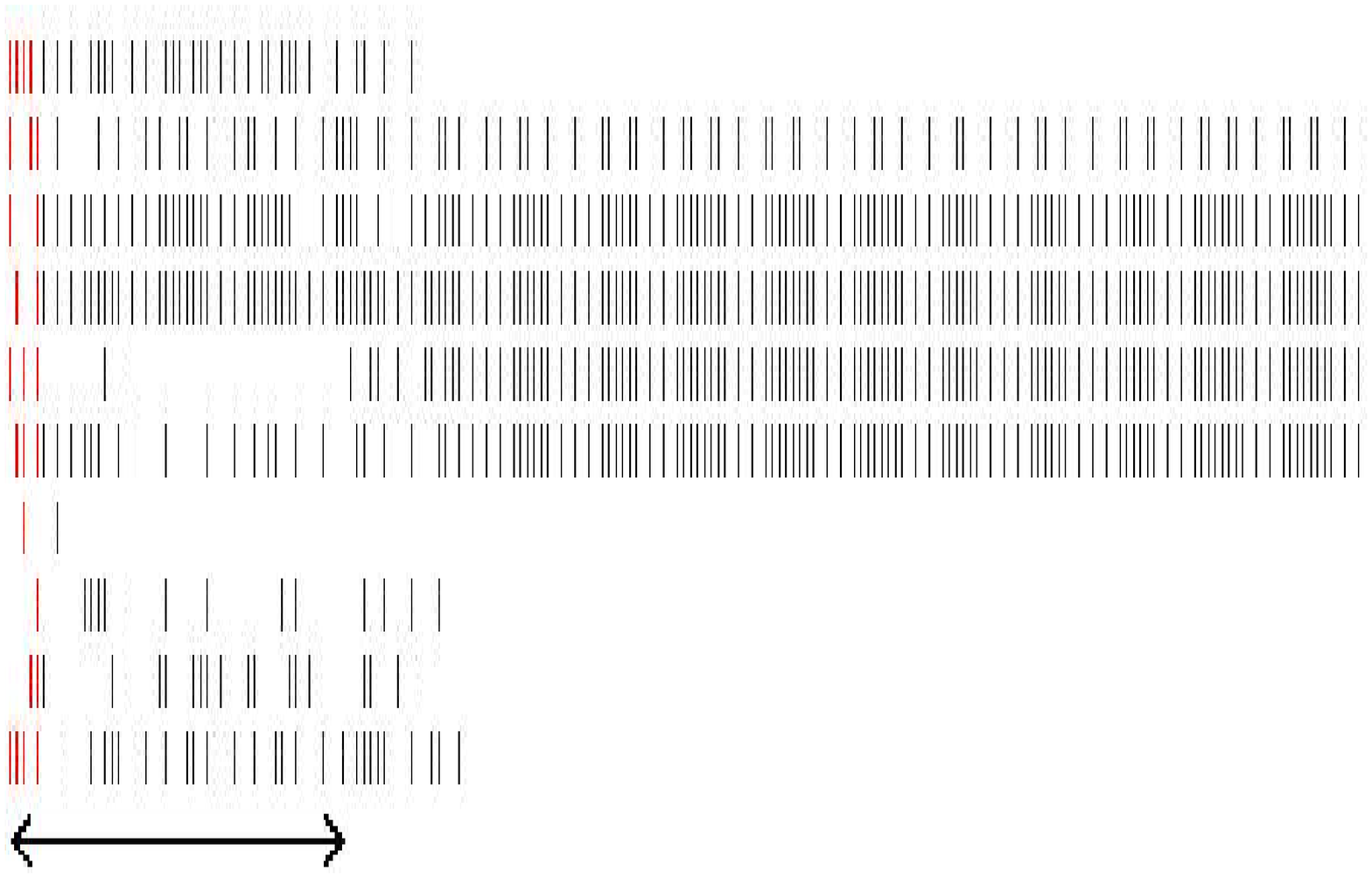}}\\
\centering
\subfigure[]{\includegraphics[width = 1\linewidth, height=4cm]{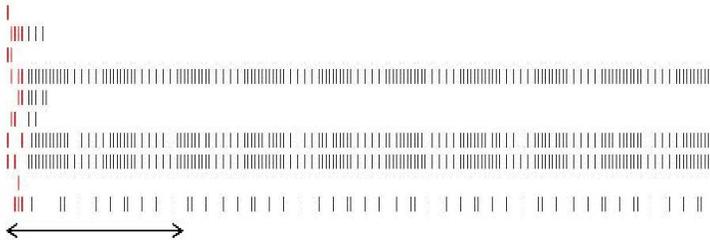}} 
\centering
\caption{\label{fig:Raster04D} Two examples of observation of the raster period, on the slave network, observing the raster 
after the delay $T$ where it matches the master raster (shown by an arrow in the figure).
(a) With $N = 10$, $\gamma = 0.9$, $\sigma = 3$, $D = 5$, $T = 50$, a periodic regime of periode $P=4$ is installed after a change in the dynamics.
(b) With $N = 10$, $\gamma = 0.9$, $\sigma = 6$, $D = 5$, $T = 50$, a periodic regime of periode $P=9$ corresponds to the master periodic regime.
}
\end{figure}

If the original raster is a-periodic, for small $N$ and $T$, we have observed that the generated master is still periodic, as illustrated in Fig.~\ref{fig:Raster04D},
and this number roughly exponentially increases with $N$ and $T$ as predicted by the theory \cite{cessac:08}. We however, have not observed any further regularity, 
for instance changes of regime can occur after the $T$ delay, huge period can be observed for relatively small numbers of $N$ and $T$, etc..
Further investigating this aspect is a perspective of the present work.

\subsection{Retrieving delayed weights from the observation of spikes, using hidden units}

In this last set of numerical experiments we want to verify that the proposed method in section~\ref{sec:reservoir} is ``universal'' and 
to evaluate the number of hidden neurons to be recruited in order to exactly simulate the required raster. If the answer is positive,
this means that we have here available a new ``reservoir computing'' mechanism.

\subsubsection*{Considering Bernoulli distribution}

We start we a completely random input, drawn from a uniform Bernoulli distribution. This corresponds to an input with maximal entropy. 
Here the master/servant trick is no more used. 
Thus, the raster to reproduce has no chance to verify the neural network dynamics constraints induced by~(\ref{BMS}), 
unless we add hidden neurons as proposed in section~\ref{sec:reservoir}.

\begin{figure}[!htbp]\begin{center}\begin{tabular}{cc}
(a) & (b) \\
\parbox[u]{0.4\textwidth}{\includegraphics[width = 0.4\textwidth, height=1cm]{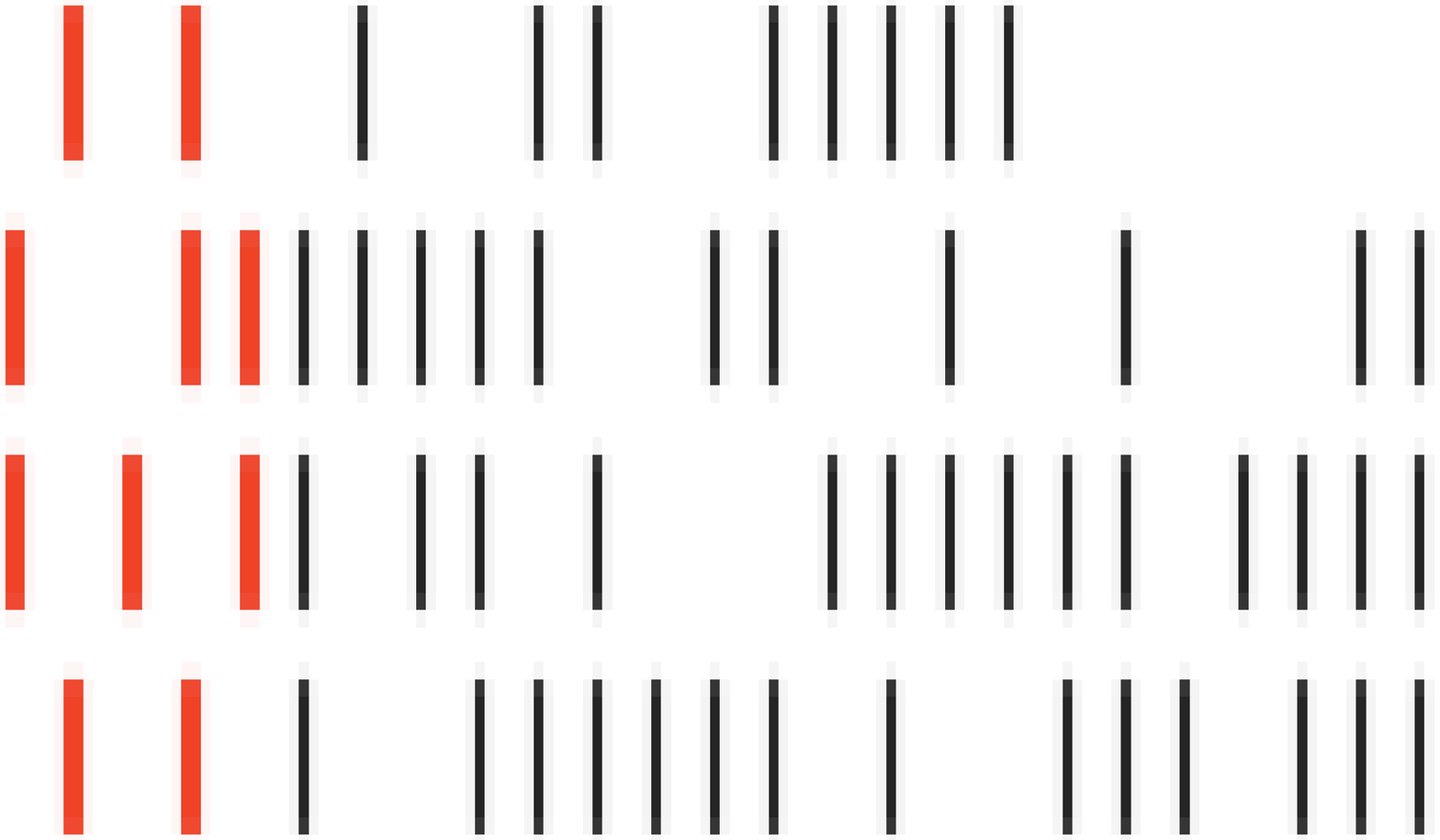}}
&
\parbox[u]{0.4\textwidth}{\includegraphics[width = 0.4\textwidth, height=1cm]{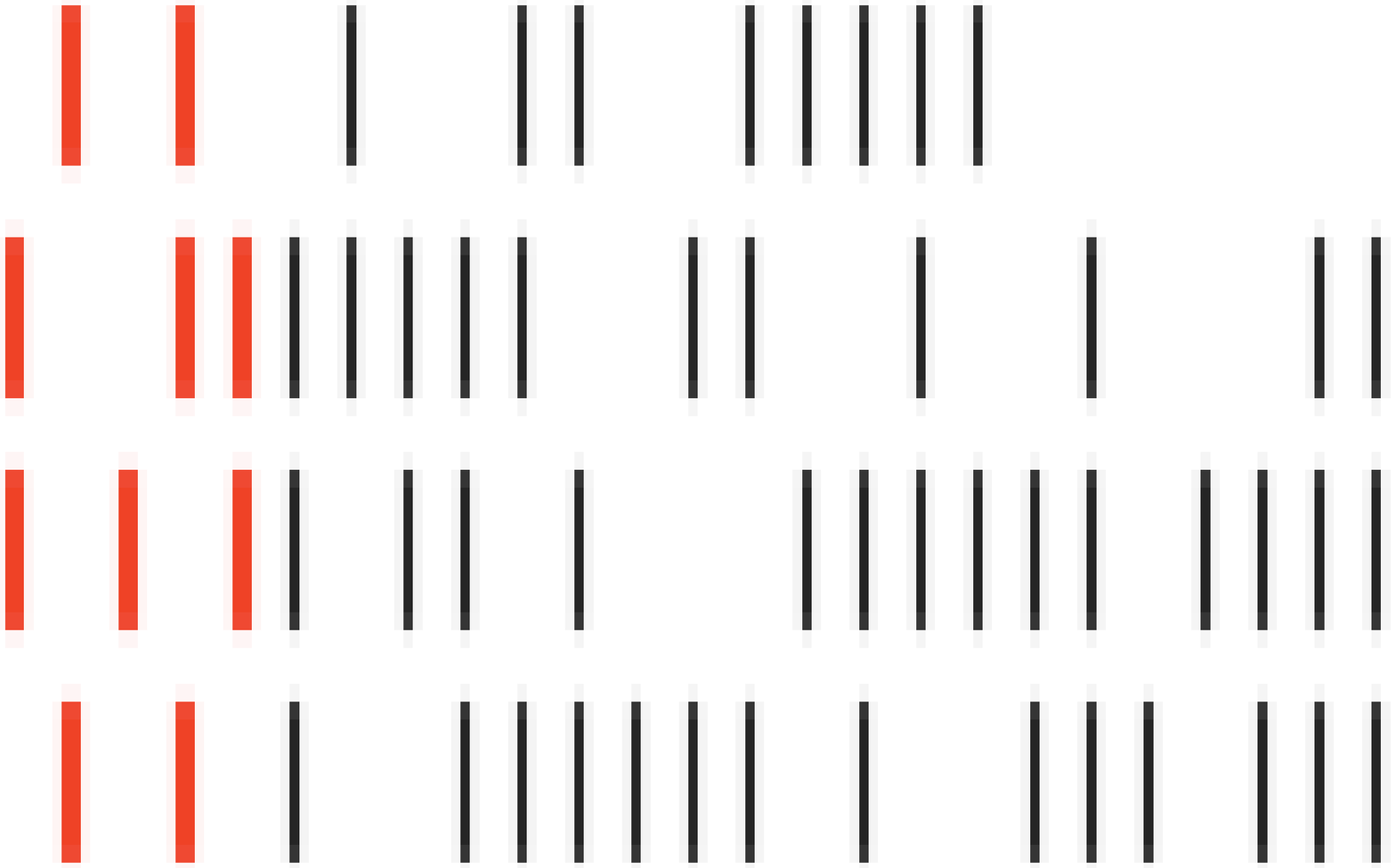}} 
\\
(c) & (d) \\
\parbox[u]{0.4\textwidth}{\includegraphics[width = 0.4\textwidth, height=1cm]{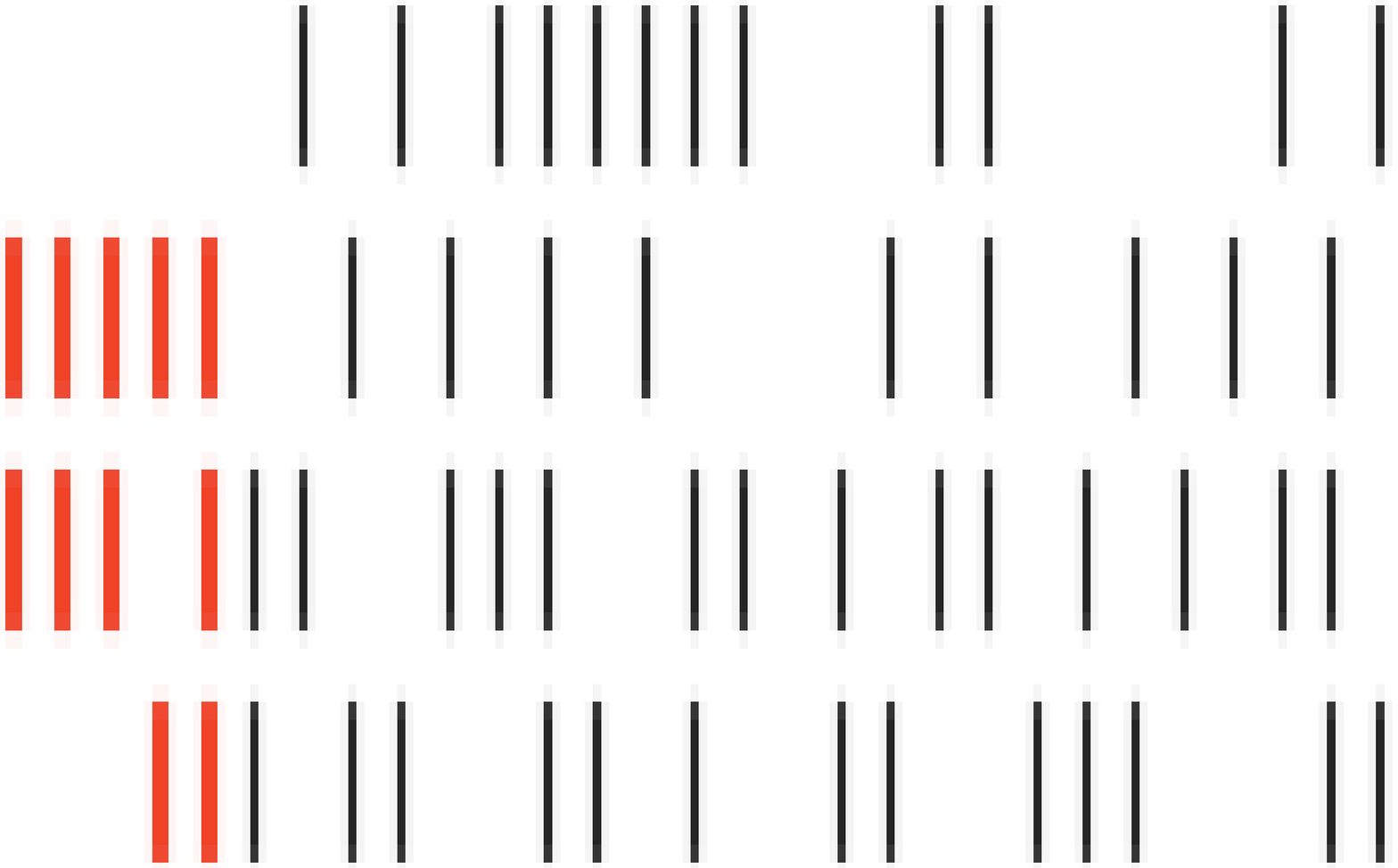}~\vspace{0.25cm}}
&
\parbox[u]{0.4\textwidth}{\includegraphics[width = 0.4\textwidth, height=1.25cm]{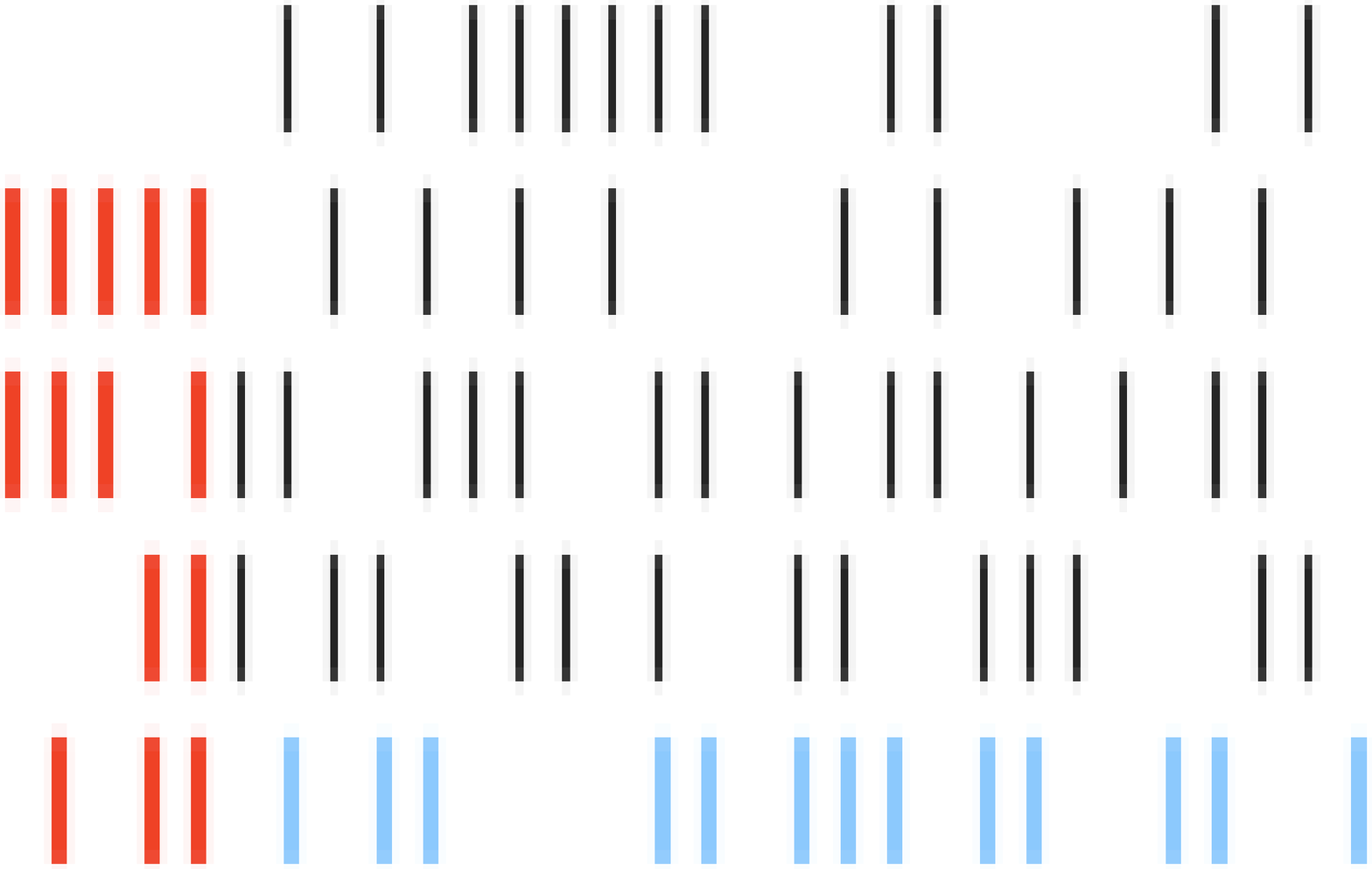}} 
\\
(e) & (f) \\
\parbox[u]{0.4\textwidth}{\includegraphics[width = 0.4\textwidth, height=1cm]{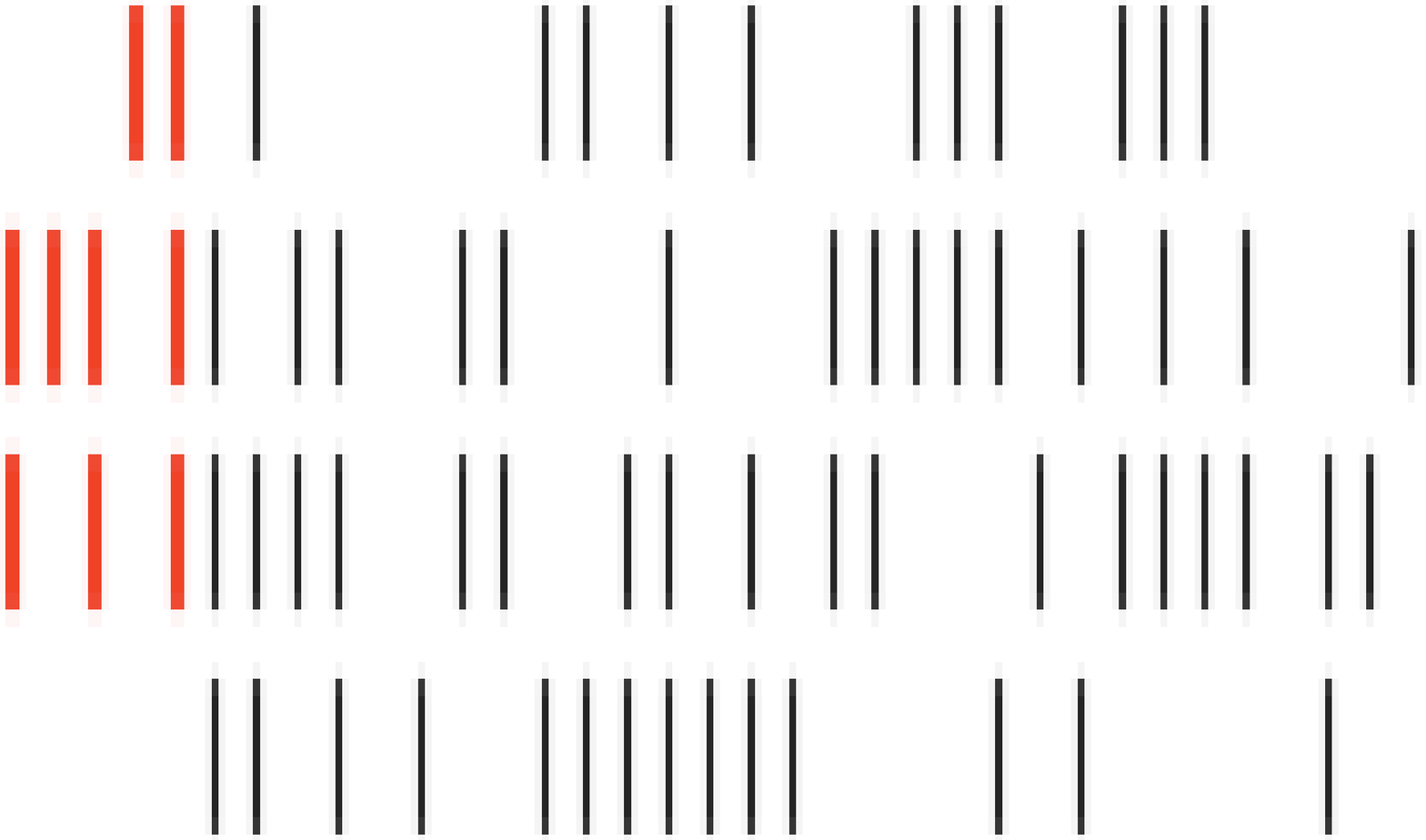}~\vspace{0.50cm}}
& 
\parbox[u]{0.4\textwidth}{\includegraphics[width = 0.4\textwidth, height=1.5cm]{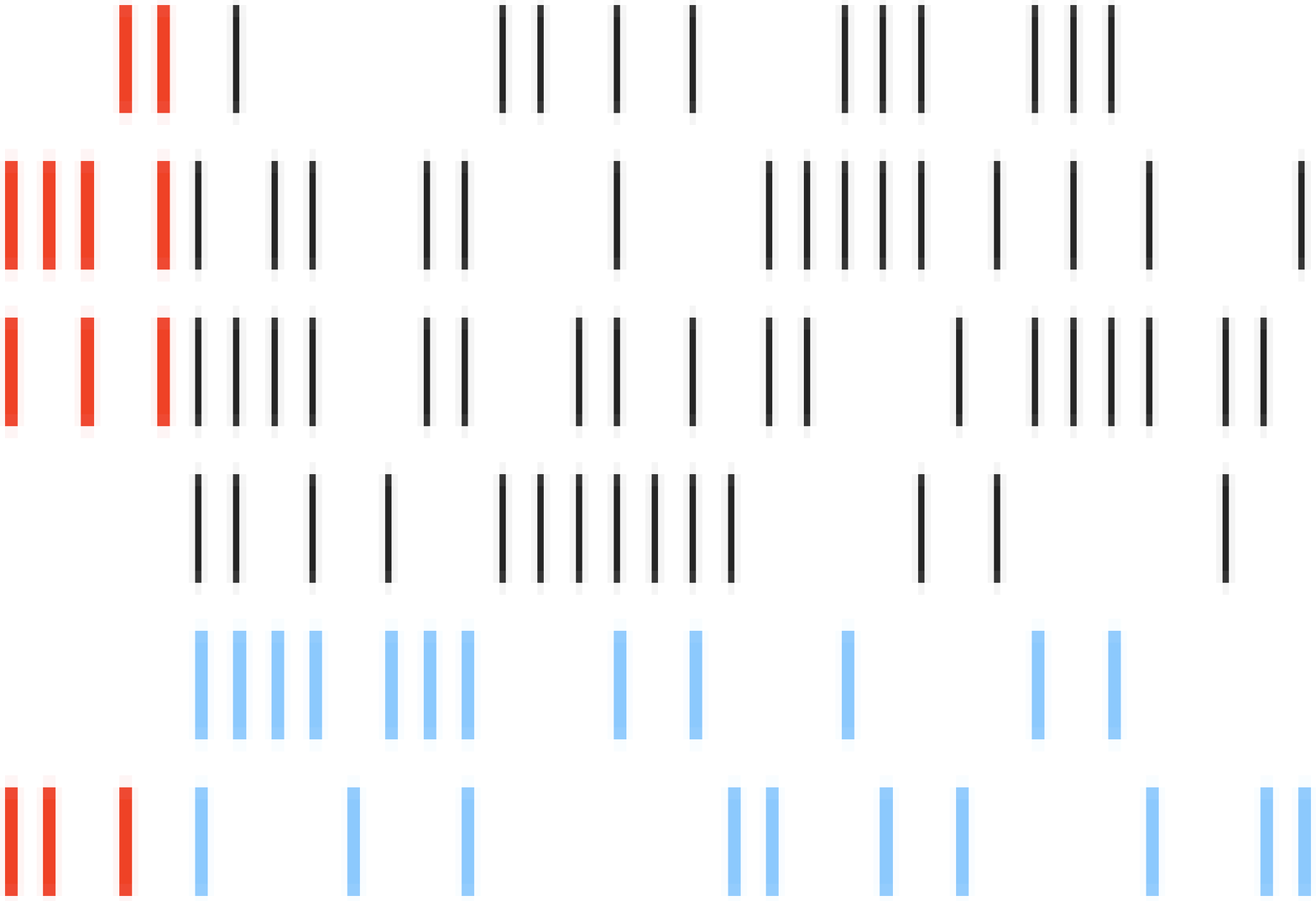}} 
\\
(g) & (h) \\
\parbox[u]{0.4\textwidth}{\includegraphics[width = 0.4\textwidth, height=1cm]{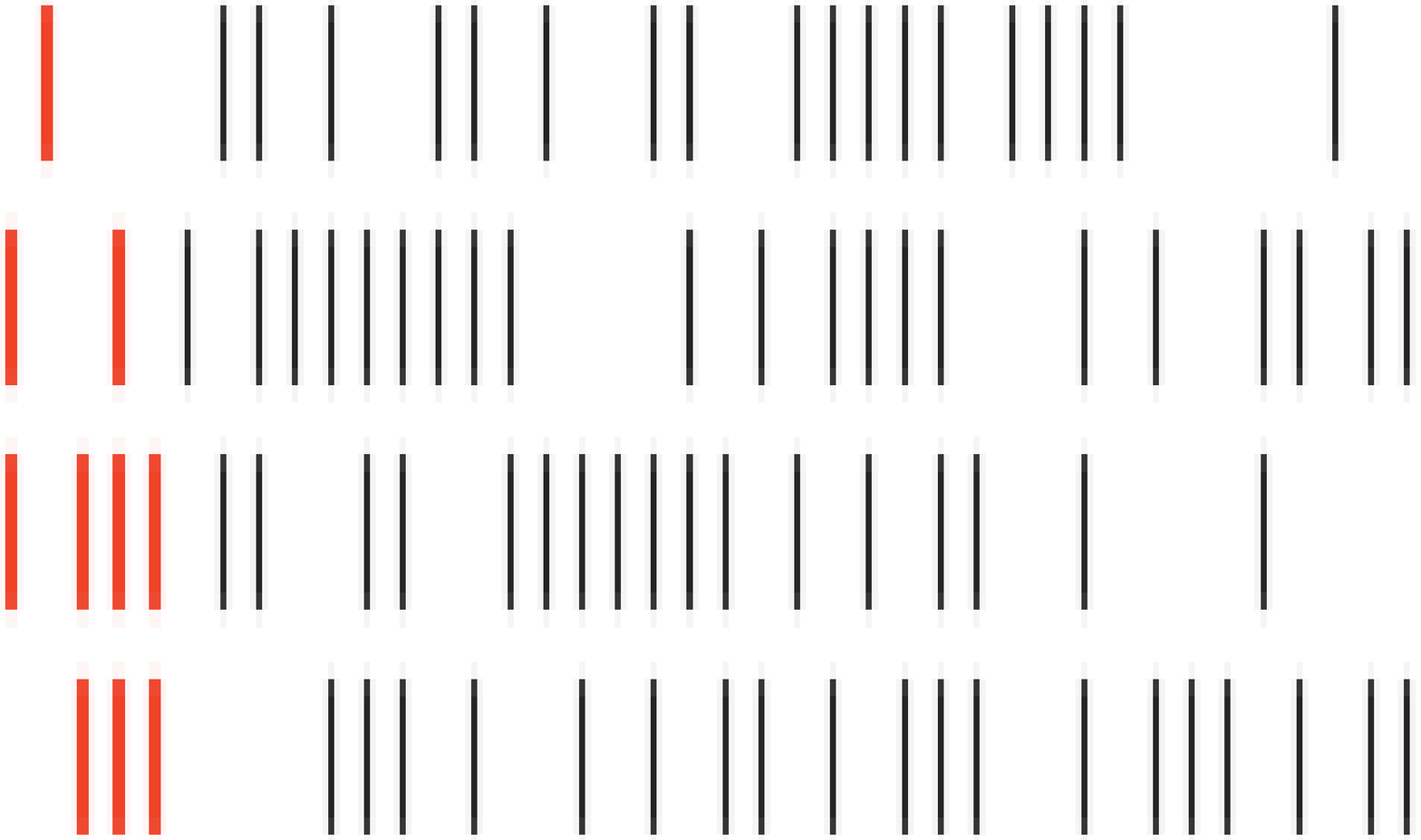}~\vspace{0.75cm}}
& 
\parbox[u]{0.4\textwidth}{\includegraphics[width = 0.4\textwidth, height=1.75cm]{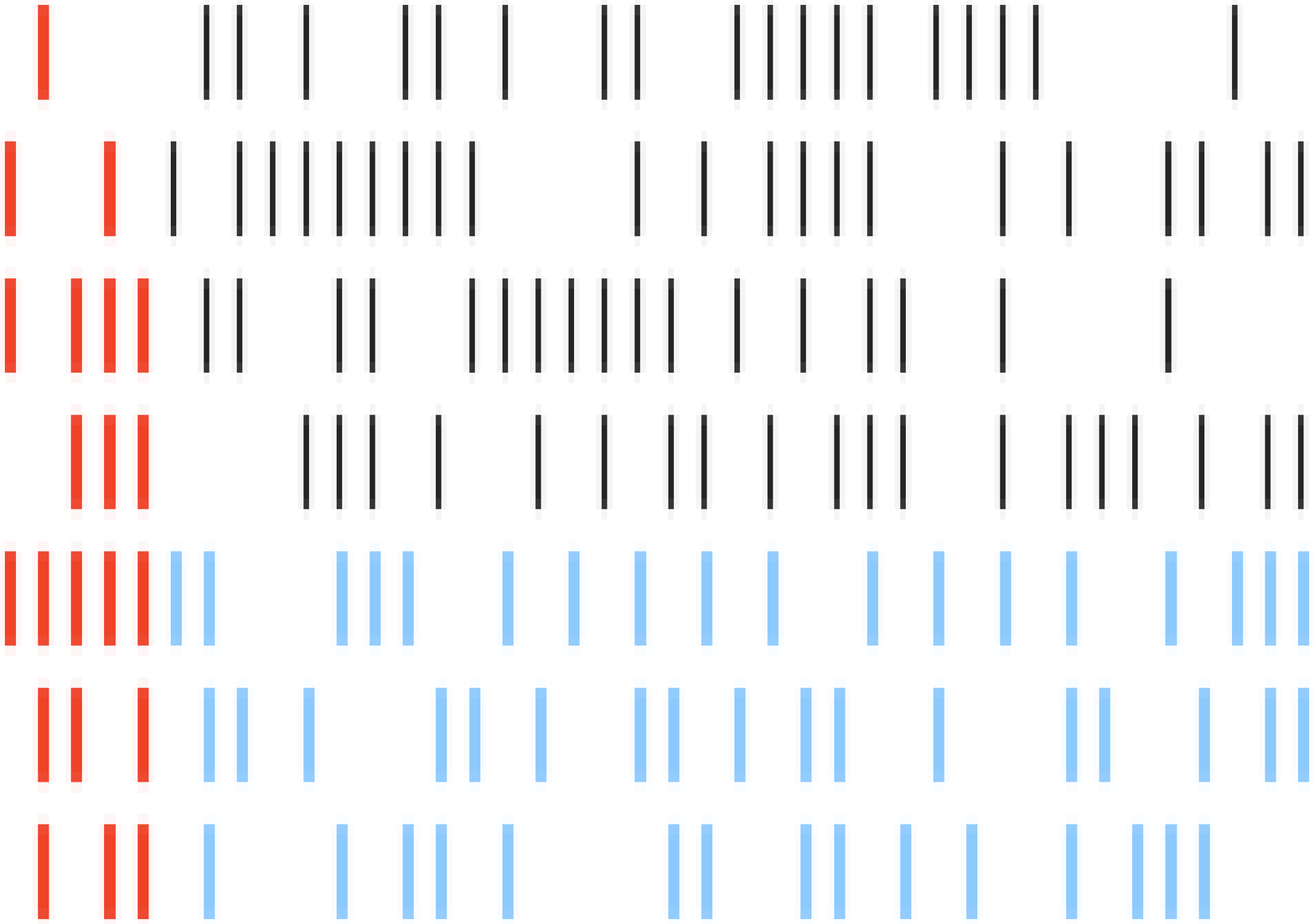}} 
\\
\end{tabular}
\caption{\label{fig:RasterB01} Finding the expected dynamics from a raster with uniform distribution. (a), (c), (e) and (g) correspond to different raster with Bernoulli Distribution in addition (b), (d), (f) and (h) show the raster calculated by the methodology proposed. The red lines correspond to initial conditions (initial raster), the black ones are the spikes calculated by the method and the blues ones are the spikes in the hidden layer obtained with a Bernoulli Distribution. We can also observe that the number of neurons in the hidden layer increases, 1 by 1, between (b), (d), (f) and (h), this is because the observation time T is augmented, 5 by 5, as predicted.
Here $N = 4$, $\gamma = 0.95$, $D = 5$; 
in (a)(b) $T = 25$ with $S = 0$,
in (c)(d) $T = 30$ with $S = 1$,
in (e)(f) $T = 35$ with $S = 2$,
in (g)(h) $T = 40$ with $S = 3$, while $S$ correspond to the number of neurons in the hidden layer, detailed in the text.}
\end{center}
\end{figure}

\begin{figure}[!htbp]
\begin{center}
\includegraphics[width=0.4\textwidth, totalheight = 0.2\textheight]{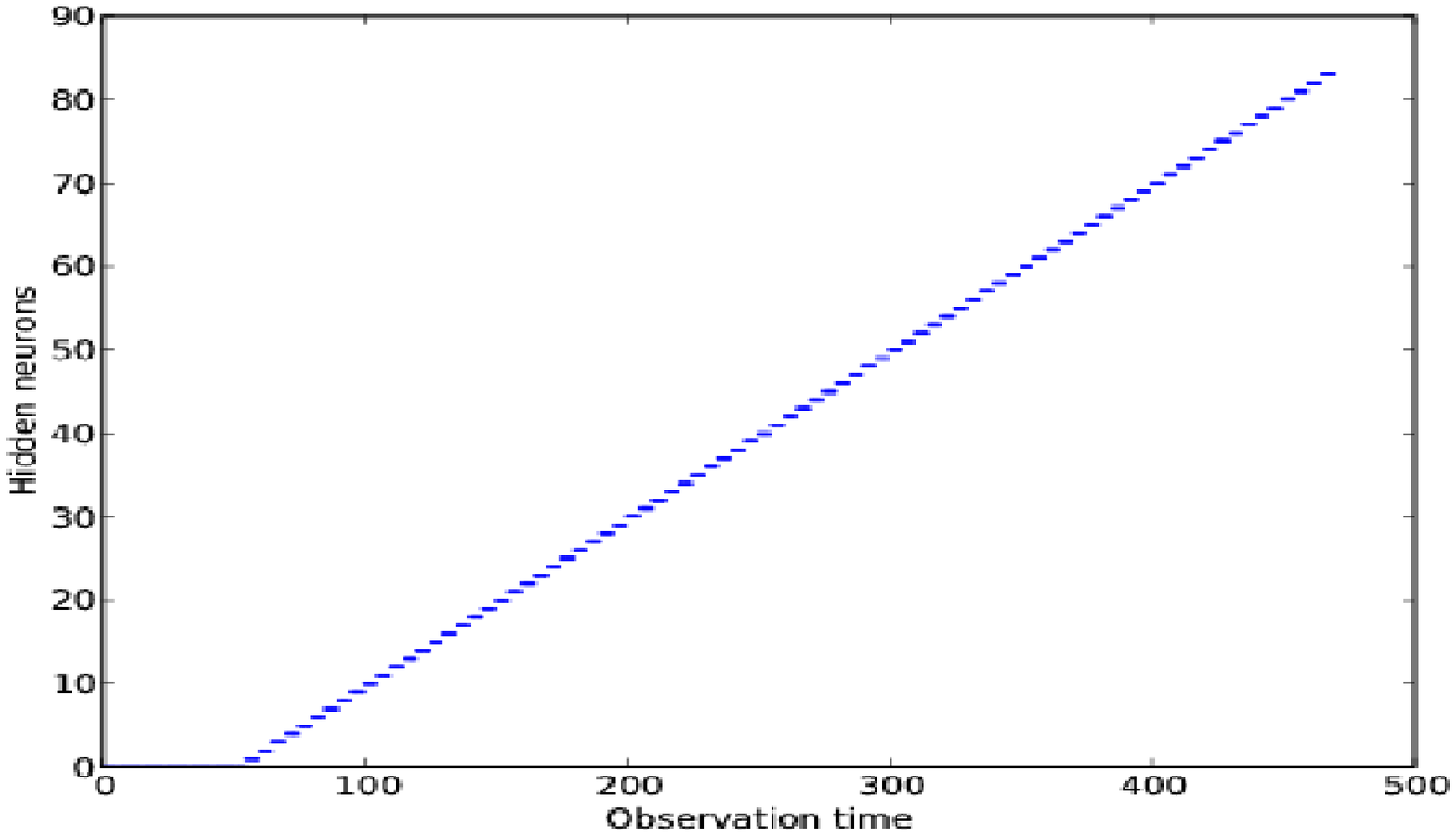} ~~
\includegraphics[width=0.4\textwidth, totalheight = 0.2\textheight]{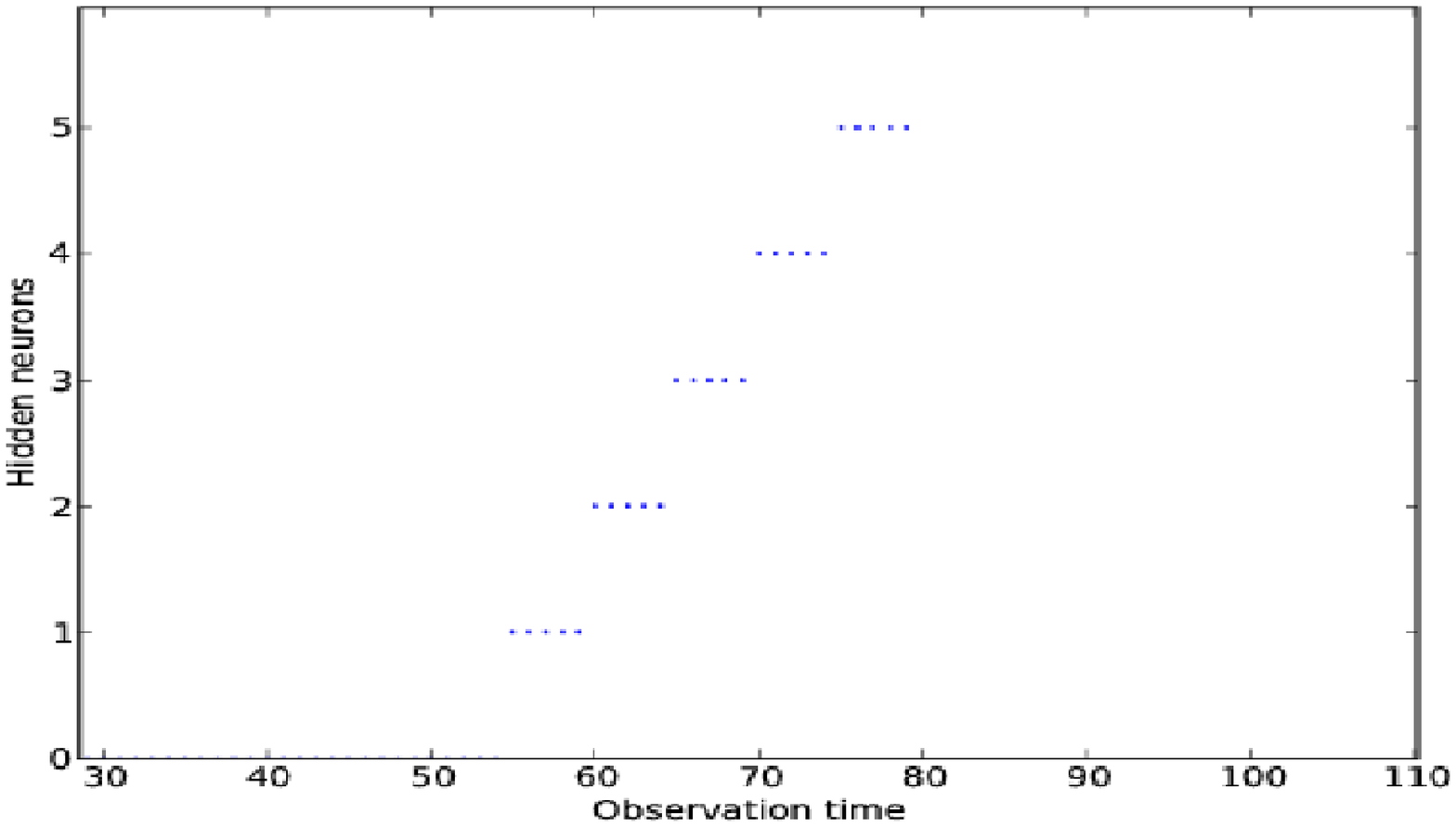}
\caption{\label{fig:RelNhT} Relationship between the number of hidden neurons $S$ and the observation time $T$, 
here $N = 10$, $T = 470$, $D = 5$, $\gamma = 0.95$ for this simulation. The right-view is a zoom of the left view.
This curves shows the required number of hidden neurons, using the proposed algorithm, in order to obtain an exact simulation.
We observe that $S = \frac{T}{D} - N$, thus that an almost maximal number of hidden neuron is required.
This curve has been drawn from $45000$ independent randomly selected inputs.}
\end{center}
\end{figure}

As observed in Fig.~\ref{fig:RasterB01}, we obtain as expected an exact reconstruction of the raster, while as reported in Fig.~\ref{fig:RelNhT}, 
we need an almost maximal number of hidden neurons for this reconstruction, as expected since we are in a situation of maximal randomness,

\subsubsection*{Considering correlated distributions}

We now consider a correlated random input, drawn from a Gibbs distribution \cite{chazottes-et-al:98,cessac-etal:08b}. To make it simple, the raster input is drawn
from a Gibbs distribution, i.e. a parametrized rank $R$ Markov conditional probability of the form:
\\ \centerline{$P(\{Z_i[k], 1 \leq i \leq N\}|\{Z_i[k - l], 1 \leq i \leq N, 1 \leq l < R\}) = 
 \frac{1}{Z} \, exp\left(\Phi_\lambda(\{Z_i[k - l], , 1 \leq i \leq N, 0 \geq l > -R\})\right)$} \\
where $\Phi_\lambda()$ is the related Gibbs potential parametrized by $\lambda$ and $Z$ a normalization constant.

This allows to test our method on highly-correlated rasters. We have chosen a potential of the form:
\\ \centerline{$\Phi_\lambda(\left.Z\right|_{k = 0}) = r \, \sum_{i = 1}^N Z_i[0] + C_t \, \sum_{i = 1}^N \prod_{l = 0}^R Z_i[l] + C_i \, \prod_{l = 0}^R Z_i[0]$} \\
thus with a term related to the firing rate $r$, a term related to temporal correlations $C_t$, and a term related to inter-unit correlation $C_i$.

\begin{figure}[!htbp]
\begin{center}\includegraphics[width = 1\linewidth,]{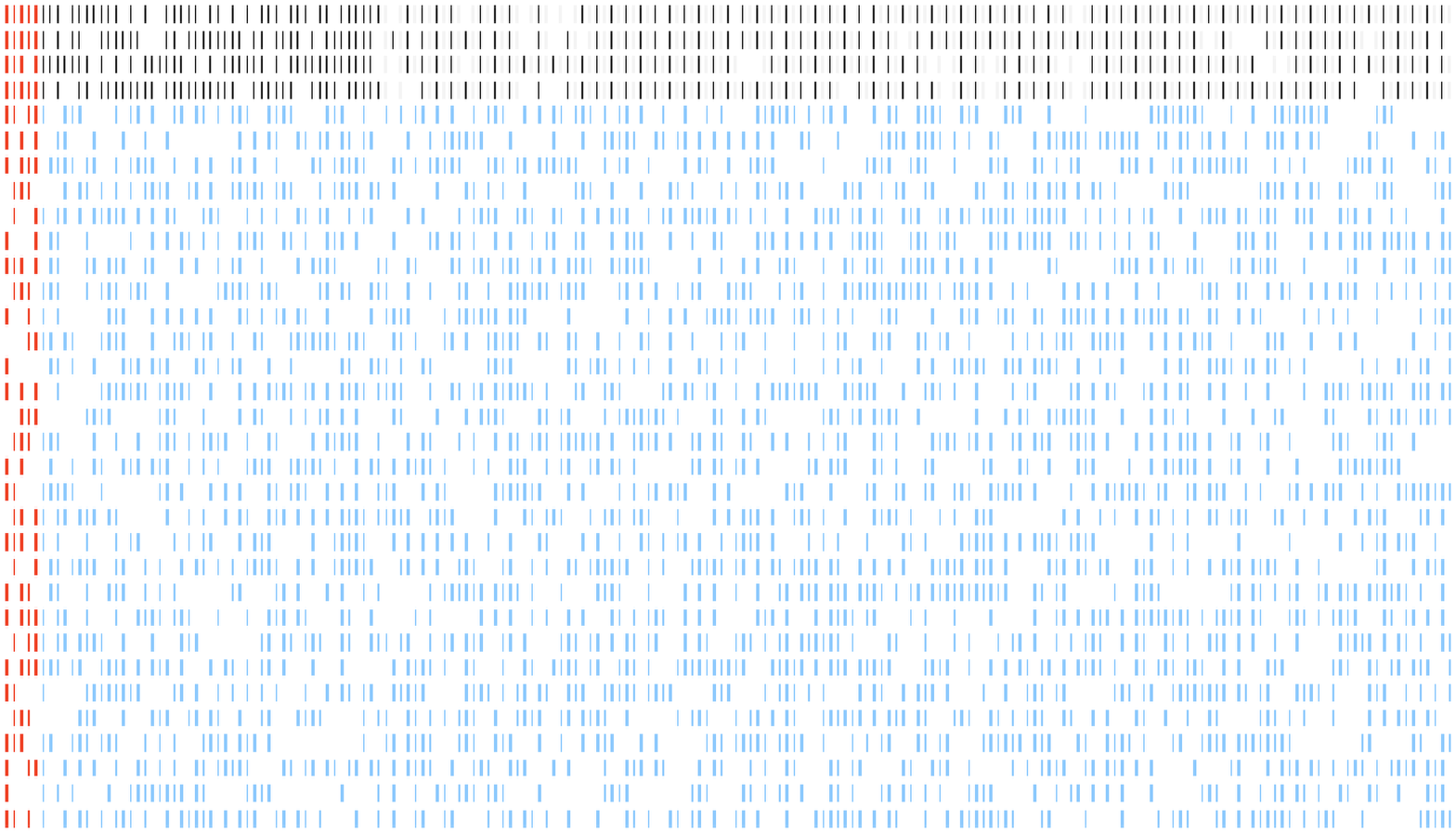}
\caption{\label{fig:GibbsC0} Raster calculated, by the proposed method, from a highly correlated Gibbs distribution. Here $r = 1$, $C_t = 0.5$ and $C_i = 1$.
The other parameters are $N = 4$, $\gamma = 0.95$, $D = 6$, $T = 200$ with $S = 29$. 
The red lines correspond to initial conditions (initial raster), the black ones are the input/output spikes and the blues ones are the spikes in the hidden layer.}
\end{center}
\end{figure}

We obtain a less intuitive result in this case, as illustrated in Fig.~\ref{fig:GibbsC0}: event strongly correlated (but aperiodic) rasters are reproduced only if 
using as many hidden neurons as in the non-correlated case. In fact we have drawn the number $S$ of hidden neurons against the observation period $T$
randomly selecting $45000$ inputs and have obtained the same curve as in Fig.~\ref{fig:RelNhT}.

This result is due to the fact that since the raster is aperiodic, thus non predictable changes occur in the general case, at any time. 
The raster must thus be generated by a maximal number of degrees of freedom, as discussed in the previous sections.

\begin{figure}[!htbp]
\begin{center}
\includegraphics[width=1\textwidth]{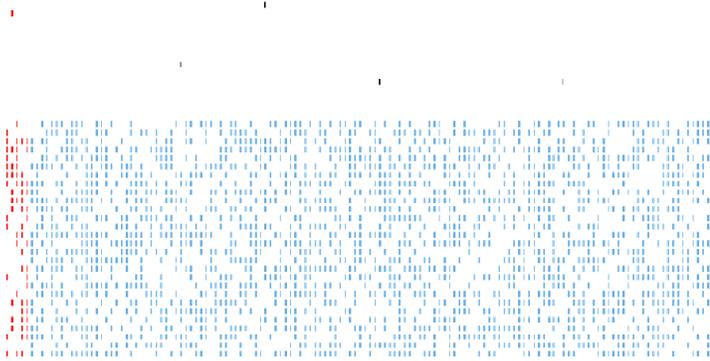}
\caption{\label{fig:Bio03out} Raster calculated, by the proposed method, from a very sparse raster, with $N = 20$, $\gamma = 0.98$, $D = 5$, $T = 130$ and $S = 28$.
The hidden neurons derived by the present algorithm simply allow to maintain the network activity in order to fire the spare spikes at the right time. 
Color codes are the same as previously.}
\end{center}
\end{figure}

In order to further illustrate this aspect, we also show in Fig.~\ref{fig:Bio03out} how a very sparse raster is simulated. We again obtain a solution with the same ratio of 
hidden neurons. Clearly the number of hidden neurons could have been less, as discussed in the previous sections. This shows that the algorithm is very general,
but not optimal in terms of number of hidden neurons.

\subsubsection*{Considering biological data}

As a final numerical experiment, we consider two examples of biological data set borrowed from \cite{riehle-etal:00} by the courtesy of the authors.
Data are related to spike synchronization in a population of motor cortical neurons in the monkey, during preparation for movement,
in a movement direction and reaction time paradigm. 
Raw data are presented trial by trial (without synchronization on the input), 
for different motion directions and the input signal is not represented, since meaningless for the purpose.
Original data resolution was $0.1ms$ while we have have considered a $1ms$ scale here. 

\begin{figure}[!htbp]
\begin{center}
\includegraphics[width=1\textwidth]{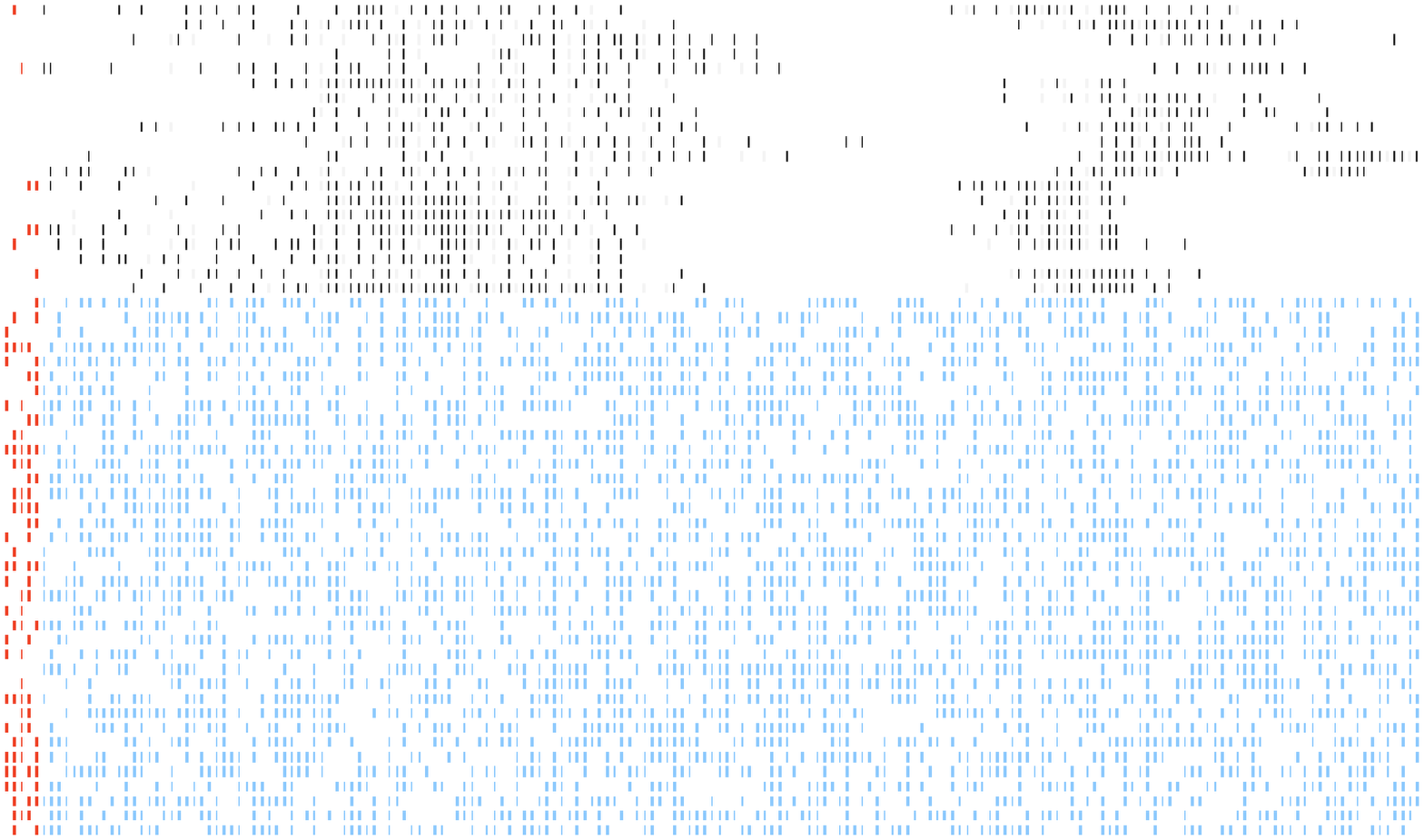}
\caption{\label{fig:Bio01out} Raster calculated, by the proposed method, from a biological data set, with $N = 20$, $\gamma = 0.95$, $D = 5$ $T = 190$ and $S = 38$.
Color codes are the same as previously. See text for details. From \cite{riehle-etal:00} by the courtesy of the authors.}
\end{center}
\end{figure}

\begin{figure}
\begin{center}
\includegraphics[width=1\textwidth]{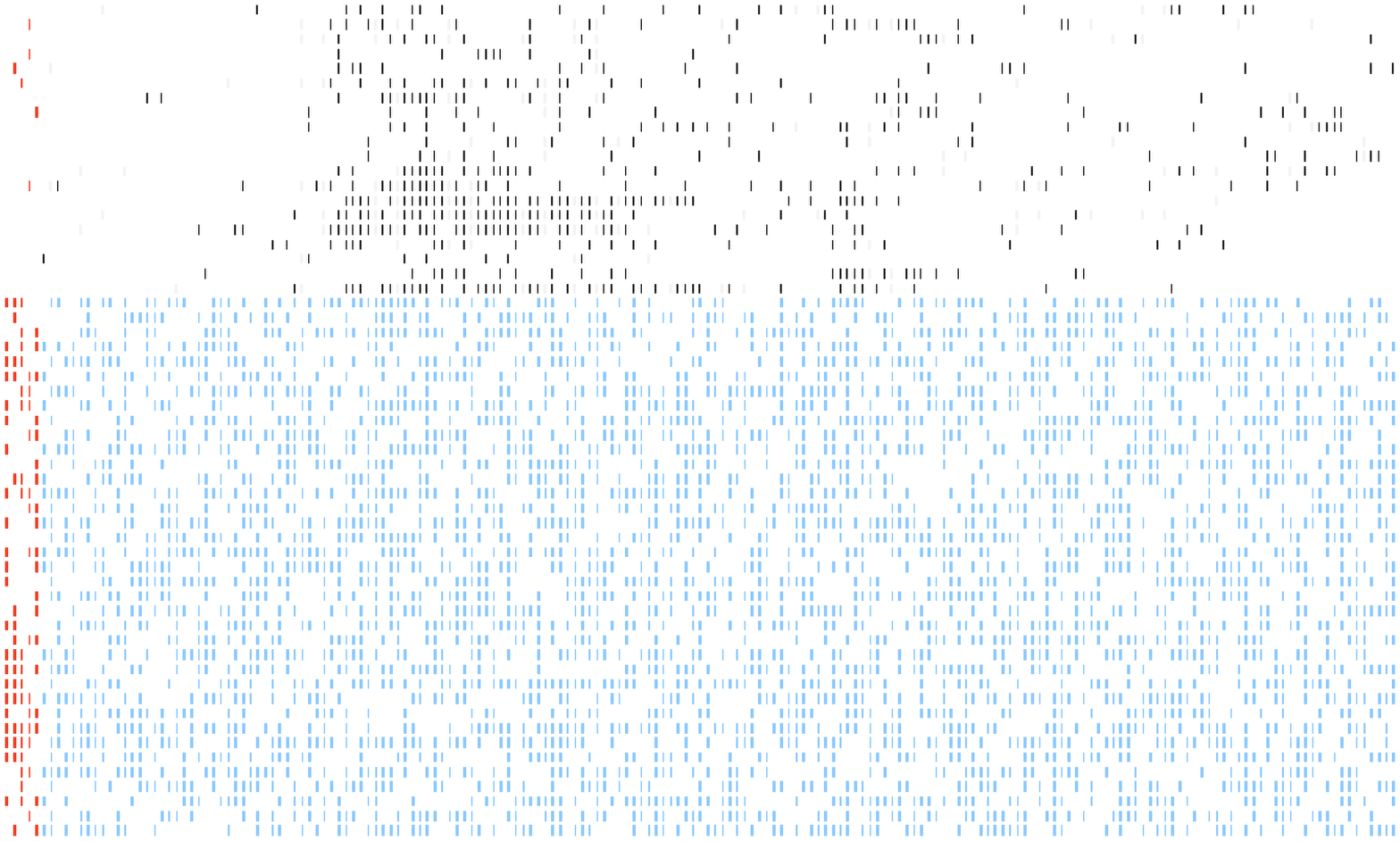}	
\caption{\label{fig:Bio06out} Raster calculated, by the proposed method, from a biological data set, with $N = 20$, $\gamma = 0.95$, $D = 5$ $T = 190$ and $S = 38$.
Color codes are the same as previously. See text for details. From \cite{riehle-etal:00} by the courtesy of the authors.}
\end{center}
\end{figure}

What is interesting here is that we can apply the proposed method on non-stationary rasters, qualitatively very different,
such as a very sparse raster, similar to the one shown in Fig.~\ref{fig:Bio03out}, 
a raster with two activity phases (presently movement preparation and execution) in Fig.~\ref{fig:Bio01out} and
a raster with a rich non-stationary activity  in Fig.~\ref{fig:Bio06out}.
In fact a dozen of such data sets have been tested, with the same result: exact raster reconstruction, with the same hidden unit ratio.

\subsubsection*{On the computation time}

 Since the computation time is exclusively the LP problem resolution computation time we have simply verify that we roughly obtain 
what is generaly observed with this algorithm, i.e. that the computation time order of magnitude is:
\\ \centerline{$O\left(S \, T\right)$} \\
when $N \ll T$, which is the case in our experimenation. On a standard laptop computer, this means a few seconds.

\section{Conclusion}
\label{sec:conclusion}

 Considering a deterministic time-discretized spiking network of neurons with connection weights having delays, we have been able to investigate in details to which extend
it is possible to back-engineer the networks parameters, i.e., the connection weights. 
Contrary to the known NP-hard problem which occurs when weights and delays are to be calculated, the present reformulation, now expressed as a Linear-Programming (LP) problem,
provides an efficient resolution and we have discussed extensively all the potential applications of such a mechanism, 
including regarding what is known as reservoir computing mechanisms, with or without a full connectivity, etc..

 At the simulation level, this is a concrete instantiation of the fact rasters produced by the simple model proposed here,
can produce any rasters produced by more realistic models such as Hodgkin-Huxley, for a finite horizon.  
At a theoretical level, this property is reminiscent of the shadowing lemma of dynamical systems theory \cite{katok-hasselblatt:98},
stating that chaotic orbits produced by a uniformly hyperbolic system can be approached arbitrary close by periodic orbits.

 At the computational level, we are here in front of a method which allows to ``program'' a spiking network, 
i.e. find a set of parameters allowing us to exactly reproduce the network output, given an input.
Obviously, many computational modules where information is related to ``times'' and ``events'' are now easily programmable using the present method.
A step further, if the computational requirement is to both consider ``analog'' and ``event'' computations, the fact that we have studied both the unit analog state 
and the unit event firing back-engineering problems (corresponding to the L and LP problems), tends to show that we could generalize this method to 
networks where both ``times'' and ``values'' have to be taken into account. The present equations are to be slightly adapted, yielding to a LP problem
with both equality and inequality constraints, but the method is there.

 At the modeling level, the fact that we do not only statistically reproduce the spiking output, but reproduce it \textit{exactly},
corresponds to the computational neuroscience paradigm where ``each spike matters'' \cite{guyonneau-vanrullen-etal:04,delorme-perrinet-etal:01}.
The debate is far beyond the scope of the present work, but interesting enough is the fact that, when considering natural images, 
the primary visual cortex activity seems to be very sparse and deterministic, contrary to what happens with unrealistic stimuli \cite{baudot:07}.
This means that it is not a nonsense to address the problem of estimating a raster exactly.

 As far as modeling is concerned, the most important message is in the ``delayed weights design: the key point, in the present context is not to have one weight
or  one weight and delay but several weights at different delays''. We have seen that this increase the computational capability of the network. 
In other words, more than the connection's weight, the connection's profile matters.

 Furthermore, we point out how the related LP adjustment mechanism is distributed and has the same structure as an ``Hebbian'' rule. 
This means that the present learning mechanism corresponds to a local plasticity rule, adapting the unit weights, from only the unit and output spike-times.
It has the same architecture as another spike-time dependent plasticity mechanism. 
However, this is supervised learning mechanisms, whereas usual STDP rules are unsupervised ones, while the rule implementations is entirely different.

To which extends this LP algorithm can teach us something about how other plasticity mechanisms is an interesting perspective of the present work.
Similarly, better understanding the dynamics of the generated networks is another issue to investigate, as pointed out previously.

 We consider the present approach as very preliminary and point out that it must be further investigated at three levels: 
 
 \begin{itemize}
\item[i] {\em optimal number of hidden units:} we have now a clear view of the role of these hidden units, used to span the linear space corresponding to the expected raster,
as detailed in section~\ref{sec:reservoir}. This opens a way, not only to find a correct set of hidden units, but a minimal set of hidden units. 
This problem is in general NP-hard, but efficient heuristics may be found considering greedy algorithms.
We have not further discussed this aspect in this paper, because quite different non trivial algorithms have to be considered, 
with the open question of their practical algorithmic complexity. But this is an ongoing work.
\item[ii] {\em approximate raster matching:} we have seen that, in the deterministic case using, e.g., alignment metric, approximate matching is a much more challenging problem, 
since the distance to minimize are not differentiable, thus not usable without a combinatorial explosion of the search space.
However, if we consider other metric (see \cite{schrauwen:07,cessac-etal:08a} for a review), the situation may be more easy to manage, and this is to be further investigated.
\item[iii] {\em application to unsupervised or reinforcement learning:} though we deliberately have considered, here, the simplest paradigm of supervised learning 
in order to separate the different issues, it is clear that the present mechanism must be studied in a more general setting of, e.g., reinforcement learning
\cite{sutton-barto:98}, for both computational and modeling issues. Since the specification is based on a variational formulation, such a generalization
considering criterion related to other learning paradigms, seems possible to develop.
\end{itemize}

Though we are still far from solving the three issues, the present study is completed in the sense that we not only propose theory and experimentation, 
but a true usable piece of software\footnote{\label{enas} Source code available at {\tt http://enas.gforge.inria.fr}.}.

\begin{acknowledgements}
Partially supported by the ANR MAPS, the MACCAC ARC projects and the CONACYT of Mexico.
\end{acknowledgements}

\bibliographystyle{plain}
\bibliography{string, biblio, odyssee}

\begin{thebibliography}{10}

\bibitem{riehle-etal:00}
A.~Riehle A, F.~Grammont, M.~Diesmann, and S.~Grün.
\newblock Dynamical changes and temporal precision of synchronized spiking
  activity in monkey motor cortex during movement preparation.
\newblock {\em J. Physiol (Paris)}, 94:569--582, 2000.

\bibitem{albers-sprott-2:06}
D.~J. Albers and J.~C. Sprott.
\newblock Routes to chaos in high-dimensional dynamical systems : A qualitative
  numerical study.
\newblock {\em Physica D}, 223:194--207, 2006.

\bibitem{albers-sprott:06}
D.~J. Albers and J.~C. Sprott.
\newblock Structural stability and hyperbolicity violation in high-dimensional
  dynamical systems.
\newblock {\em Non linearity}, 19:1801--1847, 2006.

\bibitem{aronov:03}
Dmitriy Aronov.
\newblock Fast algorithm for the metric-space analysis of simultaneous
  responses of multiple single neurons.
\newblock {\em Journal of Neuroscience Methods}, 124(2), 2003.

\bibitem{baudot:07}
P.~Baudot.
\newblock {\em Nature is the code: high temporal precision and low noise in
  V1}.
\newblock PhD thesis, Univ. Paris 6, 2007.

\bibitem{bixby:92}
Robert~E. Bixby.
\newblock Implementing the simplex method: The initial basis.
\newblock {\em J. on Computing}, 4(3), 1992.

\bibitem{bohte-mozer:07}
S.~M. Bohte and M.~C. Mozer.
\newblock Reducing the variability of neural responses: A computational theory
  of spike-timing-dependent plasticity.
\newblock {\em Neural Computation}, 19(2):371--403, 2007.

\bibitem{cessac:08}
B.~Cessac.
\newblock A discrete time neural network model with spiking neurons. rigorous
  results on the spontaneous dynamics.
\newblock {\em J. Math. Biol.}, 56(3):311--345, 2008.

\bibitem{cessac-etal:08b}
B.~Cessac, H.Rostro-Gonzalez, J.C. Vasquez, and T.~Vi\'eville.
\newblock Statistics of spikes trains, synaptic plasticity and gibbs
  distributions.
\newblock In {\em Neurocomp 2008}, 2008.

\bibitem{cessac-etal:08a}
B.~Cessac, H.Rostro-Gonzalez, J.C. Vasquez, and T.~Vi\'eville.
\newblock To which extend is the "neural code" a metric ?
\newblock In {\em Neurocomp 2008}, 2008.

\bibitem{cessac-rostro-etal:08b}
B.~Cessac, H.~Rostro, J.-C. Vasquez, and T.~Vi{\'e}ville.
\newblock To which extend is the ``neural code'' a metric ?
\newblock In {\em Deuxi{\`e}me conf{\'e}rence fran{\c{c}}aise de Neurosciences
  Computationnelles}, 2008.

\bibitem{cessac-vieville:08b}
B.~Cessac and T.~Vi\'eville.
\newblock Introducing numerical bounds to improve event-based neural network
  simulation.
\newblock {\em Frontiers in neuroscience}, 2008.
\newblock submitted.

\bibitem{cessac-vieville:08}
B.~Cessac and T.~Vi{\'e}ville.
\newblock On dynamics of integrate-and-fire neural networks with adaptive
  conductances.
\newblock {\em Frontiers in neuroscience}, 2(2), jul 2008.

\bibitem{chazottes-et-al:98}
J.R. Chazottes, E.~Floriani, and R.~Lima.
\newblock Relative entropy and identification of gibbs measures in dynamical
  systems.
\newblock {\em J. Statist. Phys.}, 90(3-4):697--725, 1998.

\bibitem{chechik:03}
G.~Chechik.
\newblock Spike-timing-dependent plasticity and relevant mutual information
  maximization.
\newblock {\em Neural Computation}, 15(7):1481--1510, 2003.

\bibitem{cooper-intrator-etal:04}
L.N. Cooper, N.~Intrator, B.S. Blais, and H.Z. Shouval.
\newblock {\em Theory of Cortical Plasticity}.
\newblock World Scientific Publishing, 2004.

\bibitem{darst:90}
Richard~B. Darst.
\newblock {\em Introduction to Linear Programming—Applications and
  Extensions}.
\newblock Marcel Dekker Ltd, New-York, 1990.

\bibitem{delorme-perrinet-etal:01}
A.~Delorme, L.~Perrinet, and S.~Thorpe.
\newblock Network of integrate-and-fire neurons using rank order coding b:
  spike timing dependant plasticity and emergence of orientation selectivity.
\newblock {\em Neurocomputing}, 38:539--545, 2001.

\bibitem{destexhe:97}
Alain Destexhe.
\newblock Conductance-based integrate and fire models.
\newblock {\em Neural Computation}, 9:503--514, 1997.

\bibitem{gantmatcher:77}
F.R. Gantmatcher.
\newblock {\em Matrix Theory}.
\newblock Chelsea, New-York, 1977.

\bibitem{gerstner-kistler:02}
W.~Gerstner and W.~M. Kistler.
\newblock Mathematical formulations of hebbian learning.
\newblock {\em Biological Cybernetics}, 87:404--415, 2002.

\bibitem{guyonneau-vanrullen-etal:04}
R.~Guyonneau, R.~vanRullen, and S.J. Thorpe.
\newblock Neurons tune to the earliest spikes through stdp.
\newblock {\em Neural Computation}, 2004.
\newblock In review.

\bibitem{hornik-etal:89}
K.~Hornik, M.~Stinchcombe, and H.~White.
\newblock Multilayer feedforward networks are universal approximators.
\newblock {\em Neural Networks}, 2:359--366, 1989.

\bibitem{tropp-etal:06}
M.~J.~Strauss J.~A.~Tropp, A. C.~Gilbert.
\newblock Algorithms for simultaneous sparse approximation. part i: Greedy
  pursuit.
\newblock {\em Signal Processing}, 86:572--588, 2006.

\bibitem{jaeger:03}
H.~Jaeger.
\newblock Adaptive nonlinear system identification with {E}cho {S}tate
  {N}etworks.
\newblock In S.~Becker, S.~Thrun, and K.~Obermayer, editors, {\em NIPS*2002,
  Advances in Neural Information Processing Systems}, volume~15, pages
  593--600. MIT Press, 2003.

\bibitem{katok-hasselblatt:98}
A.~Katok and B.~Hasselblatt.
\newblock {\em Introduction to the modern theory of dynamical systems}.
\newblock Kluwer, 1998.

\bibitem{kornprobst-vieville-etal:06}
P.~Kornprobst, T.~Vieville, S.~Chemla, and O.~Rochel.
\newblock Modeling cortical maps with feed-backs.
\newblock In {\em 29th European Conference on Visual Perception}, page~53, aug
  2006.

\bibitem{maass:97}
W.~Maass.
\newblock Fast sigmoidal networks via spiking neurons.
\newblock {\em Neural Computation}, 9:279--304, 1997.

\bibitem{maass:01}
W.~Maass.
\newblock On the relevance of time in neural computation and learning.
\newblock {\em Theoretical Computer Science}, 261:157--178, 2001.
\newblock (extended version of ALT'97, in LNAI 1316:364-384).

\bibitem{maass-natschlager:97}
W.~Maass and T.~Natschlager.
\newblock Networks of spiking neurons can emulate arbitrary hopfield nets in
  temporal coding.
\newblock {\em Neural Systems}, 8(4):355--372, 1997.

\bibitem{maass-etal:02}
W.~Maass, T.~Natschl\"ager, and H.~Markram.
\newblock Real-time computing without stable states: A new framework for neural
  computation based on perturbations.
\newblock {\em Neural Computation}, 14(11):2531--2560, 2002.

\bibitem{maass-bishop:03}
Wolfgang Maass and Christopher~M. Bishop, editors.
\newblock {\em Pulsed Neural Networks}.
\newblock MIT Press, 2003.

\bibitem{paugam-moisy-etal:08}
H\'el\`ene Paugam-Moisy, R\'egis Martinez, and Samy Bengio.
\newblock Delay learning and polychronization for reservoir computing.
\newblock {\em Neurocomputing}, 71:1143--1158, 2008.

\bibitem{rudolph-destexhe:06}
M.~Rudolph and A.~Destexhe.
\newblock Analytical integrate and fire neuron models with conductance-based
  dynamics for event driven simulation strategies.
\newblock {\em Neural Computation}, 18:2146--2210, 2006.

\bibitem{schafer-zimmermann:06}
Anton~Maximilian Sch\"afer and Hans~Georg Zimmermann.
\newblock Recurrent neural networks are universal approximators.
\newblock {\em Lecture Notes in Computer Science}, 4131:632--640, 2006.

\bibitem{schrauwen:07}
Benjamin Schrauwen.
\newblock {\em Towards Applicable Spiking Neural Networks}.
\newblock PhD thesis, Universiteit Gent, Belgium, 2007.

\bibitem{soula-chow:07}
H.~Soula and C.~C. Chow.
\newblock Stochastic dynamics of a finite-size spiking neural networks.
\newblock {\em Neural Computation}, 19:3262--3292, 2007.

\bibitem{strata-harvey:99}
P.~Strata and R.Harvey.
\newblock Dale's principle.
\newblock {\em Brain Res. Bull.}, 50:349—50, 1999.

\bibitem{sutton-barto:98}
Richard~S. Sutton and Andrew~G. Barto.
\newblock {\em Reinforcement Learning: An Introduction}.
\newblock MIT Press, Cambridge, MA, 1998.

\bibitem{toyoizumi-etal:07}
T.~Toyoizumi, J.-P. Pfister, K.~Aihara, and W.~Gerstner.
\newblock Optimality model of unsupervised spike-timing dependent plasticity:
  Synaptic memory and weight distribution.
\newblock {\em Neural Computation}, 19:639--671, 2007.

\bibitem{toyoizumi-etal:05}
Taro Toyoizumi, Jean-Pascal Pfister, Kazuyuki Aihara, and Wulfram Gerstner.
\newblock Generalized bienenstock-cooper-munro rule for spiking neurons that
  maximizes information transmission.
\newblock {\em Proceedings of the National Academy of Science}, 102:5239--5244,
  2005.

\bibitem{tropp:06}
J.~A. Tropp.
\newblock Algorithms for simultaneous sparse approximation. part ii: Convex
  relaxation.
\newblock {\em Sparse approximations in signal processing}, 86:589--602, 2006.

\bibitem{tropp:04a}
Joel~A. Tropp.
\newblock Greed is good: Algorithmic results for sparse approximation.
\newblock {\em IEEE Trans. Inform. Theory}, 50:2231--2242, 2004.

\bibitem{tropp:04b}
Joel~A. Tropp.
\newblock Just relax: Convex programming methods for subset selection and
  sparse approximation.
\newblock Technical report, Texas Institute for Computational Engineering and
  Sciences, 2004.

\bibitem{verstraeten-etal:07}
D.~Verstraeten, B.~Schrauwen, M.~D’Haene, and D.~Stroobandt.
\newblock An experimental unification of reservoir computing methods.
\newblock {\em Neural Networks}, 20(3):391--403, 2007.

\bibitem{victor:05}
J.D. Victor.
\newblock Spike train metrics.
\newblock {\em Current Opinion in Neurobiology}, 15(5):585--592, 2005.

\bibitem{victor-purpura:96}
J.D. Victor and K.P. Purpura.
\newblock Nature and precision of temporal coding in visual cortex: a
  metric-space analysis.
\newblock {\em J Neurophysiol}, 76:1310--1326, 1996.

\bibitem{vieville-lingrand-etal:01}
T.~Vi\'eville, D.~Lingrand, and F.~Gaspard.
\newblock Implementing a multi-model estimation method.
\newblock {\em The International Journal of Computer Vision}, 44(1), 2001.

\bibitem{vieville-rochel:06}
T.~Vi\'eville and O.~Rochel.
\newblock One step towards an abstract view of computation in spiking
  neural-networks.
\newblock In {\em International Conf. on Cognitive and Neural Systems}, 2006.

\bibitem{sima-sgall:05}
Ji\v{r}\'i \v{S}\'ima and Ji\v{r}\'i Sgall.
\newblock On the nonlearnability of a single spiking neuron.
\newblock {\em Neural Computation}, 17(12):2635--2647, 2005.

\end{thebibliography}

\normalsize
\newpage

%%%%%%%%%%%%%%%%%%%%%%%% FIGURES %%%%%%%%%%%%%%%%%%

\begin{figure}[!htbp]
\begin{center}
\includegraphics[width = 0.5\textwidth, height=2cm]{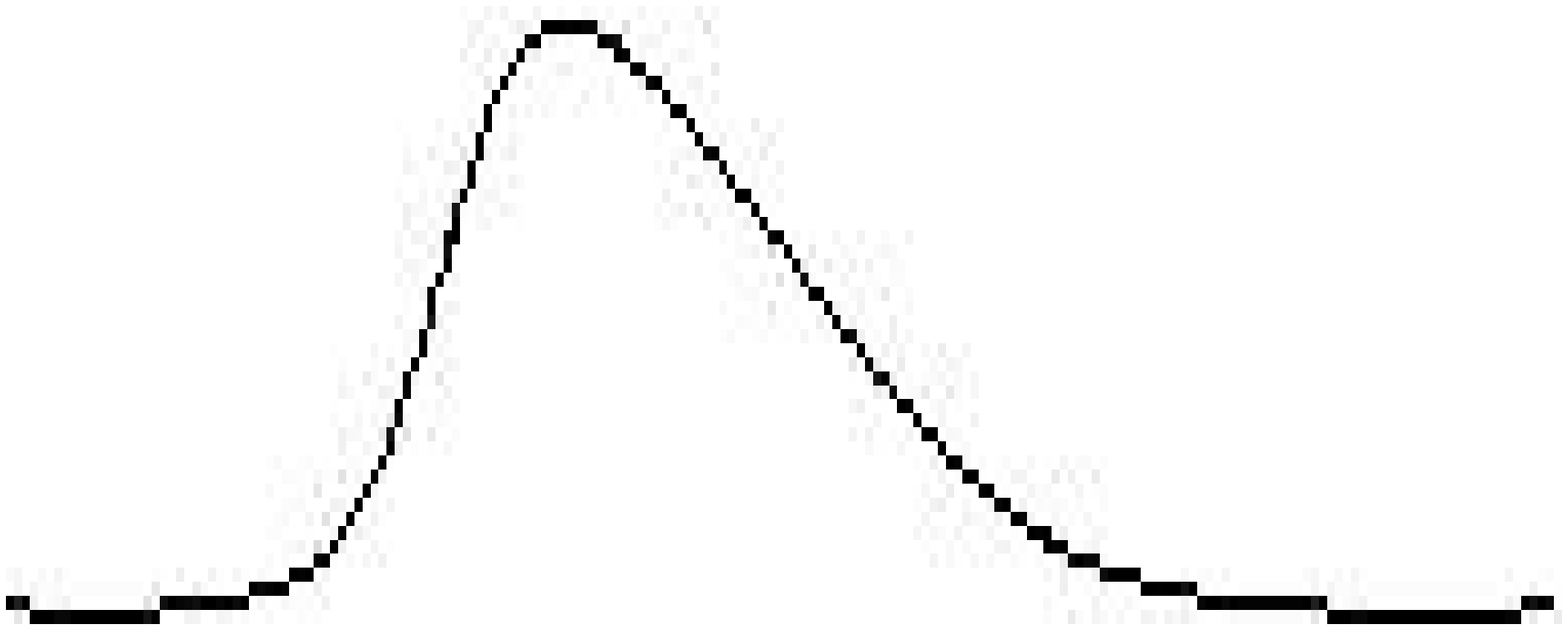}
\label{fig:alp}
\end{center}
\end{figure} 

\clearpage

%%%%%%%%%%%%%%%%%%%%%%%%%%%%

\end{document}